\documentclass[a4paper,11pt]{article}
\pdfoutput=1 
\usepackage{jheppub} 
\usepackage{subfig}
\usepackage[T1]{fontenc} 
\usepackage{graphicx}
\usepackage{epsfig}
\usepackage{axodraw}
\usepackage{amsmath}
\usepackage{amssymb}
\usepackage{float}
\usepackage{placeins}
\usepackage[utf8]{inputenc}
\usepackage{changepage}
\usepackage{bm}

\newcommand{\be}{\begin{equation}}
\newcommand{\ee}{\end{equation}}

\newcommand{\bea}{\begin{eqnarray}}
\newcommand{\eea}{\end{eqnarray}}
\newcommand{\bfig}{\begin{figure}}
\newcommand{\efig}{\end{figure}}
\newcommand{\bc}{\begin{center}}
\newcommand{\ec}{\end{center}}

\newcommand{\bigzero}{\mbox{\normalfont\Large\bfseries 0}}
\newcommand{\bigasterisk}{\mbox{\normalfont\Large\bfseries *}}
\newcommand{\img}[1]{\includegraphics[scale=0.29]{figures/Plots/Plot_#1.pdf}}

\allowdisplaybreaks[4]

\title{\boldmath Evaluating a family of  two-loop non-planar master integrals for Higgs + jet production  with full heavy-quark mass dependence}

\preprint{MPP-2019-157\\
	\phantom{~} \hfill  TCDMATH 19-16}

\author[a,b]{R.~Bonciani,}\author[c,d]{V.~Del Duca,}\author[e,f]{H.~Frellesvig,}\author[g]{J.M.~Henn,}\author[h,i]{M.~Hidding,}
\author[g]{L.~Maestri,}\author[c]{F.~Moriello,}\author[c,j,k]{G.~Salvatori,}\author[l]{V.A.~Smirnov}

\affiliation[a]{Sapienza - Universit\`a di Roma, Dip. di Fisica, Piazzale Aldo Moro 5, 00185, Roma, Italy}

\affiliation[b]{INFN Sezione di Roma, Piazzale Aldo Moro 2, 00185, Roma, Italy}

\affiliation[c]{ETH Zurich, Institut fur theoretische Physik, Wolfgang-Paulistr. 27, 8093, Zurich, Switzerland}

\affiliation[d]{INFN Laboratori Nazionali di Frascati, 00044 Frascati (Roma), Italy}

\affiliation[e]{Dipartimento di Fisica e Astronomia, Universit\`a di Padova, Via Marzolo 8, 35131 Padova, Italy}

\affiliation[f]{INFN, Sezione di Padova, Via Marzolo 8, 35131 Padova, Italy}

\affiliation[g]{Max-Planck-Institut f\"ur Physik, Werner-Heisenberg-Institut, D-80805 M\"unchen, Germany}

\affiliation[h]{Hamilton Mathematics Institute, Trinity College, Dublin 2, Ireland}

\affiliation[i]{School of Mathematics, Trinity College, Dublin 2, Ireland}

\affiliation[j]{Dipartimento di Fisica, Universit\`a degli Studi di Milano, Via Celoria 16, 20133 Milano, Italy}

\affiliation[k]{INFN, Sezione di Milano, Via Celoria 16, 20133 Milano, Italy}

\affiliation[l]{Skobeltsyn Inst. of Nuclear Physics of Moscow State University, 119991 Moscow, Russia}

\emailAdd{roberto.bonciani@roma1.infn.it}
\emailAdd{delducav@itp.phys.ethz.ch}
\emailAdd{hjalte.frellesvig@pd.infn.it}
\emailAdd{henn@mpp.mpg.de}
\emailAdd{hiddingm@tcd.ie}
\emailAdd{maestri@mpp.mpg.de}
\emailAdd{fmoriell@phys.ethz.ch}
\emailAdd{giulio.salvatori@unimi.it}
\emailAdd{smirnov@theory.sinp.msu.ru}

\abstract{We present the analytic computation of a family of non-planar master integrals which contribute to the two-loop scattering amplitudes for Higgs plus one jet production, with full heavy-quark mass dependence. These are relevant for the NNLO corrections to  inclusive Higgs production and for the NLO corrections to Higgs production in association with a jet, in QCD. The computation of the integrals is performed with the method of differential equations. We provide a choice of basis for the polylogarithmic sectors, that puts the system of differential equations in canonical form. Solutions up to weight 2 are provided in terms of logarithms and dilogarithms, and 1-fold integral solutions are provided at weight 3 and 4. There are two elliptic sectors in the family, which are computed by solving their associated set of differential equations in terms of generalized power series. The resulting series may be truncated to obtain numerical results with high precision. The series solution renders the analytic continuation to the physical region straightforward. Moreover, we show how the series expansion method can be used to obtain accurate numerical results for all the master integrals of the family in all kinematic regions.}

\begin{document} 
\maketitle
\flushbottom

\section{Introduction}
\label{sec:intro}
After the discovery of the Higgs boson~\cite{Aad:2012tfa,Chatrchyan:2012xdj} at the Large Hadron Collider (LHC) at CERN,
the LHC physics program has been centered around measuring the properties, couplings and quantum numbers of the Higgs boson,
and looking for footprints of New Physics (NP) effects. A process which may be useful in this respect is the production of a Higgs
boson in association with a jet. At the LHC, the main production mode of the Higgs boson is via gluon-gluon fusion. 
However, the Higgs boson does not couple directly to the gluons, the interaction being mediated by a heavy-quark loop. Thus, leading-order production requires the evaluation of one-loop amplitudes, the next-to-leading order (NLO) QCD corrections involve the evaluation of two-loop amplitudes, and so on. 

In many NP models, the high $p_T$ tail of the Higgs $p_T$ distribution is sensitive to deviations of the Higgs-top
coupling~\cite{Harlander:2013oja,Banfi:2013yoa,Azatov:2013xha,Grojean:2013nya,Schlaffer:2014osa,Buschmann:2014twa,Dawson:2014ora,Buschmann:2014sia,Ghosh:2014wxa,Dawson:2015gka,Langenegger:2015lra,Azatov:2016xik,Grazzini:2016paz,Grazzini:2018eyk,Gorbahn:2019lwq} from the Standard Model (SM) coupling. In the full theory, Higgs production in
association with one jet~\cite{Ellis:1987xu} and the Higgs $p_T$ distribution~\cite{Kauffman:1991jt} are known only at leading order;
at NLO they are known~\cite{Jones:2018hbb} when including only the dependence on the top-quark mass and neglecting the bottom-quark mass.
The computations are made more tractable by using the Higgs Effective Field Theory (HEFT), i.e. by replacing the loop-mediated Higgs-gluon coupling 
by a tree-level effective coupling, which lowers by one loop the amplitudes to be computed. In the HEFT, Higgs production in association with one
jet~\cite{Boughezal:2015dra,Boughezal:2015aha} and the Higgs $p_T$ distribution~\cite{Chen:2016zka} are known at next-to-next-to-leading order 
(NNLO) accuracy. However, for Higgs production in association with one jet or for the Higgs $p_T$ distribution one can show that the HEFT can be applied when the Higgs mass 
is smaller than the heavy-quark mass, $m_H \lesssim m_Q$, and when the jet or Higgs transverse momenta are smaller
than the heavy-quark mass, $p_T \lesssim m_Q$~\cite{Baur:1989cm,deFlorian:1999zd}, by using the leading-order results~\cite{Ellis:1987xu,Kauffman:1991jt} as a benchmark. In the $p_T \lesssim m_Q$ region, the HEFT approximation
may be improved by including the top-bottom interference at NLO, which is estimated by interfering a top-quark loop computed in HEFT 
with a bottom-quark loop computed as an expansion in a small bottom-quark mass~\cite{Lindert:2017pky}.

When the jet or Higgs transverse momenta are of the order or larger than the heavy-quark mass, $p_T \gtrsim m_Q$, the HEFT is not a viable
approximation. In that region, approximate computations at NLO exist~\cite{Lindert:2018iug,Neumann:2018bsx}, based on
the two-loop amplitudes for Higgs plus three partons, computed in the limit $p_T \gg m_Q$~\cite{Kudashkin:2017skd},
and, as outlined above, a numerical NLO computation in the full theory~\cite{Jones:2018hbb},
which includes only the dependence on the top-quark mass.

To this date, no NLO computation of the whole $p_T$ spectrum for Higgs production in association with one jet, where quark-mass effects 
are included for all flavours, is available. In Ref.~\cite{Bonciani:2016qxi} an analytic computation is presented of the planar master integrals which contribute to the two-loop scattering amplitudes for Higgs+jet production, with full heavy-quark mass dependence.
In this paper, we compute analytically one of the two remaining families of non-planar master integrals.

The remainder of the paper is organized as follows. In Section \ref{ref:familiesection} we review the integral families which are necessary to compute the two-loop Higgs+jet amplitudes in QCD with full heavy quark mass dependence, and we set up the notation for the non-planar family of integrals that is computed in this paper. In Section \ref{sec:PolylogarithmicSectors} we present the computation of the master integrals of the polylogarithmic sectors. We briefly review some properties of the canonical basis, and then we discuss how to obtain a minimal-size alphabet for the canonical master integrals. We then discuss how to obtain a region where the canonical basis and the alphabet is manifestly real-valued, and we provide results in this region in terms of logarithms and dilogarithms up to weight 2, and 1-fold integrals over polylogarithms up to weight 4. In Section \ref{sec:ellipticsectors} we describe how (generalized) power series expansions are obtained for the polylogarithmic and the elliptic sectors. Series expansions are applied to obtain results below and above the physical threshold. Lastly in Section \ref{sec:conclusion} we conclude and summarize the results.

\section{The Higgs + jet integral families}
\label{ref:familiesection}
\begin{figure}[!ht]
\centering
\includegraphics[width=2.3cm]{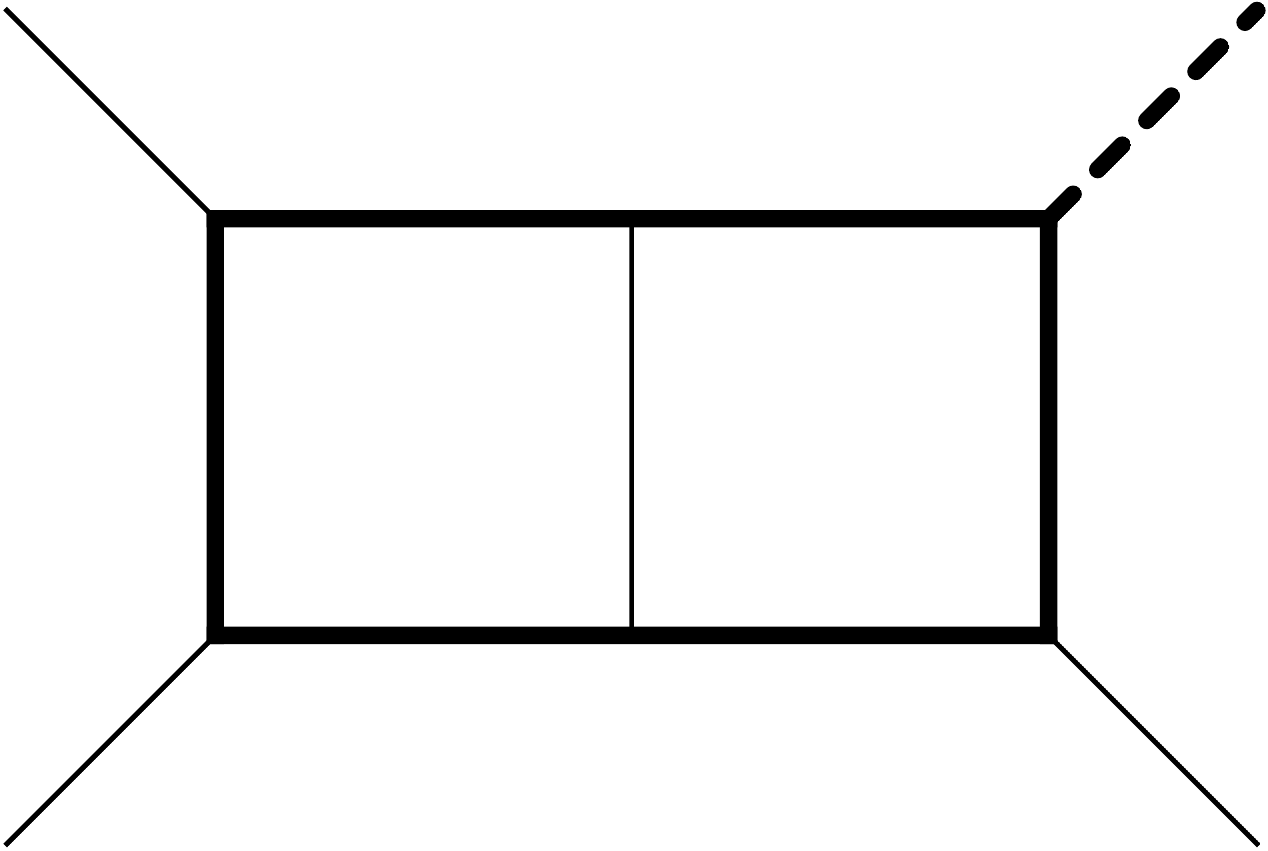} \;
\includegraphics[width=2.3cm]{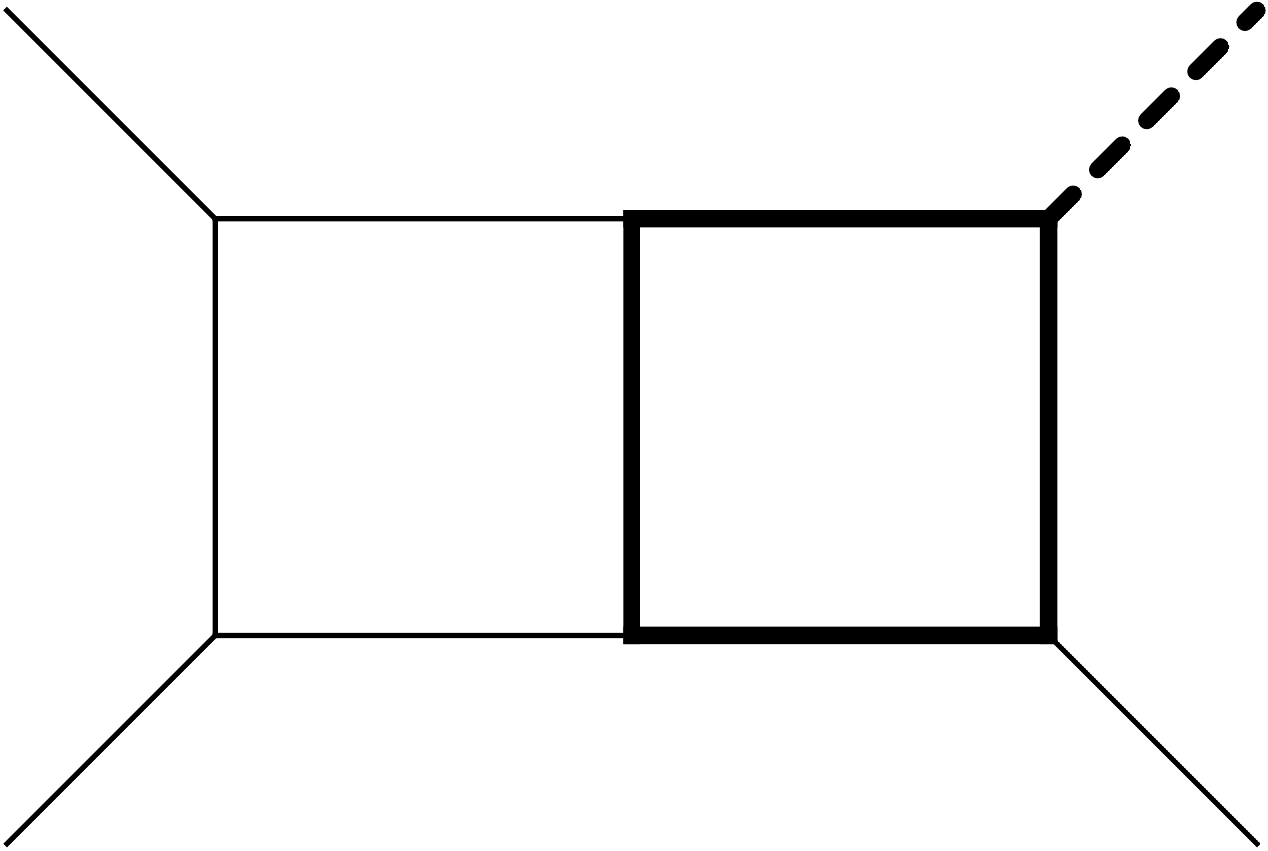} \;
\includegraphics[width=2.3cm]{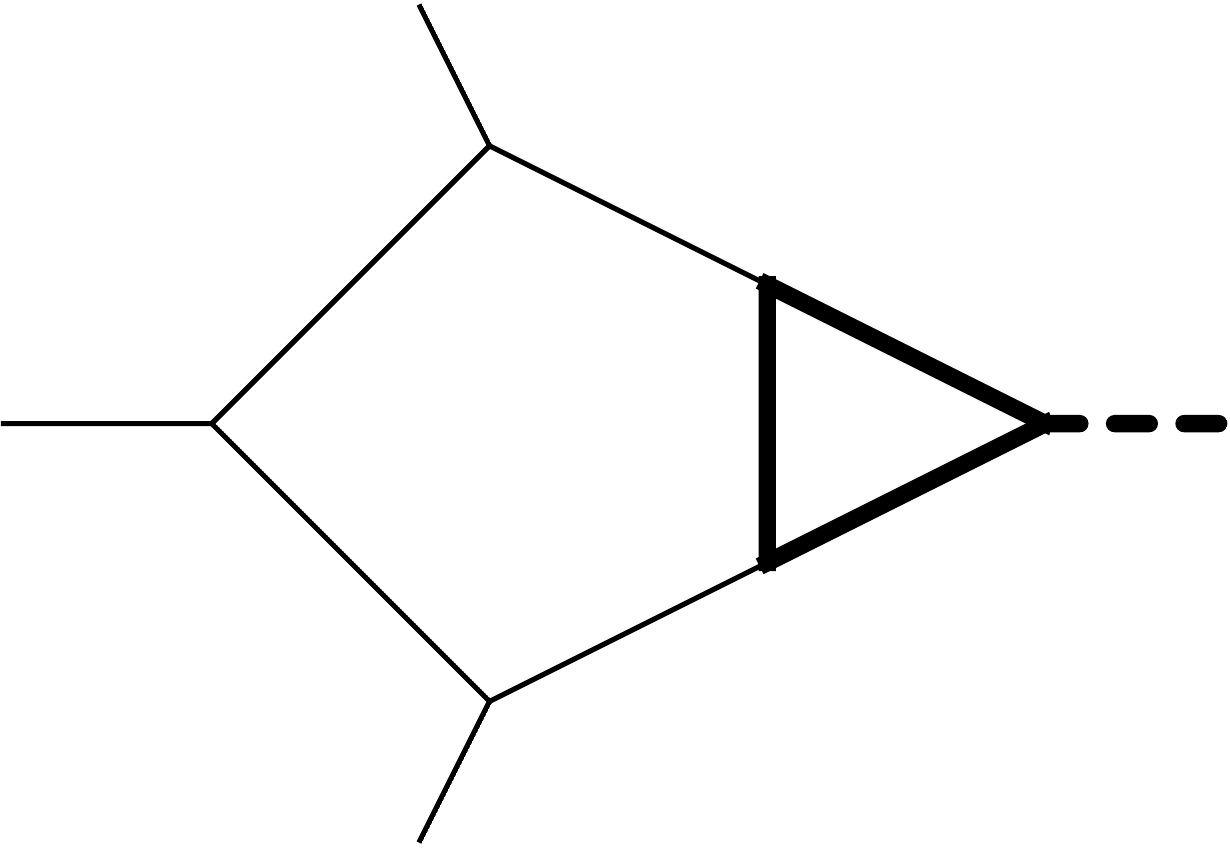} \;
\includegraphics[width=2.3cm]{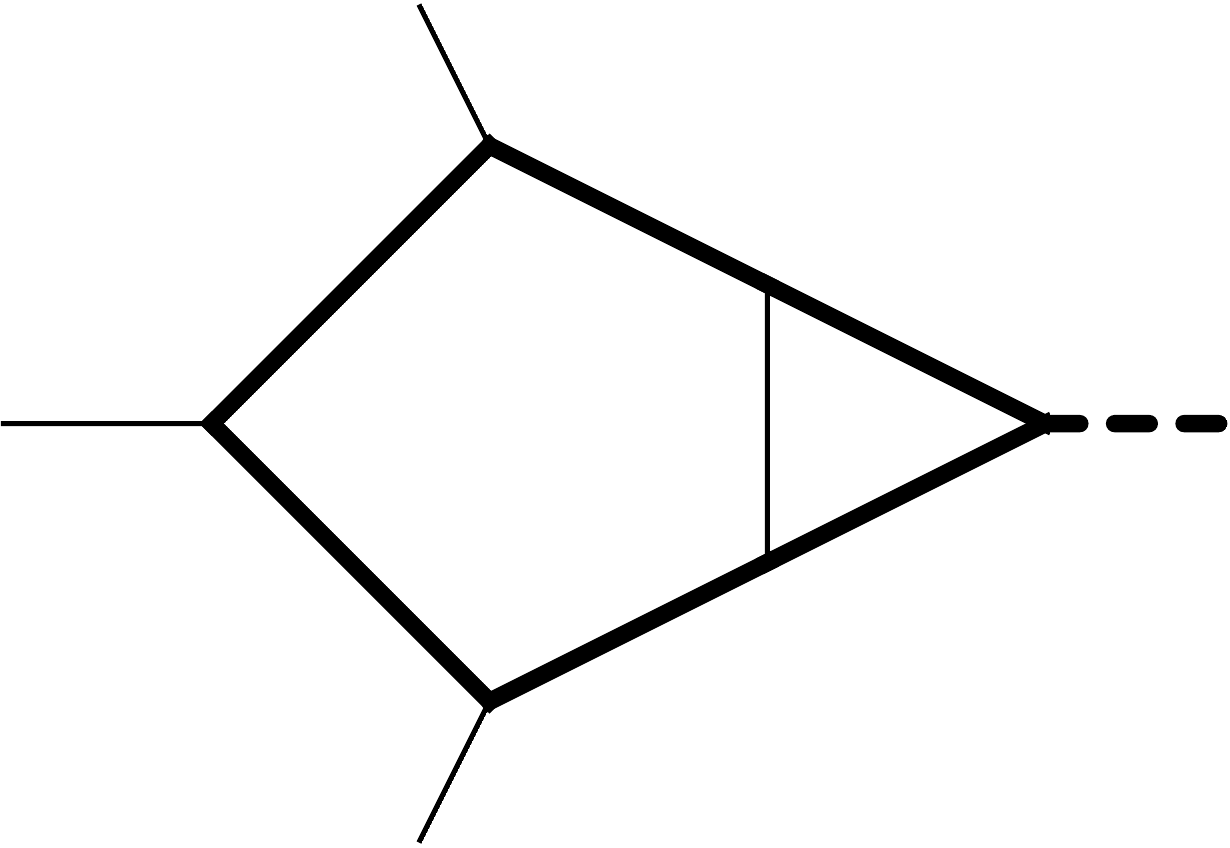} \\
\begin{large}
\vspace{-4mm}
A $\qquad\qquad\quad\,$ B $\qquad\qquad\quad\;\;\,$ C $\qquad\qquad\quad$ D $\!\!\!\!\!\!$ \\
\end{large}
\vspace{3mm}
\includegraphics[width=2.3cm]{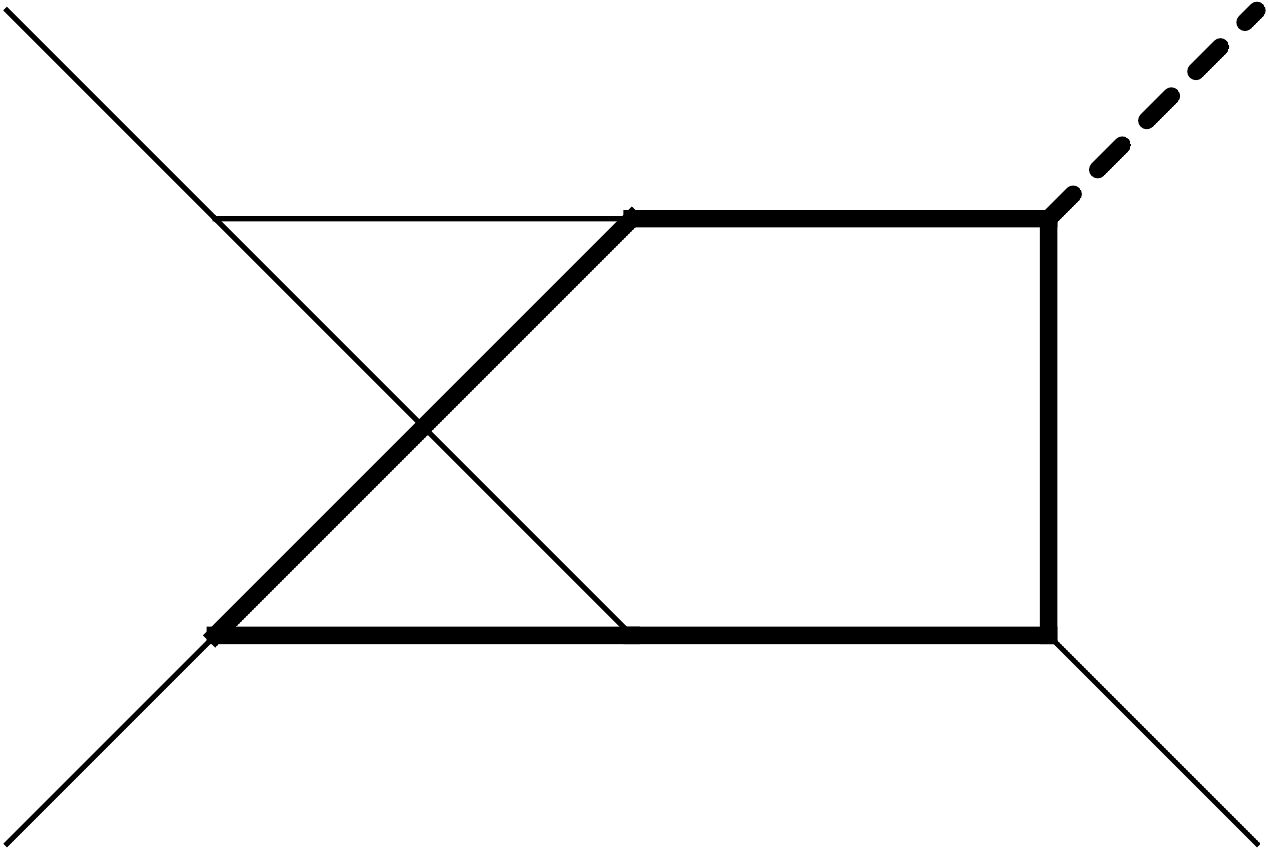} \;
\includegraphics[width=2.3cm]{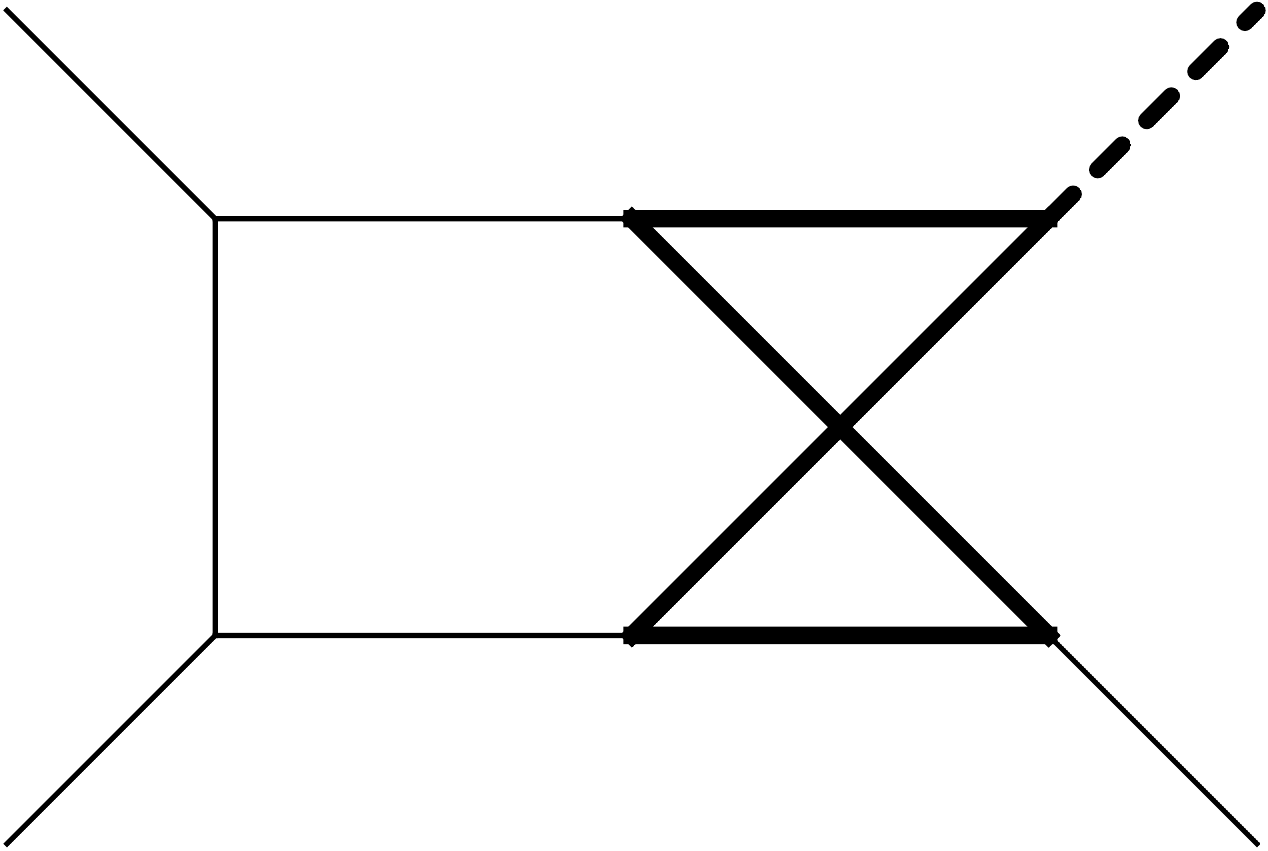} \;
\includegraphics[width=2.3cm]{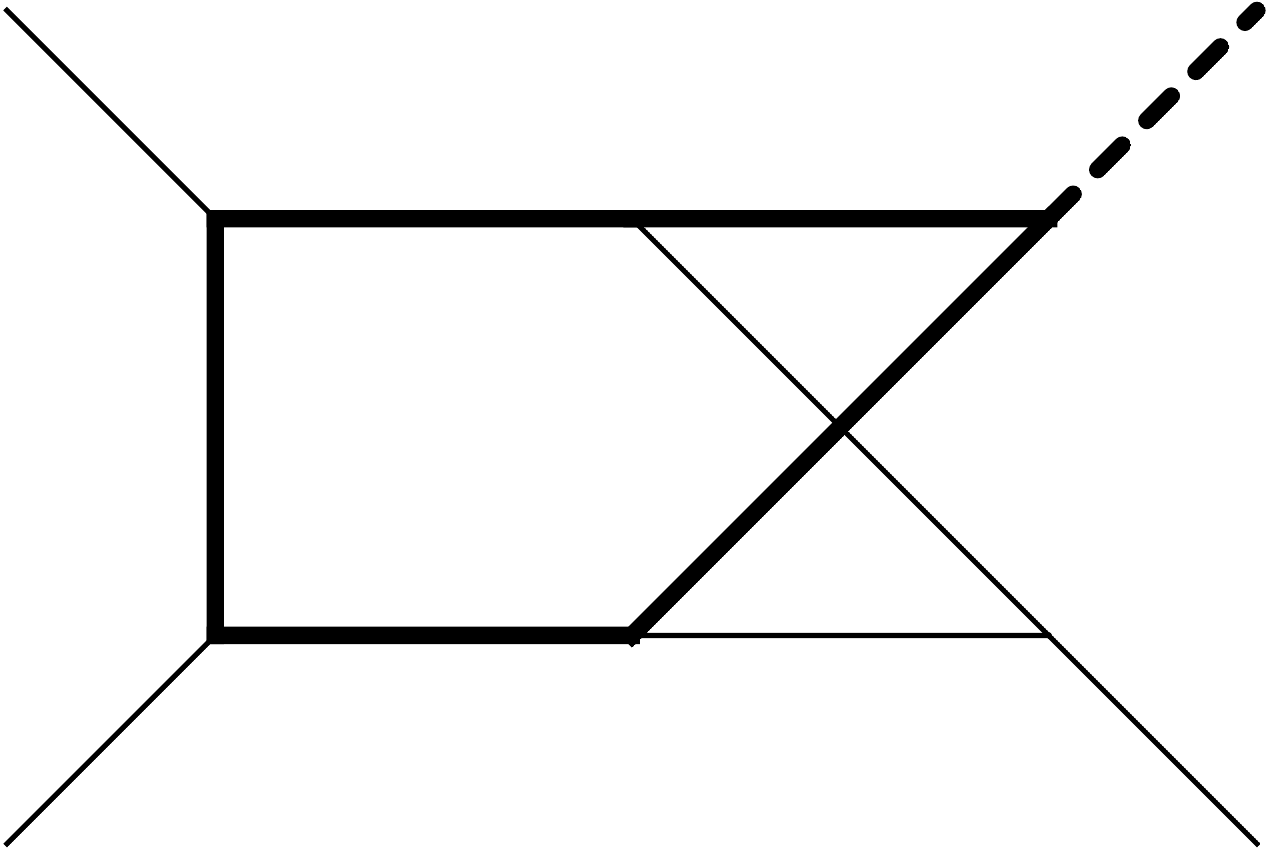} \\
\begin{large}
\vspace{-4mm}
E $\qquad\qquad\quad\,$ F $\qquad\qquad\quad$ G $\!\!\!\!\!$
\end{large}
\caption{The seven integral families contributing to two-loop $H{+}j$-production in QCD.}
\label{fig:families}
\end{figure}

The evaluation of the two-loop QCD contribution to $gg\rightarrow gH$, for instance with the package FeynArts~\cite{Hahn:2000kx}, results in $\mathcal{O}(300)$ Feynman diagrams. After performing the Dirac algebra, these diagrams get expressed in terms of $\mathcal{O}(20000)$ scalar Feynman integrals. Using symmetries and IBP identities, the integrals can all be expressed in terms of a basis of master integrals which are members of one of the seven-propagator integral families depicted on Figure \ref{fig:families}, that we name with consecutive letters from A to G. The quark channels $q \bar{q} \rightarrow gH$, $qg \rightarrow qH$, and $\bar{q}g \rightarrow \bar{q}H$ map into a subset of the same seven families. 

Of these families, the planar families A to D were computed in Ref. \cite{Bonciani:2016qxi}. The non-planar family E turns out to not contribute to the amplitude, as each diagram in that family gets multiplied by the vanishing color-structure,
\begin{align}
f^{ade} \, \text{tr} \big( T^d T^b T^e T^c \big) = 0\,.
\end{align}
Thus, only non-planar families F and G need still to be computed. This paper is devoted to the integrals in family F, while family G is postponed to a future publication.

\subsection{Definition of the integral family}
\label{sec:familydefinitions}

\begin{figure}[h]
\centering
\vspace{1cm}
\includegraphics[width=0.45\textwidth]{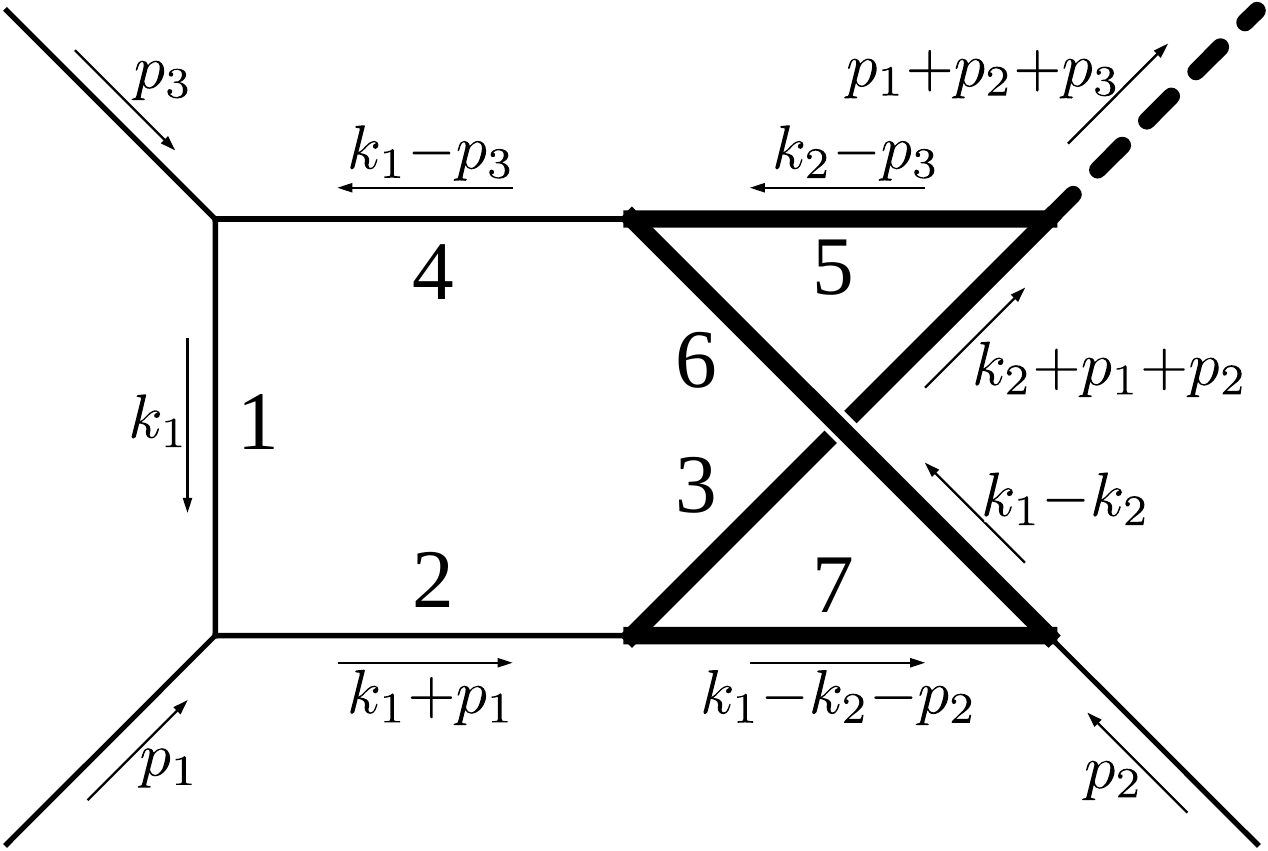}
\caption{The integral family F with momenta and propagator labels.}
\label{fig:familydef}
\end{figure}

The integral family we are considering is depicted in Figure \ref{fig:familydef}, and is given by
\begin{align}
\label{eq:fintdef}
I_{a_1 a_2 a_3 a_4 a_5 a_6 a_7 a_8 a_9} &= e^{2 \gamma_E \epsilon}\int \!\! \int \frac{d^d k_1 d^d k_2}{(i \pi^{d/2})^2} \frac{P_8^{-a_8} \, P_9^{-a_9}}{P_1^{a_1} P_2^{a_2} P_3^{a_3} P_4^{a_4} P_5^{a_5} P_6^{a_6} P_7^{a_7}}\,,
\end{align}
where $\gamma_E=-\Gamma'(1)$ is the Euler-Mascheroni constant, and where
\begin{align}
P_1 &= -k_1^2, & P_4 &= -(k_1{-}p_3)^2, & P_7 &= m^2 - (k_1{-}k_2{-}p_2)^2, \nonumber \\
P_2 &= -(k_1{+}p_1)^2, & P_5 &= m^2 - (k_2{-}p_3)^2, & P_8 &= m^2 - k_2^2, \\
P_3 &= m^2 - (k_2{+}p_1{+}p_2)^2, & P_6 &= m^2 - (k_1{-}k_2)^2, & P_9 &= m^2 - (k_1{-}k_2{-}p_1{-}p_2)^2. \nonumber
\end{align}
In the above, $P_1$-$P_7$ are propagators while $P_8$ and $P_9$ are numerator factors, so $a_8$ and $a_9$ are restricted to non-positive integers. The kinematics is such that $p_1^2 = p_2^2 = p_3^2 = 0$, and additionally,
\begin{align}
s&=(p_1{+}p_2)^2, & t &= (p_1{+}p_3)^2, & u &= (p_2{+}p_3)^2, & p_4^2 &= (p_1{+}p_2{+}p_3)^2 = s{+}t{+}u.
\end{align}
where $p_4^2$ is the squared Higgs-mass, $p_4^2=m_H^2$, and $m^2$ is the squared mass of the massive quark coupling to the Higgs.

The family contains 73 master integrals, defined in Appendix  \ref{app:CanonicalBasis}. It is shown in Appendix \ref{app:ellipticcurveappendix} that the maximal cuts associated to integral sectors 126 and 127 are elliptic integrals. We refer to these sectors as the \emph{elliptic} sectors, while the remaining sectors will be referred to as the \emph{polylogarithmic} sectors. The starting point for solving all master integrals will be the differential equation method. To obtain closed form systems of differential equations for the master integrals, we have to perform IBP-reductions, for which we used the programs FIRE \cite{Smirnov:2008iw, Smirnov:2013dia, Smirnov:2014hma, Smirnov:2019qkx} and Kira \cite{Maierhoefer:2017hyi, Maierhofer:2018gpa}.

\section{The polylogarithmic sectors}
\label{sec:PolylogarithmicSectors}
This section contains the computation of the master integrals of the polylogarithmic sectors, which relies on expressing the integrals in terms of a canonical basis. The integration of the polylogarithmic sectors is performed at weight 2 in terms of manifestly real-valued logarithms and dilogarithms in a sub-region $\mathcal{R}$ of the Euclidean region, where the canonical basis integrals and the alphabet are real-valued. The weight 3 and 4 expressions are then obtained as 1-fold integrals over the weight 2 result.

\subsection{Canonical basis}
\label{sec:PolylogCanonicalBasis}
Suppose that $\vec{\text{I}}$ is a set of master integrals, each of which is given by a Feynman integral times an overall normalization factor that does not depend on the external scales. We may then write down a closed form system of first order linear differential equations,
\begin{align}
    \frac{\partial}{\partial s_i}\vec{\text{I}} = \mathbf{M}_{s_i} \vec{\text{I}}\,,
\end{align}
where $\{s_i\}$ denotes the set of external scales, and where the matrix $\mathbf{M}_{s_i}$ consists of rational functions of the external scales and the dimensional regulator $\epsilon$. We parametrize our kinematics by the external scales $\{s_i\} = \{s,t,m^2,p_4^2\}$ defined in Section \ref{sec:familydefinitions}.

One may perform a change of basis $\vec{B} = \mathbf{T} \vec{\text{I}}$, to find a new system of differential equations,
\begin{align}
    \label{eq:canformpartials1}
    \frac{\partial}{\partial s_i}\vec{B} = \left[\left(\partial_{s_i}\mathbf{T}\right)\mathbf{T}^{-1}+\mathbf{T}\mathbf{M}_{s_i}\mathbf{T}^{-1}\right] \vec{B}\,.
\end{align}
It has been conjectured in Ref. \cite{Henn:2013pwa} that for some matrix $\mathbf{T}$, it is possible to put the system of differential equations in a canonical form. In this form all dependence on the dimensional regulator $\epsilon$ is in an overall prefactor,
\begin{align}
    \label{eq:canformpartials}
    \left(\partial_{s_i}\mathbf{T}\right)\mathbf{T}^{-1}+\mathbf{T}\mathbf{M}_{s_i}\mathbf{T}^{-1} = \epsilon \mathbf{A}_{s_i}\,,
\end{align}
where the matrices $\mathbf{A}_{s_i}$ do not depend on $\epsilon$. In the polylogarithmic case it is conjectured that $\mathbf{T}$ is a matrix of rational or algebraic functions, and that it is possible to combine the partial derivative matrices into a $d\log$-form. The system of differential equations then becomes,
\begin{align}
    d \vec{B} = \epsilon d\tilde{\mathbf{A}} \vec{B}\,,
\end{align}
where the entries of the matrix $\tilde{\mathbf{A}}$ are $\mathbb{Q}$-linear combinations of logarithms of rational or algebraic functions, and where in particular $\partial_{s_i}\tilde{\mathbf{A}} = \mathbf{A}_{s_i}$. The canonical basis of the polylogarithmic sectors consists of 65 integrals, which are given in Appendix \ref{app:CanonicalBasis}.

\subsubsection{Deriving the \texorpdfstring{$\tilde{\mathbf{A}}$}{A}-matrix}
Starting from equations (\ref{eq:canformpartials1}), (\ref{eq:canformpartials}), the matrix $\tilde{\mathbf{A}}$ may be found by defining the following matrices,
\begin{equation}
\begin{aligned}
& \tilde{\mathbf{A}}_1 := \int \mathbf{A}_{s_1} ds_1 \,,\\
& \tilde{\mathbf{A}}_i := \int \bigg( \mathbf{A}_{s_i} - \partial_{s_i} \sum_{j=1}^{i-1} \tilde{\mathbf{A}}_j \bigg) ds_i \,,~~~~ i=2,...,4\,.
\end{aligned}
\label{eq:Atildefrompdes}
\end{equation}
The matrix $\tilde{\mathbf{A}}$ is then obtained as the sum of the matrices $\tilde{\mathbf{A}}_i$. The ordering of the variables can be chosen arbitrarily. Hence, trying this procedure for different orderings might lead to simpler expressions. We benefited from the fact that $\tilde{\mathbf{A}}_i$ should not depend on the variables $s_j$, with $j<i$ and used numerical evaluations on those variables for which the simplification was not explicit in each of the intermediate steps. Further simplifications were done by rationalizing different roots, changing the differential equation matrices in accordance with the corresponding chain rule of the substitution and replacing back the square roots in the end. In this way we obtained $\tilde{\mathbf{A}}$ as a $\mathbb{Q}$-linear combination of logarithms, for which the correctness of all the partial derivatives was checked analytically.

\subsubsection{Simplifying the alphabet}
The form of $\tilde{\mathbf{A}}$ that is found by repeatedly integrating and projecting out components, in the manner of Eq. (\ref{eq:Atildefrompdes}), contains many linearly dependent combinations of logarithms. It is desirable to express these logarithms in terms of a linearly independent set. We will generally use the term \lq letter\rq\, to refer to the arguments of the logarithms, and `alphabet' to a set of such arguments. However, when we speak about linear (in-)dependence of a set of letters we refer to the linear (in-)dependence of the logarithms of the letters. For example, if $\mathcal{A}^{\text{example}}=\{a,b\}$, then we let $\text{Span}_{\mathbb{Q}} \mathcal{A}^{\text{example}} = \text{Span}_{\mathbb{Q}}\{\log a,\log b\}$.

We seek to express $\tilde{\mathbf{A}}$ in terms of a linearly independent alphabet, which is also manifestly symmetric under flipping signs of the square roots. As a first step we enumerate the irreducible factors of the arguments of the logarithms that appear in $\tilde{\mathbf{A}}$. An alphabet consisting of these letters spans the space of logarithms in $\tilde{\mathbf{A}}$, but is still overcomplete. Let us denote this overcomplete set by $\mathcal{A}^{\text{oc}}$. We may obtain a linearly independent basis of $\text{Span}_{\mathbb{Q}}\mathcal{A}^{\text{oc}}$ by starting with an empty set $\emptyset=\mathcal{A}^{\text{idp}}$, and iteratively adding to $\mathcal{A}^{\text{idp}}$ a letter from $\mathcal{A}^{\text{oc}}$ that is independent of the ones already contained in $\mathcal{A}^{\text{idp}}$. After adding a letter to $\mathcal{A}^{\text{idp}}$ we remove from $\mathcal{A}^{\text{oc}}$ the elements that are contained in $\text{Span}_{\mathbb{Q}}\mathcal{A}^{\text{idp}}$. For simplicity we let each choice of new letter be the one that is of smallest (notational) size in $\mathcal{A}^{\text{oc}}$.

After iterating this procedure we are left with a linearly independent alphabet $\mathcal{A}^{\text{idp}}$, such that all elements of $\tilde{\mathbf{A}}$ lie in $\text{Span}_{\mathbb{Q}}\mathcal{A}^{\text{idp}}$. Next we seek to find an alphabet whose letters are manifestly symmetric under flipping signs of their square roots. In particular, we seek to write the algebraic letters in the form,
\begin{align}
    \label{eq:alglettergen}
    \frac{a+\text{Alg}}{a-\text{Alg}} 
\end{align}
where $a$ is some rational function of the external scales, and $\text{Alg}$ denotes a product of square roots, and inverses of square roots. Flipping a sign of a square root in $\text{Alg}$ sends the letter to its reciprocal, and hence the logarithm of the letter to its negative. Note that there is the freedom to combine the square roots in $\text{Alg}$ together, but we prefer to split up the roots in terms of irreducible factors. This can be done unambiguously if we work in a region where each irreducible radicand has a definite sign, which is described in more detail in Section \ref{sec:manifestlyrealregion}.

To find an alphabet $\mathcal{A}^{\text{sym}}$ spanning the elements of $\tilde{\mathbf{A}}$, and in which all algebraic letters are symmetric of the form of Eq. (\ref{eq:alglettergen}), we encountered two cases. The first case consists of letters of the type $\log\left(a+ b\,\text{Alg}\right)$, which for simplicity we denote by $\log\left(a+b\sqrt{c}\right)$, as the following arguments go through in the same way for a combination of square roots or a single square root. One may substitute letters of this type in terms of symmetric ones by considering the following relation, obtained from multiplying by the conjugate,
\begin{align}
	\label{eq:conjlog1}
    \log\left(a\pm b\sqrt{c}\right) = \frac{1}{2}\left[\pm \log\left(\frac{a/b+\sqrt{c}}{a/b-\sqrt{c}}\right) + \log\left(a^2-b^2 c\right)\right]\,\pmod{i\pi}\,.
\end{align}
If the term $a^2-b^2 c$ contains irreducible factors that are linearly independent of the original alphabet, we add these to the alphabet as well.

The second case consists of letters of the form $\left(a+b\sqrt{c}+d\sqrt{e}\right)$, where $\sqrt{c}$ and $\sqrt{e}$ may again denote a combination of square roots. Multiplying by the conjugate with respect to the first square root, i.e. $\left(a-b \sqrt{c}+d \sqrt{e}\right)$, leads to the following relation,
\begin{align}
    \log \left(a+b \sqrt{c}+d \sqrt{e}\right) = \frac{1}{2} \bigg[&\log \Big[\left(a+b \sqrt{c}+d \sqrt{e}\right) \left(a-b \sqrt{c}+d \sqrt{e}\right)\Big]\nonumber\\&+\log \left(\frac{a+b \sqrt{c}+d \sqrt{e}}{a-b \sqrt{c}+d \sqrt{e}}\right)\bigg]\,\pmod{i\pi}\,.
\end{align}
Furthermore, multiplying the fraction by the term $1=\left(a+b \sqrt{c}-d \sqrt{e}\right)/\left(a+b \sqrt{c}-d \sqrt{e}\right)$, composed of the conjugate with respect to the second square root, yields the relation,
\begin{align}
 	\log \left(\frac{a+b \sqrt{c}+d \sqrt{e}}{a-b \sqrt{c}+d \sqrt{e}}\right) &= \log \Big[\left(a+b \sqrt{c}+d \sqrt{e}\right) \left(a+b \sqrt{c}-d \sqrt{e}\right)\Big]\nonumber\\&-\log \Big[\left(a-b \sqrt{c}+d \sqrt{e}\right) \left(a+b \sqrt{c}-d \sqrt{e}\right)\Big]\,\pmod{i\pi}\,.
\end{align}
After expanding the products of conjugate terms, their algebraic dependence is captured in a single term of the sum. For example,
\begin{align}
	\left(a+b \sqrt{c}-d \sqrt{e}\right) \left(a+b \sqrt{c}+d \sqrt{e}\right)=a^2+b^2 c-d^2 e+2 a b \sqrt{c}\,.
\end{align}
Such terms can be dealt with in the same manner as Eq. (\ref{eq:conjlog1}). Putting everything together leads to the final relation,
\begin{align}
\label{eq:conjlog2}
\log \bigg(a+b \sqrt{c}+d \sqrt{e}\bigg)&=\frac{1}{4} \log \left(\frac{a^2+b^2 c-d^2 e+2 a b \sqrt{c}}{a^2+b^2 c-d^2 e-2 a b \sqrt{c}}\right)\nonumber\\&-\frac{1}{4} \log \left(\frac{a^2-b^2 c-d^2 e+2 b d \sqrt{c} \sqrt{e}}{a^2-b^2 c-d^2 e-2 b d \sqrt{c} \sqrt{e}}\right)\nonumber\\&+\frac{1}{4} \log \left(\frac{a^2-b^2 c+d^2 e+2 a d \sqrt{e}}{a^2-b^2 c+d^2 e-2 a d \sqrt{e}}\right)\nonumber\\&+\frac{1}{4} \log \left(a^4-2 a^2 \left(b^2 c+d^2 e\right)+\left(b^2 c-d^2 e\right)^2\right)\,\pmod{i\pi}\,.
\end{align}
It can be verified that the letters on the right-hand side are in fact sufficient to rewrite every term of the form $\log \left(\pm a \pm b \sqrt{c}\pm d \sqrt{e}\right)$, where the plus and minus signs may differ from each other.

At this point the alphabet $\mathcal{A}^{\text{sym}}$ consists of linearly independent letters with manifest symmetry properties under flip of sign of any square root, and the alphabet covers the entries of $\tilde{\mathbf{A}}$, i.e.  $\text{Span}_{\mathbb{Q}}\{\tilde{\mathbf{A}}_{ij}\}\subseteq \text{Span}_{\mathbb{Q}}\mathcal{A}^{\text{sym}}$. The alphabet is still larger than necessary when $\text{Span}_{\mathbb{Q}}\{\tilde{\mathbf{A}}_{ij}\}$ is a proper subspace. In this case some letters only appear in fixed combinations in $\{\tilde{\mathbf{A}}_{ij}\}$.

Indeed, we found at this stage an alphabet $\mathcal{A}^{\text{sym}}$ that contains 75 letters, while the rank of the vector space spanned by the entries of the canonical matrix $\{\tilde{\mathbf{A}}_{ij}\}$ is equal to 69. To reduce the alphabet to 69 independent letters, we sorted $\{\tilde{\mathbf{A}}_{ij}\}$ by (notational) complexity and picked out the first 69 independent entries. We could identify that 12 letters of $\mathcal{A}^{\text{sym}}$ only appear together in pairs of two in these entries. Combining these pairs into 6 letters yields the final alphabet $\mathcal{A}$ which is written out fully in Appendix \ref{app:AlphabetPolylogarithmic}.

Letters $l_{63},l_{64},l_{65},l_{67},l_{68}$ and $l_{69}$ are the result of combining pairs of letters of $\mathcal{A}^{\text{sym}}$. Each pair contains the same square roots, so that the algebraic dependence of the \lq combined\rq\, letters is still symmetric: changing the sign of a square root sends the letter to its reciprocal, and hence changes the overall sign of its logarithm.

\subsection{A manifestly real region}
\label{sec:manifestlyrealregion}
It is convenient to work in a kinematic region in which the integrals are real-valued and free of branch cuts. Such a region can be found for the Feynman integrals in Eq. (\ref{eq:fintdef}) by requiring that their second Symanzik polynomial is positive in the whole integration domain, i.e. $\mathcal{F} > 0$. It is sufficient to consider the scalar integral with maximal number of propagators, which provides the region
\begin{align}
    \mathcal{E}: \;\; m^2>0 \;\; \& \;\; p_4^2<2 m^2 \;\; \& \;\; p_4^2{-}4 m^2<t<0 \;\; \& \;\; {-}2 m^2{+}p_4^2{-}t<s<2 m^2\,,
\end{align}
which we refer to as the Euclidean region. The canonical basis integrals may be complex-valued in the Euclidean region as they are algebraic combinations of Feynman integrals, and the square roots in the prefactors may be evaluated at negative argument. The alphabet also contains these square roots, and we seek to work in a region where the letters are manifestly real-valued as well.

In Appendix \ref{app:CanonicalBasis} we label 15 square roots whose radicands are irreducible polynomials. These roots appear in the canonical basis and the alphabet of the polylogarithmic sectors, and are given explicitly in Eq. (\ref{eq:famFroots}). As the roots only appear in certain fixed combinations, they may be combined together into just 10 roots. We decided to keep them separated. The letters are real-valued when the ratios of roots which they contain are real-valued. To find a region where this is the case for all letters, let us first decompose the phase space into $2^{15}$ regions $\tilde{\mathcal{R}}_{\sigma_1,\ldots,\sigma_{15}}$, depending on the signs of the radicands of the 15 roots in Eq. (\ref{eq:famFroots}),
\begin{align}
    \tilde{\mathcal{R}}_{+,+,\ldots, +}&:\quad -p_4^2 \geq 0 \quad \& \quad -s \geq 0 \quad \& \quad \ldots  \quad \& \quad \left(p_4^2\right)^2 \left(-p_4^2+s+t\right)-4 m^2 s t \geq 0 \nonumber\\
    \tilde{\mathcal{R}}_{-,+,\ldots, +}&:\quad -p_4^2 \leq 0 \quad \& \quad -s \geq 0 \quad \& \quad \ldots  \quad \& \quad \left(p_4^2\right)^2 \left(-p_4^2+s+t\right)-4 m^2 s t \geq 0 \nonumber\\
    \vdots\quad\,&\quad\quad\vdots\nonumber\\
    \tilde{\mathcal{R}}_{-,-,\ldots, -}&:\quad -p_4^2 \leq 0 \quad \& \quad -s \leq 0 \quad \& \quad \ldots  \quad \& \quad \left(p_4^2\right)^2 \left(-p_4^2+s+t\right)-4 m^2 s t \leq 0\,.
\end{align}
Next, we filter out regions where (some of) the letters are complex valued. For example, consider the letter
\begin{align}
    l_{25} = \frac{1+\frac{\sqrt{4 m^2-s}}{\sqrt{-s}}}{1-\frac{\sqrt{4 m^2-s}}{\sqrt{-s}}}\,.
\end{align}
Looking at the ratio of two roots in the letter, we see that the letter is real-valued if and only if
\begin{align}
    (4m^2{-}s\geq 0 \;\; \& \;\; {-}s\geq 0) \quad || \quad (4m^2{-}s\leq 0 \;\; \& \;\; {-}s\leq 0)\,,
\end{align}
since the external scales are real-valued and with any other assignment of signs to the radicands, the ratio evaluates to an imaginary number. Note that this is also the case if the roots in $l_{25}$ had been combined together. Hence, we filter out all the regions with $(\sigma_2,\sigma_7)=(+-)$ or $(\sigma_2,\sigma_7)=(-+)$.

After selecting all regions where the alphabet is manifestly real-valued, we consider their intersection with the Euclidean region. Just one region has a non-empty intersection, which is the one where all the radicands are non-negative: $\mathcal{R} = \tilde{\mathcal{R}}_{+\ldots+}\cap \mathcal{E}$. In this region the canonical basis is real-valued as well. After simplifying, we find that the region is specified by the following constraints,
\begin{align}
    \label{eq:regionR}
    \mathcal{R}:\quad t\leq &-4 m^2\,\&\, s\leq -4 m^2\,\&\, \bigg(\left(s\leq t\,\&\, \frac{4 m^2 (s+t)-s t}{4 m^2}\leq p_4^2\leq \frac{-4 m^2 s+s t+t^2}{t}\right)\,||\,\nonumber\\& \left(t<s\,\&\, \frac{4 m^2 (s+t)-s t}{4 m^2}\leq p_4^2\leq \frac{-4 m^2 t+s^2+s t}{s}\right)\bigg)\,\&\, m^2\geq 0\,.
\end{align}

\subsection{Analytic integration at weight 2}
\label{sec:analyticintegrationweight2}
The solutions of a canonical form system of differential equations may be written, order by order in $\epsilon$, in terms of Chen iterated integrals~\cite{Chen}. Because our family of integrals has multiple square roots that cannot be simultaneously rationalized through a variable change, it is not straightforward to rewrite these iterated integrals in terms of multiple polylogarithms, although such a representation is not necessarily excluded \cite{Heller:2019gkq}. In the following section it is outlined how to perform the integration of the basis integrals at weight 2, in terms of a basis of manifestly real-valued logarithms and classical polylogarithms. The results are provided in the region $\mathcal{R}$ defined in the previous section, where the canonical basis integrals and the letters are real-valued.

It is useful to first consider the symbol \cite{Brown:2009qja, Goncharov.A.B.:2009tja, Goncharov:2010jf} of the canonical basis integrals, and integrate the differential equations up to terms that lie in the kernel of the symbol, which will be referred to as \lq integrating the symbol\rq. Terms in the kernel of the symbols may be fixed afterwards by imposing that our solutions satisfy the system of differential equations, and by fixing overall transcendental constant from boundary conditions. Let us expand the basis integrals in $\epsilon$ as $\vec{B} = \sum_{k=0}^\infty B^{(k)} \epsilon^k$. The symbol of the $i$-th basis integral at order $k$ may then be obtained from the following recursive formula
\begin{align}
    \mathcal{S}\left(B_i^{(k)}\right) = \sum_j \mathcal{S}\left(B_j^{(k-1)}\right) \otimes \tilde{\mathbf{A}}_{ij}\,.
\end{align}
Note that the leading order of the canonical integrals $\vec{B}^{(0)}$ is constant and hence equal to the leading order of the boundary term given in Eq. (\ref{eq:boundarytermspolylogarithmic}). 

Consider next the weight 2 integration of the symbol. At weight 2, i.e. order $\epsilon^2$, some canonical integrals are identically zero. Furthermore, the symbols of the remaining 40 nonzero integrals can be expressed in terms of the symbols of basis integrals
\begin{align}
    \label{eq:independentsymbolsweight2}
    \{1,2,3,4,5,6,7,8,9,10,11,12,16,20,23,26,33,35,38,40,49\}\,.
\end{align}
By considering permutations of $p_1,p_2,p_3$ the number of independent symbols can be reduced even further, decreasing the number of integrals that need to be studied. We aim to integrate the symbol by writing a sufficiently general ansatz of logarithms and dilogarithms with undetermined prefactors, in the spirit of Ref. \cite{Duhr:2011zq}. We then equate the symbol of the ansatz with the symbol of the individual master integrals and solve the resulting linear system. This can be done unambiguously since we express the symbol in terms of the linearly independent alphabet of Appendix \ref{app:AlphabetPolylogarithmic}. We pick the basis of logarithms and dilogarithms in the ansatz such that they are manifestly real-valued in the region $\mathcal{R}$. Thus we require the arguments of the dilogarithms to lie in the range $(-\infty,1]$ for all of $\mathcal{R}$, while we require the arguments of the logarithms to be positive. This also guarantees that no branch-cuts will be crossed, such that the final expressions will be valid in at least the region $\mathcal{R}$. If one moves outside of $\mathcal{R}$ the basis functions may cross spurious branch cuts, which leads to incorrect results.

We denote the set of letters that appear at weight 2 by $\mathcal{A}_2$. In the region $\mathcal{R}$ the signs of these letters are completely fixed in the following way,
\begin{align}
\begin{array}{rrrrrrrr}
 l_1>0\,, & l_2<0\,, & l_3<0\,, & l_4<0\,, & l_5<0\,, & l_6<0\,, & l_7<0\,, & l_8<0\,,  \\
 l_9>0\,, & l_{10}>0\,, & l_{11}>0\,, & l_{13}>0\,, & l_{25}<0\,, & l_{26}<0\,, & l_{27}<0\,, & l_{28}>0\,,  \\
 l_{29}<0\,, & l_{39}>0\,, & l_{40}>0\,, & l_{43}>0\,, & l_{44}>0\,, & l_{46}>0\,, & l_{48}>0\,, & l_{49}>0\,,  \\
 l_{53}>0\,, & l_{54}>0\,, & l_{55}>0\,, & l_{56}>0\,, & l_{60}>0\,, & l_{61}>0\,.
\end{array}
\end{align}
For the logarithmic terms in the ansatz we therefore consider products of the type $\log(\pm l_i)\log(\pm l_j)$ with $l_i,l_j \in \mathcal{A}_2$, where a minus may be included to fix a positive sign for the argument. Furthermore, we include dilogarithms with the following arguments in the ansatz,
\begin{align}
    \label{eq:dilogansatz}
    \text{Li}_2\left(\pm l_i l_j\right), \text{Li}_2\left(\pm \frac{l_i}{l_j}\right), \text{Li}_2\left(\pm \frac{1}{l_i l_j}\right)\quad \text{for }l_i,l_j \in \mathcal{A}_2 \cup \{l_{33}, l_{38},l_{41}\}\,,
\end{align}
where we filter out dilogarithms whose argument does not lie between $(-\infty,1]$ in the region $\mathcal{R}$. We included the spurious letters $l_{33}, l_{38}$ and $l_{41}$ in the ansatz, which do not appear in the symbol at weight 2, but are necessary for the ansatz to be sufficiently general. We identified these letters by using direct integration methods, outlined at the end of this section. Without knowledge of these spurious letters, we could have proceeded with an ansatz that includes all letters in $\mathcal{A}$.

After equating the symbol of the ansatz with the symbol of the canonical integrals, and solving the resulting system of equations, the following products of logarithms survive,
\begin{align}
    \begin{array}{llll}
     \log ^2\left(l_1\right)\,, & \log ^2\left(-l_4\right)\,, & \log ^2\left(-l_{25}\right)\,, & \log ^2\left(-l_{26}\right)\,, \\
     \log ^2\left(-l_{27}\right)\,, & \log ^2\left(l_{28}\right)\,, & \log \left(l_1\right) \log \left(-l_4\right)\,, & \log \left(-l_3\right) \log \left(-l_{25}\right)\,, \\
     \log \left(-l_4\right) \log \left(-l_{25}\right)\,, & \log \left(-l_4\right) \log \left(-l_{26}\right)\,, & \log \left(-l_2\right) \log \left(-l_{27}\right)\,, & \log \left(-l_5\right) \log \left(-l_{27}\right)\,, \\
     \log \left(-l_7\right) \log \left(-l_{27}\right)\,, & \log \left(-l_8\right) \log \left(-l_{27}\right)\,, & \log \left(-l_{25}\right) \log \left(-l_{27}\right)\,, & \log \left(-l_4\right) \log \left(l_{28}\right)\,, \\
     \log \left(l_9\right) \log \left(l_{28}\right)\,, & \log \left(-l_{27}\right) \log \left(l_{28}\right)\,, & \log \left(-l_{25}\right) \log \left(l_{43}\right)\,, & \log \left(-l_{26}\right) \log \left(l_{44}\right)\,, \\
     \log \left(l_{28}\right) \log \left(l_{48}\right)\,, & \log \left(l_{28}\right) \log \left(l_{55}\right)\,, & \log \left(-l_{26}\right) \log \left(l_{56}\right)\,, & \log \left(-l_{27}\right) \log \left(l_{60}\right)\,, \\
     \log \left(-l_{27}\right) \log \left(l_{61}\right)\,.
    \end{array}
\end{align}
Furthermore, the following dilogarithms are contained in the final result,
\begin{align}
\begin{array}{lllll}
 \text{Li}_2\left(\frac{1}{l_{25}}\right)\,, & \text{Li}_2\left(-\frac{1}{l_{25}}\right)\,, & \text{Li}_2\left(\frac{1}{l_{26}}\right)\,, & \text{Li}_2\left(-\frac{1}{l_{26}}\right)\,, & \text{Li}_2\left(-\frac{1}{l_{27}}\right)\,, \\
 \text{Li}_2\left(\frac{1}{l_{27}}\right)\,, & \text{Li}_2\left(\frac{1}{l_{25} l_{27}}\right)\,, & \text{Li}_2\left(\frac{l_{25}}{l_{27}}\right)\,, & \text{Li}_2\left(\frac{1}{l_{26} l_{27}}\right)\,, & \text{Li}_2\left(\frac{l_{26}}{l_{27}}\right)\,, \\
 \text{Li}_2\left(\frac{1}{l_{28}}\right)\,, & \text{Li}_2\left(-\frac{1}{l_{28}}\right)\,, & \text{Li}_2\left(-\frac{1}{l_{27} l_{28}}\right)\,, & \text{Li}_2\left(-\frac{l_{28}}{l_{27}}\right)\,, & \text{Li}_2\left(\frac{1}{l_{27} l_{29}}\right)\,, \\
 \text{Li}_2\left(\frac{l_{29}}{l_{27}}\right)\,, & \text{Li}_2\left(-\frac{1}{l_{33}}\right)\,, & \text{Li}_2\left(\frac{1}{l_{25} l_{33}}\right)\,, & \text{Li}_2\left(\frac{l_{25}}{l_{33}}\right)\,, & \text{Li}_2\left(\frac{1}{l_{26} l_{33}}\right)\,, \\
 \text{Li}_2\left(\frac{l_{26}}{l_{33}}\right)\,, & \text{Li}_2\left(\frac{1}{l_{27} l_{33}}\right)\,, & \text{Li}_2\left(\frac{l_{27}}{l_{33}}\right)\,, & \text{Li}_2\left(-\frac{1}{l_{38}}\right)\,, & \text{Li}_2\left(\frac{1}{l_{25} l_{38}}\right)\,, \\
 \text{Li}_2\left(\frac{l_{25}}{l_{38}}\right)\,, & \text{Li}_2\left(\frac{1}{l_{27} l_{38}}\right)\,, & \text{Li}_2\left(\frac{l_{27}}{l_{38}}\right)\,, & \text{Li}_2\left(-\frac{1}{l_{28} l_{38}}\right)\,, & \text{Li}_2\left(-\frac{l_{28}}{l_{38}}\right)\,, \\
 \text{Li}_2\left(-\frac{1}{l_{41}}\right)\,, & \text{Li}_2\left(\frac{1}{l_{26} l_{41}}\right)\,, & \text{Li}_2\left(\frac{l_{26}}{l_{41}}\right)\,, & \text{Li}_2\left(\frac{1}{l_{27} l_{41}}\right)\,, & \text{Li}_2\left(\frac{l_{27}}{l_{41}}\right)\,, \\
 \text{Li}_2\left(-\frac{1}{l_{28} l_{41}}\right)\,, & \text{Li}_2\left(-\frac{l_{28}}{l_{41}}\right)\,. 
\end{array}
\end{align}
It remains to fix terms that lie in the kernel of the symbol. At weight 2 these are transcendental constants and terms that have the form of a transcendental number times a logarithm, such as $i\pi\log(l_i)$ or $\log(2)\log(l_i)$. Since the canonical basis is real-valued in $\mathcal{R}$, it is already guaranteed there will be no contributions of the form $i \pi \log(\ldots)$. Furthermore, if terms of the type $\log(2)\log(l_i)$ were missing in the solutions that we determined at the symbol level, they would not satisfy the system of differential equations. However, we find that our solutions already satisfy the differential equations without adding such terms, so that only additive transcendental constants are left undetermined. Note that to check if our weight 2 solutions satisfy the differential equations, we first need to derive the basis integrals at weight 1, which can be done in a similar manner as described here for the weight 2 case.

Next, we fix the overall transcendental constants from boundary conditions. It is not directly possible to use the heavy mass limit for this purpose, derived in Section \ref{sec:boundaryterms}, as it lies outside of region $\mathcal{R}$ where our solution is valid. Using the series solution strategy described in Section \ref{sec:ellipticsectors}, we can transport the heavy mass limit up to high numerical precision to a regular point in $\mathcal{R}$, and use this to fix the remaining constants in our solution. We find in this way that some integrals carry additive constants proportional to $\pi^2$. The final result is provided in Appendix \ref{app:PolylogsWeight2}.

Lastly, we note that the weight 2 integration of the symbol may also be performed by direct integration methods. Firstly, note that the most complicated Feynman integrals $I_{a_1,a_2,\ldots}$ appearing in Eq. (\ref{eq:independentsymbolsweight2}) have 5 propagators but at weight 2 we only need to compute them at order $1/\epsilon$ because they carry a prefactor proportional to $\epsilon^3$ in the canonical basis. Setting up the Feynman parametrization for these integrals, and regularizing them using analytic regularization \cite{Panzer:2014gra}, we find that they are linearly reducible \cite{Brown:2009ta, Panzer:2015ida} at order $1/\epsilon$ and can be computed algorithmically by direct integration, for example using HyperInt \cite{Panzer:2014caa}. All other Feynman integrals appearing in the basis elements of Eq. (\ref{eq:independentsymbolsweight2}) may also be computed using direct integration up to the required order in $\epsilon$. The resulting solutions are then given in terms of Goncharov multiple polylogarithms \cite{Goncharov:1998kja}. One may use the Mathematica package of Ref. \cite{Frellesvig:2016ske}, to convert these Goncharov polylogarithms to classical polylogarithms. The resulting expressions are generally larger than those obtained from a suitable ansatz, but they can be used to identify the spurious letters needed for the ansatz, such as the letters $l_{33}, l_{38}$ and $l_{41}$ in Eq. (\ref{eq:dilogansatz}).

\subsection{One-fold integrals for weights 3 and 4}
\label{sec:analyticintegrationweight34}
At weights 3 and 4, the polylogarithmic sectors have not been integrated out analytically. For those weights,
we use the method of Ref. \cite{Caron-Huot:2014lda} (see also Appendix E of \cite{Bonciani:2016qxi}) to represent the result in terms of one-fold integrals. The method works as follows. Firstly, it is clear that
\begin{align}
    \label{eq:1foldweight3}
    \int_\gamma d \vec{B} = \vec{B}(\gamma(1)) - \vec{B}(\gamma(0))= \epsilon \int_\gamma d\tilde{\mathbf{A}} \vec{B}\,,
\end{align}
where $\gamma: [0,1] \rightarrow \mathbb{C}^4$ is some path in the phase space of $(s,t,m^2,p_4^2)$. Order by order in $\epsilon$ we may write
\begin{align}
    \label{eq:1foldpolylog1}
    \vec{B}^{(i)}(\gamma(1)) = \int_\gamma d\tilde{\mathbf{A}} \vec{B}^{(i-1)} + \vec{B}^{(i)}(\gamma(0))\,.
\end{align}
Since the polylogarithmic sectors were integrated up to weight 2, we directly obtain the weight 3 expression as a one-fold integral over the weight 2 expression. By performing integration by parts, it is also possible to write the $i$-th order as a one-fold integral over the $(i-2)$-th order result,
\begin{align}
    \label{eq:1foldw4}
    \vec{B}^{(i)}(\gamma(1)) &=  \left[ \tilde{\mathbf{A}} \vec{B}^{(i-1)}\right]_{\gamma(0)}^{\gamma(1)} - \int_\gamma \tilde{\mathbf{A}} d\vec{B}^{(i-1)} + \vec{B}^{(i)}(\gamma(0))\,,\nonumber\\
    &=  \int_\gamma \left(\tilde{\mathbf{A}}(\gamma(1)) d\tilde{\mathbf{A}}- \tilde{\mathbf{A}}  d\tilde{\mathbf{A}} \right)\vec{B}^{(i-2)} + [\tilde{\mathbf{A}}]_{\gamma(0)}^{\gamma(1)}  \vec{B}^{(i-1)}(\gamma(0)) + \vec{B}^{(i)}(\gamma(0))\,,
\end{align}
where $[\tilde{\mathbf{A}}]_{\gamma(0)}^{\gamma(1)} = \tilde{\mathbf{A}}(\gamma(1)) - \tilde{\mathbf{A}}(\gamma(0))$. In this way one may obtain the weight 4 result as a one-fold integral, without needing an analytic expression of the basis integrals at weight 3.

For the base point of the integration we pick the point
\begin{align}
    (s,t,m^2,p_4^2) = (-4,-4,1,-12) = r^{\mathcal{R}}\,,
\end{align}
at which we may obtain high precision results for the canonical integrals using the series solution strategy of Section \ref{sec:ellipticsectors}. We note that the diagonal entries of $\tilde{\mathbf{A}}$ at positions $38,40,42,45,$ and $49$ are divergent in the point $r^{\mathcal{R}}$. However, we may read off from the canonical basis that these integrals are identically zero in $r^{\mathcal{R}}$, so that the term $[\tilde{\mathbf{A}}]_{\gamma(0)}^{\gamma(1)}  \vec{B}^{(i-1)}(\gamma(0))$ in Eq. (\ref{eq:1foldw4}) is well-defined.

Lastly, let us explicitly define a path $\gamma$ with basepoint $r^{\mathcal{R}}$, some endpoint in $\mathcal{R}$, and which lies fully inside region $\mathcal{R}$. A simple choice would be a straight line, but $\mathcal{R}$ is not convex. To ensure that $\gamma$ lies fully in $\mathcal{R}$, we consider a path that moves along a straight line in the $s,t$ and $m^2$ direction, but averages the $p_4^2$-coordinate between the upper and lower bounds in Eq. (\ref{eq:regionR}). We work this out in detail next. 

Let us assume that $s\leq t$, so that the bounds on $p_4^2$ in region $\mathcal{R}$ are given by
\begin{align}
    p_{\text{down}}^2 \equiv \frac{4 m^2 (s+t)-s t}{4 m^2}\leq p_4^2\leq \frac{-4 m^2 s+s t+t^2}{t} \equiv p_{\text{up}}^2\,.
\end{align}
Defining a path for $s\geq t$ can be done in the same manner, if we change to the corresponding upper bound. Let us assume the endpoint of the path $\gamma$ is given by $(s',t',m^{\prime 2},p_4^{\prime 2}) \in \mathcal{R}$. Next, write $p_4^{\prime 2}$ as a combination of the upper and lower bounds evaluated at $(s', t', m^{\prime 2})$,
\begin{align}
    p_4^{\prime 2} = y p_{\text{up}}^{\prime 2}  + (1-y) p_{\text{down}}^{\prime 2} \,,
\end{align}
which yields
\begin{align}
    y = \frac{t' \left(4 m^{\prime 2} (s'+t'-p_4^{\prime 2})-s' t'\right)}{s' \left(16 m^{\prime 4}-(t')^2\right)}\,.
\end{align}
Next, consider the following straight line path in the phase space of $(s,t,m^2)$,
\begin{align}
    \gamma^{(3)}: \lambda \mapsto \lambda(s',t',m^{\prime 2})+(1-\lambda)(-4,-4,1)\,.
\end{align}
We may then define the integration path by
\begin{align}
    \gamma&: \lambda \mapsto \left(\gamma^{(3)}(\lambda), y p_{\text{up}}^2(\gamma^{(3)}(\lambda)) + (1-y)p_{\text{down}}^2(\gamma^{(3)}(\lambda))\right)\,,
\end{align}
where $p_{\text{up}}^2(\gamma^{(3)}(\lambda))$ indicates the upper bound of $p_4^2$ evaluated at the point given by $\gamma^{(3)}(\lambda)$, and similarly for $p_{\text{down}}^2(\gamma^{(3)}(\lambda))$. By construction $\gamma$ has the endpoint $(s',t',m^{\prime 2},p_4^{\prime 2})$ at $\lambda = 1$, and lies inside $\mathcal{R}$ for all $\lambda \in [0,1]$.

\subsection{Series solutions}
\label{sec:seriessolutionspolylog}
In Section \ref{sec:seriessolutionsdeqns} a strategy is outlined to obtain series solutions (truncated at high order) for Feynman integrals from their system of differential equations. Here we briefly point out that the method can be straightforwardly applied to compute the polylogarithmic sectors. Furthermore, the strategy can be used to obtain results in the physical region.


\section{The elliptic sectors}
\label{sec:ellipticsectors}
There is a lot of recent progress in the theory of Feynman integrals evaluating to elliptic polylogarithms \cite{BrownLevin,Broedel:2014vla,Adams:2016xah,Broedel:2017kkb}  (for applications of related classes of functions to high energy physics see e.g.~\cite{Laporta:2004rb,Kniehl:2005bc,Adams:2013kgc,Bloch:2013tra,Bloch:2014qca,Adams:2014vja,Adams:2015gva,Adams:2015ydq,Remiddi:2016gno,Primo:2016ebd,Bonciani:2016qxi,Adams:2016xah,Passarino:2016zcd,Harley:2017qut,vonManteuffel:2017hms,Ablinger:2017bjx,Chen:2017pyi,Hidding:2017jkk,Bogner:2017vim,Bourjaily:2017bsb,Broedel:2017siw,Laporta:2017okg,Broedel:2018iwv,Mistlberger:2018etf,Lee:2018jsw,Broedel:2018qkq,Adams:2018bsn,Adams:2018kez,Broedel:2019hyg,Bogner:2019lfa,Kniehl:2019vwr,Broedel:2019kmn}). However, most of these cases deal with relatively restricted kinematic dependence. More work would be needed to extend these methods to our situation with several independent square roots. Moreover, analytic continuation to physical regions is not yet very well understood. For this reason, we adopt here an approach based on series expansions methods. 
\subsection{Series solution of the differential equations}
\label{sec:seriessolutionsdeqns}
It is well known that when a closed form solution of a Feynman integral is not available, it is often possible to find a series expansion which provides a good approximation of the full solution. For an application of series expansion methods to single scale integrals see e.g. \cite{Pozzorini:2005ff,Aglietti:2007as,Mueller:2015lrx,Lee:2017qql,Lee:2018ojn,Bonciani:2018uvv,Mistlberger:2018etf}. For integrals depending on several scales,  multivariate series expansions have been used for special kinematic configurations (typically small or large energy limits, see e.g. \cite{Melnikov:2016qoc,Melnikov:2017pgf,Bonciani:2018omm, Bruser:2018jnc,Davies:2018ood,Davies:2018qvx}). In this paper we reduce the multivariate problem to a single scale problem by defining a contour $\gamma(\lambda)$ that connects a boundary point to a fixed point of the phase space.
We then find generalized power series solutions by solving the (single-scale) differential equations with respect to $\lambda$ (all the other variables are replaced by numbers).

More specifically, given a boundary point and a fixed target point, we define a contour $\gamma(\lambda)$ which connects the two points. It is then possible to define the differential equations with respect to $\lambda$. The problem is now one dimensional, and we can apply e.g. the methods of~\cite{Wasow,Lee:2018ojn}, which provide a formula for the solution of the differential equations near a singular point, whose coefficients are fixed by solving recurrence relations. Here we proceed in a different way, by directly integrating the differential equations in terms of (generalized) power series. Note that in order to find series expansions for the elliptic sectors we also need to series expand the polylogarithmic sectors, which contain the majority of the integrals. For the polylogarithmic sectors we have, along a generic contour $\gamma(\lambda)$, differential equations of the form
\begin{equation}
    \frac{\partial \vec{B}^{(i)}(\lambda)}{\partial \lambda} =  \mathbf{A}_\lambda\vec{B}^{(i-1)}(\lambda).
\end{equation}
A series solution around a singular or regular point $\lambda_0$ is obtained in the following way. We first series expand the matrix elements of $\mathbf{A}_\lambda$ around $\lambda_0$, and then we recursively integrate the right hand side until the desired order of the dimensional regularization parameter is reached. Note that at each step one needs to perform only integrals of the form 
\begin{equation}
\label{eq:GenSeriesInt}
    \int (\lambda-\lambda_0)^w \log(\lambda-\lambda_0)^n\,,
\end{equation}
for $w\in \mathbb{Q}$ and $n\in \mathbb{Z}_{\geq 0}$, which can be solved again in terms of powers of $\lambda$ and of $\log(\lambda-\lambda_0)$.  

For the elliptic sectors the differential equations are not fully decoupled. However it is possible to take multiple derivatives of the differential equations and obtain higher order differential equations for single integrals. For an integral at order $\epsilon^i$ , say $g^{(i)}(\lambda)$, the differential equation has the form
\begin{equation}
\label{eq:ODE order N}
\frac{\partial^{k}g^{(i)}(\lambda)}{\partial \lambda^{k}}+a_{1}(\lambda)\frac{\partial^{k-1}g^{(i)}(\lambda)}{\partial \lambda^{k-1}}+\cdots+a_k(\lambda)g^{(i)}(\lambda)=\beta^{(i)}(\lambda)
\end{equation}
where by construction the inhomogeneous term is known at each iteration. In order to solve the equation we need to find a complete set of homogeneous (series) solutions. These can be found by using the Frobenius method~\cite{Coddington}.  Once the homogeneous solutions $g_{h,j}(\lambda), \; j\in\{1,\dots,k\}$ are known, we obtain a series solution for the full equation by using the formula
\begin{equation}
\label{eq:ODE N particular}
g^{(i)}(\lambda)=\sum_{j=1}^k \left( g_{h,j}(\lambda) \chi^{(i)}_{j}+ g_{h,j}(\lambda)\int \frac{W_j( {}\lambda{})}{W({}\lambda{})}\beta^{(i)}( {}\lambda{})d {}\lambda{}\right),
\end{equation}
where $\chi^{(i)}_{j},\; j\in\{1,\dots,k\}$ is a set of boundary constants, $W( {}\lambda{})$ is the Wronskian determinant, and $W_j( {}\lambda{})$ is the determinant obtained by replacing the $j$-th column of the Wronskian matrix by $(0,\dots,0,1)$. Since the homogeneous solutions are a power series in the vicinity of the expansion point $\lambda_0$, order by order in the dimensional regularization parameter, we only need to consider integrals of the form (\ref{eq:GenSeriesInt}), which are performed analytically in terms of  powers of $\lambda$ and of $\log(\lambda-\lambda_0)$ as explained above.

In principle the choice of the contour $\gamma(\lambda)$ is arbitrary, however we observed that choosing the contour to be a straight line renders the series expansion faster when compared to non-linear contours. In general the series expansion around a given point will converge only until the next singular point, therefore the choice of the expansion points requires some care.  We start by considering real singular points, i.e. singular points corresponding to real values of $\lambda$. Given the set of real singular points along $\gamma(\lambda)$, we perform a series expansion around each singular point. Moreover, we define the domain of these expansions to be a segment centered at the expansion point, with radius equal to half the distance from the closest singular point (this ensures that the series converge sufficiently fast in their domain of definition). Some singularities might be complex, i.e. of the form $\lambda=\lambda_r+i \lambda_i$. For each complex singularity we expand around the three real points $\lambda_r-\lambda_i, \lambda_r,\lambda_r+\lambda_i$, and we define their domain as for the case of real singularities described before.  At the end of this procedure, there might be segments along the contour not patched by any series. We then consider the uncovered regions and we expand around the middle point of each region, and we define the domain of these expansions to be, as before, a segment centered at the expansion point, with radius equal to half the distance from the closest singular point. We repeat this procedure until the entire contour is patched. 

We remark that by singular point we mean any non-analytic point of the differential equations, i.e. points where the differential equations diverge, and branching points of possible square roots of the differential equations. We identify three classes of singularities. The first class of singularities are the so-called physical singularities, i.e. the singularities predicted by unitarity. When expanding around these singularities the solution will develop branch cuts across the expansion point, and the cut ambiguity is resolved by the Feynman prescription. Another class of singularities are the so-called non-physical singularities, i.e. singularities that are not physical, but which appear in the solution of the differential equations for a given basis of integrals (for example a canonical basis). These singularities correspond to singularities of the basis choice, but they cancel after rotating the basis back to a set of scalar Feynman integrals without prefactors. Specifically, non-physical singularities might originate from poles in the coefficients defining the integral basis, or from branching points of possible square roots appearing in the integral basis\footnote{In this work we consider only integral basis with algebraic coefficients.}. Poles do not give rise to cuts, and no care is needed in their definition. On the other hand, for square roots giving rise to non-physical branch cuts, we can arbitrarily choose an analytic continuation across their branching points, since these cuts cancel at the level of physical quantities. Finally, there are the so-called spurious singularities. These singularities are not predicted by unitarity, and they do not appear in the solution of the differential equations on the physical sheet. Therefore we choose our contours to be entirely contained in the physical sheet (see also \cite{moriello_expansions} for a detailed discussion about the singularities of differential equations in the context of our expansion strategy).

Finally, we remark that the procedure described above is entirely algorithmic, and can be used, given a boundary point, to transport the solution to any point of interest.


\subsection{Differential equations for the elliptic sectors}
Our basis for the elliptic sectors is provided in Appendix \ref{app:CanonicalBasis}. 
The corresponding differential equations have the form
\begin{equation}
\label{eq:DE A elliptic}
\frac{\partial}{\partial x_i}\vec{B}_{66-73}(\vec{x},\epsilon)=\sum_{j=0}^{\infty} \epsilon^j \mathbf{A}_{x_i}^{(j)}(\vec{x}) \vec{B}_{66-73}(\vec{x},\epsilon)+\vec{G}_{66-73}(\vec{x},\epsilon) ,
\end{equation}
where the vector $\vec{G}_{66-73}(\vec{x},\epsilon)$ depends on the polylogarithmic integrals $\vec{B}_{1-65}(\vec{x},\epsilon)$, and the homogeneous matrix has the schematic form 
\begin{equation}
\mathbf{A}_{\lambda}^{(0)}=\left(
\begin{array}{cc|cccccc}
 \;\bm{*}\; & \;\bm{*}\; & \;0\; & \;0\; & \;0\; & \;0\; & \;0\; & \;0\; \\
 \bm{*} & \bm{*} & 0 & 0 & 0 & 0 & 0 & 0 \\
  \hline
 \bm{*} & \bm{*} & 0 & 0 & 0 & 0 & 0 & 0 \\
 \bm{*} & \bm{*} & 0 & 0 & 0 & 0 & 0 & 0 \\
 \bm{*} & \bm{*} & 0 & 0 & 0 & \bm{*} & 0 & 0 \\
 \bm{*} & \bm{*} & 0 & 0 & \bm{*} & 0 & 0 & 0 \\
 \bm{*} & \bm{*} & 0 & 0 & \bm{*} & \bm{*} & 0 & 0 \\
 \bm{*} & \bm{*} & 0 & 0 & 0 & \bm{*} & 0 & 0 ,
\end{array}
\right)
\end{equation}
where the lines separate sectors $I_{0,1,1,1,1,1,1,0,0}$ and $I_{1,1,1,1,1,1,1,0,0}$. We see that the integrals $\vec{B}_{66-67}$ and $\vec{B}_{70 -71 }$ are coupled, and they satisfy second order differential equations.  By defining a contour $\gamma(\lambda)$ we obtain a single set of differential equations with respect to $\lambda$,
\begin{equation}
\label{eq:DE A elliptic}
\frac{\partial}{\partial \lambda}\vec{B}_{66-73}(\lambda,\epsilon)=\sum_{j=0}^\infty \epsilon^j \mathbf{A}_{\lambda}^{(j)}(\lambda) \vec{B}_{66-73}(\lambda,\epsilon)+\vec{G}_{66-73}(\lambda,\epsilon) ,
\end{equation}
where the matrix $\mathbf{A}_{\lambda}(\lambda,\epsilon)$ is given by,
\begin{equation}
\mathbf{A}_{\lambda}(\lambda,\epsilon)=\sum_{i}\mathbf{A}_{x_i}(\lambda,\epsilon)\frac{\partial x_i(\lambda)}{\partial \lambda}.
\end{equation}
 Given the structure of the homogeneous matrix it is clear how to perform a series expansion along a contour. We start with the expansion around a given point of the contour. For a given $\epsilon$ order we series expand the polylogarithmic sectors as described above. We then use Eq.~(\ref{eq:ODE N particular}) to expand the coupled integrals. We iterate until the required $\epsilon$ order is solved ($\epsilon^4$ in our case). We repeat the procedure for a suitably chosen set of points until the entire contour is covered. In the next section we discuss in detail the expansion along two contours of interest.

\subsection{Numerical results}

\begin{figure}[h]
\centering
\includegraphics[width=13 cm]{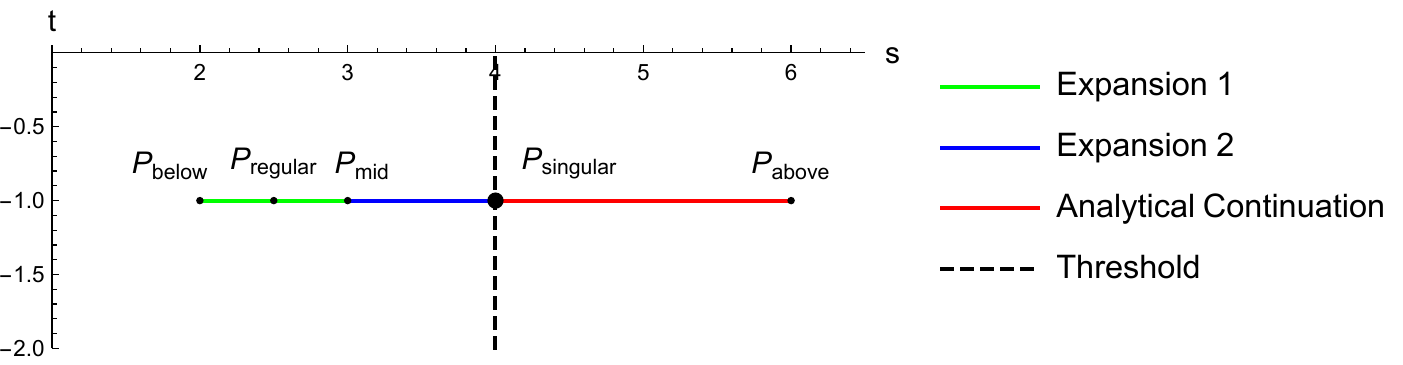}
\caption{The interval along which we plot the basis $B$, is covered by three expansions obtained by patching and analytical continuation.}
\label{fig:expansionregions}
\end{figure}
\FloatBarrier
\noindent

In this section we provide a numerical analysis of our solution. We use $p_4^2=13/25\approx (125/172)^2$, which is a good approximation to the physical value of the Higgs mass, assuming a unit mass for the top quark. However, the suitability and accuracy of the method is not related in any way to the numerical values of the kinematical points chosen. We could just as well have chosen values of $p_4^2$, which are related to the ratio of the Higgs mass to the charm or $b$-quark mass, or to anything else. 

We solve the basis integrals along a contour crossing a particle production threshold, in particular we chose a path from the kinematic point $\textrm{P}_{\textrm{below}} = (s=2, t=-1, p_4^2=13/25)$ to the kinematic point $\textrm{P}_{\textrm{above}}=(s=6,t=-1,p_4^2=13/25)$.  Our starting point is the heavy mass limit $\textbf{0} = (0,0,0)$ where we can use the results of Section \ref{sec:boundaryterms}. We perform an expansion along a straight path $\gamma_1$ from $\textbf{0}$ to the point $\textrm{P}_{\textrm{regular}}=(2.5,-1,13/25)$, which does not cross any singularity. However, by inspection of the differential equation matrices we see that they possess a singularity in $t=-\frac{52}{495}$, which spoils the convergence of the expansion in $\textbf{0}$. To obtain a better precision we then patch ten expansions on the line from $\textbf{0}$ to $\textrm{P}_{\textrm{regular}}$, by making sure that we never evaluate series at more than half the distance from the closest singularity (either physical or spurious).

\begin{figure}[!htb]
\minipage{0.5\textwidth}
\begin{center}
  \includegraphics[width=6.7 cm]{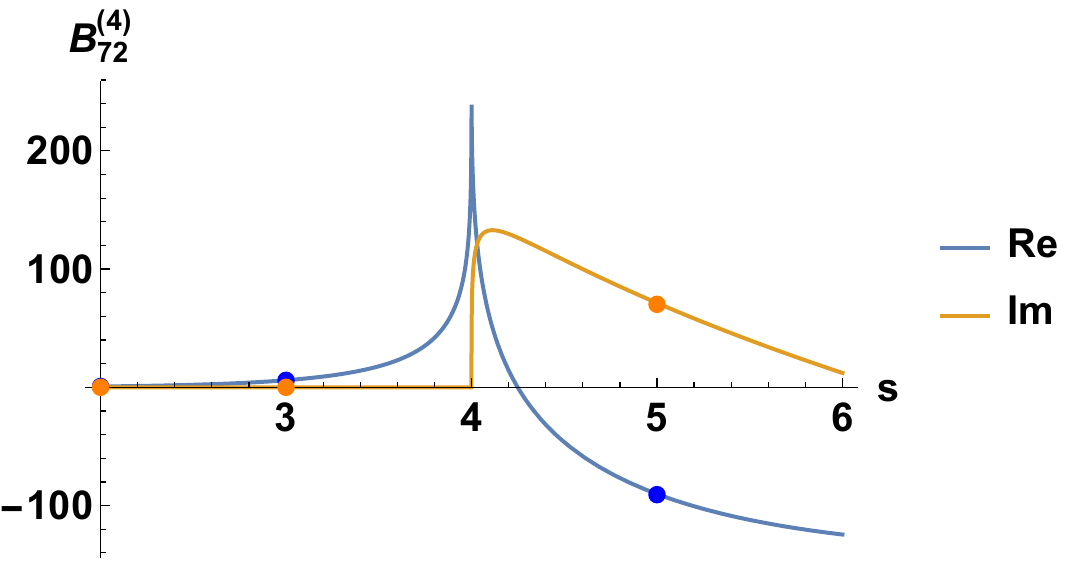}
\end{center}
\endminipage\hfill
\minipage{0.5\textwidth}
\begin{center}
  \includegraphics[width=6.7cm]{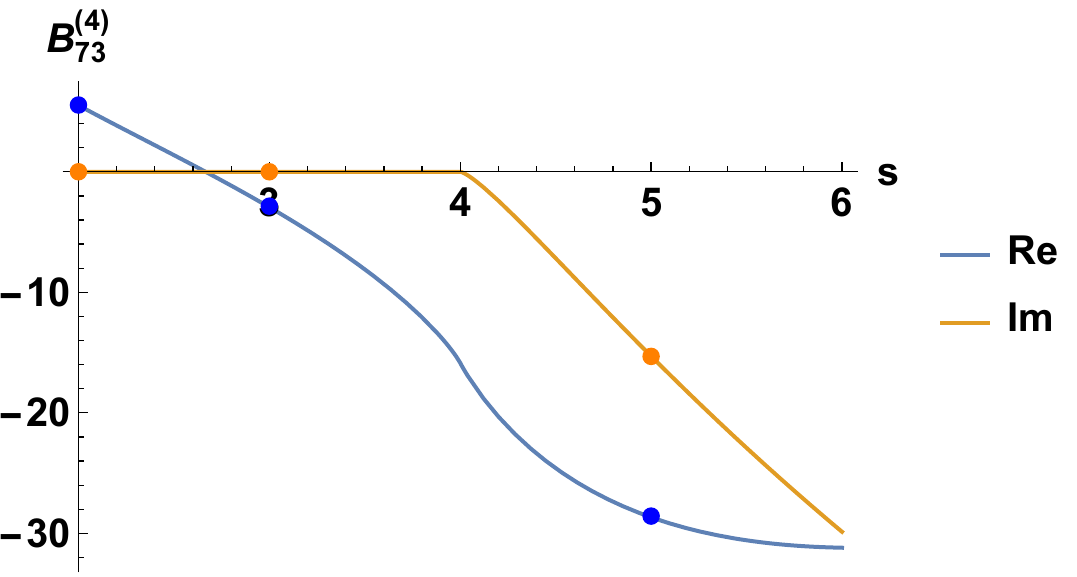}
\end{center}
\endminipage\hfill
\caption{Plot of the integrals $B^{(4)}_{72}$ and $B^{(4)}_{73}$. Note the singular behaviour at $s=4m^2, \ (m=1)$. The solid points represent values computed numerically with the software FIESTA \cite{Smirnov:2015mct}.}
\label{fig:7273}
\end{figure}

We now have high precision values at $\textrm{P}_{\textrm{regular}}$ that allows us to start expanding from here. Therefore we write an expansion along a path $\gamma_2$ joining $\textrm{P}_{\textrm{regular}}$ and $\textrm{P}_{\textrm{singular}} = (4,-1,13/25)$ which is a singular point since it lies on the threshold $s=4 m^2$. Thus, we use this expansion only up to the midpoint $\textrm{P}_{\textrm{mid}} = (3,-1,13/25)$ (Expansion 1). To go beyond this point, we perform an expansion (Expansion 2) from $\textrm{P}_{\textrm{singular}}$ to $\textrm{P}_{\textrm{mid}}$ using the values obtained there to fix the constants of integration. Finally, we perform an analytic continuation of the second expansion to obtain an expression valid above threshold up to $\textrm{P}_{\textrm{above}}$. The analytic continuation is straightforward since, when considering a series along a contour such that only one of the external scales varies ($s$ in our case), only terms of the form $\log{(s-4m^2)}$ and $\sqrt{s-4m^2}$ have to be analytically continued.  The situation is depicted in Figure \ref{fig:expansionregions}.
The expansion is carried up to order $\epsilon^4$ and the series are truncated at order $t^{50}$. In Figure \ref{fig:7273} the integrals $B^{(4)}_{72-73}$ are plotted together with some values obtained with standard numerical software. The remaining plots can be found in Appendix \ref{app:plotsappendix}.
To measure the numerical error of our method we can compare values obtained patching different contours. For example, we compared the values at $\textrm{P}_{\textrm{mid}}$ obtained expanding from $\textbf{0}$ to $\textrm{P}_{\mathrm{regular}}$ and then from $\textrm{P}_{\textrm{regular}}$ to $\textrm{P}_{\textrm{mid}}$ with those obtained expanding directly from $\textbf{0}$ to $\textrm{P}_{\textrm{mid}}$. The worst error is for $B_{72}^{(4)}$ where we get difference of $\mathcal{O}(10^{-16})$. For comparison, the numerical evaluation with FIESTA \cite{Smirnov:2015mct} of all the 73 master integrals in one point above the $s=4 m^2$ threshold requires $\mathcal{O}(10^4)$ seconds on 48 cores. On the other hand, finding a series expansion above threshold requires $\mathcal{O}(100s)$ on an ordinary laptop on 1 core, which is $\mathcal{O}(1\text{ second})$ per integral, after which the plots are virtually instantly generated.

\section{Boundary terms}
\label{sec:boundaryterms}
In this section we describe how to obtain boundary conditions for the system of differential equations associated with our basis of integrals. Similar to the computation of the planar Higgs + jet families in Ref. \cite{Bonciani:2016qxi}, we consider the heavy mass limit, where $s$, $t$ and $p_4^2$ tend to zero. For the planar families graph theoretical prescriptions~\cite{Smirnov:1990rz,Smirnov:1994tg,Smirnov:2002pj} were used to compute the limit. Although the same strategy may in principle be applied for the current non-planar family, we perform the computation of the boundary terms for the current family using the method of expansions by regions. 

The method of expansions by regions was originally developed in Ref. \cite{Beneke:1997zp} for the case of near threshold expansions, and has since been further clarified (for various limits) in various works, see for example Ref. \cite{Jantzen:2011nz}, and also Ref. \cite{Semenova:2018cwy} for a recent overview. In Ref. \cite{Smirnov:1999bza} the strategy was considered in the Feynman parametric representation, and a systematic geometry based algorithm for finding the relevant regions was presented in Refs. \cite{Pak:2010pt, Jantzen:2012mw}. We will use the formulation in the Feynman parametrization, and the package asy2.1.m from Ref. \cite{Jantzen:2012mw} to find the regions. 

First we present the results. We parametrize the heavy mass limit by a straight line,
\begin{align}
\label{eq:heavymasslimit}
(s,t,p_4^2,m^2) \rightarrow (x s,x t,x p_4^2,m^2)\,,
\end{align}
with $x \rightarrow 0$. It turns out that only the first two integrals of the polylogarithmic sectors are nonzero in the limit. More specifically, we have
\begin{align}
    \label{eq:boundarytermspolylogarithmic}
    \lim_{x\rightarrow 0} B_1 &= e^{2 \gamma_E  \epsilon } \Gamma (1+\epsilon)^2 (m^2)^{-2 \epsilon }\,,\nonumber\\
    \lim_{x\rightarrow 0} B_2 &\sim x^{-\epsilon }\left(  4^{\epsilon } e^{2 \gamma_E  \epsilon } \sqrt{\pi } (m^2)^{-\epsilon } (-t)^{-\epsilon } \frac{\Gamma (1-\epsilon ) \Gamma (\epsilon +1)^2 }{\Gamma \left(\frac{1}{2}-\epsilon \right)} \right)\,,\nonumber\\
    \lim_{x\rightarrow 0} B_i &= 0 \quad\text{for }i = 3,\ldots,65\,,
\end{align}
In the elliptic sectors all but the last integral are identically zero in the heavy mass limit. In particular, we have
\begin{align}
    \label{eq:boundarytermselliptic}
    \lim_{x\rightarrow 0} B_i &= 0 \quad\text{for }i = 66,\ldots,72\,,\nonumber\\
    \lim_{x\rightarrow 0} B_{73} &\sim x^{-\epsilon }\left(-2 e^{2 \gamma_E  \epsilon } \epsilon  \left(m^2\right)^{-\epsilon } (-t)^{-\epsilon }  \left(\frac{-p_4^2+4 s+t}{-p_4^2+2 s+t}\right)\frac{\Gamma (1-\epsilon )^2 \Gamma (1+\epsilon)^2}{\Gamma (1-2 \epsilon )}\right)\,.
\end{align}
Next, we work out the computation of the last boundary term $B_{73}$ as an illustrative example. This basis integral is given explicitly by
\begin{align}
    \label{eq:B73contribution}
    B_{73} &= t \epsilon ^4 \left(\text{I}_{1,1,1,1,1,1,1,-2,0}+\frac{4 s   \text{I}_{1,1,1,1,1,1,1,-1,-1}}{2 s+t-p_4^2}+\text{I}_{1,1,1,1,1,1,1,0,-2}\right)+\nonumber\\ &\quad- \frac{t \epsilon ^4 \left(-4 s-t+p_4^2\right)}{4}   \left( \text{I}_{1,1,1,1,1,1,1,-1,0}+ \text{I}_{1,1,1,1,1,1,1,0,-1}\right)\,.
\end{align}
All the Feynman integrals that appear in $B_{73}$ lie in the same sector, but with different numerators. Using asy2.1.m we find that the contributing regions as $x\rightarrow 0$ are
\begin{align}
    S_1&:\quad \alpha_i \rightarrow \alpha_i \,,\nonumber\\
    S_2&:\quad \alpha_i \rightarrow \alpha_i \quad\text{for }i = \{1,2,4\}\,,\,\alpha_i \rightarrow x\alpha_i \quad\text{for }i = \{3,5,6,7\}\,,
\end{align}
where the $\alpha_i$'s denote the Feynman parameters associated to each propagator in the Feynman parametrization. The asymptotic limit of the integrals then takes the form,
\begin{align}
    \label{eq:Itopsectorasymptoticmasks}
    &\lim_{x\rightarrow 0} \text{I}_{1,1,1,1,1,1,1,\sigma_1,\sigma_2}  \sim \text{I}_{1,1,1,1,1,1,1,\sigma_1,\sigma_2}^{(1)}+x^{-\epsilon -1} \text{I}_{1,1,1,1,1,1,1,\sigma_1,\sigma_2}^{(2)}\,,\nonumber\\&\quad\text{ for }(\sigma_1,\sigma_2) \in \{(-2,0),(-1,0),(-1,-1),(0,-1),(0,-2)\}\,,
\end{align}
where the superscripts denote the contributions of both regions respectively. In other words, for the integrals with superscripts the kinematics has been rescaled in the manner of Eq. (\ref{eq:heavymasslimit}), the Feynman parameters are rescaled according to Eq. (\ref{eq:boundarytermselliptic}), and the overall dependence on $x$ has been explicitly factored out of the integrand.

Rescaling the prefactors in Eq. (\ref{eq:B73contribution}), and plugging in the results of Eq. (\ref{eq:Itopsectorasymptoticmasks}) yields the asymptotic limit of $B_{73}$,
\begin{align}
    \label{eq:B73asymplimit}
    \lim_{x\rightarrow 0}B_{73} \sim \epsilon^4 x^{-\epsilon }\left[-\frac{4  s t  \text{I}_{1,1,1,1,1,1,1,-1,-1}^{(2),(x=0)}}{p_4^2-2 s-t}+ t \left(\text{I}_{1,1,1,1,1,1,1,-2,0}^{(2),(x=0)}+   \text{I}_{1,1,1,1,1,1,1,0,-2}^{(2),(x=0)}\right)\right]\,.
\end{align}
There are 3 surviving contributions, which lie in region 2. It is sufficient to compute these at leading order in $x=0$, as the higher orders vanish in the limit. The integrations themselves can be done by picking a suitable integration order and integrating one Feynman parameter at a time, while keeping full dependence on the dimensional regulator $\epsilon$.\footnote{We found that some of the integrations converge in disjoint domains of $\epsilon$. In situations without a domain of $\epsilon$, where all parts of an integral are convergent, one usually considers each part in its own domain of convergence from which it can analytically be continued to a desired common final domain after an evaluation. Still to be on the safe side, we prefer to proceed with an auxiliary analytical regularization, and assign the third propagator the exponent $\nu_3$, which serves as an additional regulator. After performing all integrations we take the limit $\nu_3 \rightarrow 1$. The end result is found to be the same.} The results are given by
\begin{align}
    \label{eq:Iregion2integrated}
    \text{I}_{1,1,1,1,1,1,1,-2,0}^{(2),(x=0)}=\text{I}_{1,1,1,1,1,1,1,-1,-1}^{(2),(x=0)}=\text{I}_{1,1,1,1,1,1,1,0,-2}^{(2),(x=0)}= \frac{e^{2 \gamma_E  \epsilon } \left(m^2\right)^{-\epsilon } (-t)^{-\epsilon -1} \Gamma (1+\epsilon)^2 \Gamma (1-\epsilon )^2}{\epsilon ^3 \Gamma (1-2 \epsilon )}\,.
\end{align}
Plugging these expressions into Eq. (\ref{eq:B73asymplimit}) yields the boundary term of $B_{73}$, stated in Eq. (\ref{eq:boundarytermselliptic}). We used the assumption $t<0$ during the integration to avoid branch cuts, and we may analytically continue the expression to the physical region by using the Feynman prescription, which tells us to interpret $t$ as having an infinitesimally small positive imaginary part.

The remaining boundary terms may be computed using the same analysis as was done here for the most complicated sector. Although we found that all basis integrals except for $B_1$, $B_2$, and $B_{73}$ are zero, showing this requires the computation of numerous integrals which cancel with each other at the end.

\section{Conclusion}
\label{sec:conclusion}
The Feynman diagrams of the $gg \rightarrow gH$ amplitudes with full heavy quark mass dependence fit into seven integral families. The computation of the planar families was presented in Ref. \cite{Bonciani:2016qxi}, leaving two non-planar integral families to be evaluated. In this paper we presented the computation of one of the two families of non-planar master integrals. The family under consideration consists of 73 master integrals, of which 65 are polylogarithmic, while the remaining 8 integrals in the top sectors are elliptic, implying that their solution may not be presented in terms of multiple polylogarithms. Therefore the paper was split up into two parts, the first dealing with the computation of the polylogarithmic integrals, and the second with the computation of the elliptic sectors.

The polylogarithmic sectors may be computed by choosing a basis that puts their associated system of differential equations in canonical form, in which all dependence on the dimensional regularization parameter is in an overall prefactor, and the partial derivative matrices are combined into a matrix that is in $d\log$-form. Although the formal solution of such a system may be immediately given in terms of Chen's iterated integrals, the presence of many square roots, that cannot be simultaneously rationalized, makes it unclear how to convert the iterated integrals to multiple polylogarithms which admit fast numerical evaluation. We showed that at weight 2 the canonical basis integrals may nonetheless be solved by starting from an ansatz of dilogarithms and logarithms. The weight 3 and 4 expressions could then be written as 1-fold integrals over polylogarithmic expressions.

Alternatively, we showed how the system of differential equations may be solved in terms of (generalized) series solutions. The approach relies on reducing the multidimensional system of differential equations to a one-dimensional system, by integrating along a line towards a given kinematic point of interest. The differential equations are then expanded in terms of a (generalized) series, and the integration is performed analytically up to very high order. This expansion based approach furthermore works to obtain high precision numerical results for both the polylogarithmic and the elliptic sectors, and in a sense trivializes the analytic continuation of the family of integrals to the physical region as well. At any stage it is only required to perform the analytic continuation of powers of logarithms across their branch cuts, and the direction in which to cross the branch cut is determined by the Feynman prescription.

To compute the Higgs + jet cross section it is necessary to perform the evaluation of another non-planar integral family, which we believe may be done using the same methods as have been presented for the current family. We leave the computation of the remaining family to a future publication.

\section*{Acknowledgements}
HF, MH, LM and GS would like to thank ETH Z\"urich for the hospitality during the preparation of this work. The work of HF, MH, LM and GS was supported by STSM Grants from the COST Action CA16201 PARTICLEFACE. The work of GS was supported by ETH Z\"urich and by the Pauli Center for Theoretical Studies, Z\"urich. The work of FM was supported by the Swiss National Science Foundation project No. 177632 (ElliptHiggs). The work of HF is part of the HiProLoop project funded by the European Union's Horizon 2020 research and innovation programme under the Marie Sk{\l}odowska-Curie grant agreement 747178. The work of VS was supported by RFBR, grant 17-02-00175.
This research received funding from the European Research Council (ERC) under the European Union's Horizon 2020 research and innovation programme, grant agreements No 725110 (Amplitudes) and 647356 (CutLoops).

\vfill
\appendix

\section{Alphabet of the polylogarithmic sectors}
\label{app:AlphabetPolylogarithmic}
The full alphabet is given by the following 69 letters,
\begin{align*}
\begin{array}{ll}
 l_1= m^2\,, & l_2= p_4^2\,, \\
 l_3= s\,, & l_4= t\,, \\
 l_5= s+t\,, & l_6= -4 m^2+p_4^2\,, \\
 l_7= -s+p_4^2\,, & l_8= -t+p_4^2\,, \\
 l_9= s+t-p_4^2\,, & l_{10}= 4 m^2-s\,, \\
 l_{11}= 4 m^2-t\,, & l_{12}= 4 m^2+t-p_4^2\,, \\
 l_{13}= 4 m^2+s+t-p_4^2\,, & l_{14}= s^2+m^2 p_4^2-s p_4^2\,, \\
 l_{15}= t^2+m^2 p_4^2-t p_4^2\,, & l_{16}= -4 m^2 s-4 m^2 t+s t+4 m^2 p_4^2\,, \\
 l_{17}= -s^2+4 m^2 t-s t+s p_4^2\,, & l_{18}= 4 m^2 s-s t-t^2+t p_4^2\,, \\
 l_{19}= m^2 s^2+2 m^2 s t+m^2 t^2-s t p_4^2\,, & l_{20}= s^2+2 s t+t^2+m^2 p_4^2-s p_4^2-t p_4^2\,, \\
 l_{21}= -4 m^2 s t-4 m^2 t^2+4 m^2 t p_4^2+s p_4^4\,,\;\;\;\; & l_{22}= 4 m^2 s t-s p_4^4-t p_4^4+p_4^6\,, \\
 l_{23}= q_{11}\,, & l_{24}= q_{12}\,, \\[0.5em]
 l_{25}= \dfrac{1+\dfrac{r_7}{r_2}}{1-\dfrac{r_7}{r_2}}\,, & l_{26}= \dfrac{1+\dfrac{r_8}{r_3}}{1-\dfrac{r_8}{r_3}}\,, \\[2em]
 l_{27}= \dfrac{-p_4^2+r_1 r_6}{-p_4^2-r_1 r_6}\,, & l_{28}= \dfrac{1+\dfrac{r_5}{r_{10}}}{1-\dfrac{r_5}{r_{10}}}\,, \\[2em]
 l_{29}= \dfrac{\left(t-p_4^2\right)+r_4 r_9}{\left(t-p_4^2\right)-r_4 r_9}\,, & l_{30}= \dfrac{\left(p_4^2-2 s\right)+r_1 r_6}{\left(p_4^2-2 s\right)-r_1 r_6}\,, \\[1.5em]
 l_{31}= \dfrac{\left(p_4^2-2 s-2 t\right)+r_1 r_6}{\left(p_4^2-2 s-2 t\right)-r_1 r_6}\,, & l_{32}= \dfrac{\dfrac{2 m^2-t}{t}+\dfrac{r_6}{r_1}}{\dfrac{2 m^2-t}{t}-\dfrac{r_6}{r_1}}\,, \\[2em]
 l_{33}= \dfrac{1+\dfrac{r_{11}}{r_2 r_3}}{1-\dfrac{r_{11}}{r_2 r_3}}\,, & l_{34}= \dfrac{\dfrac{-q_1}{t}+r_1 r_6}{\dfrac{-q_1}{t}-r_1 r_6}\,, \\[2em]
 l_{35}= \dfrac{\dfrac{1}{p_4^2}+\dfrac{r_2}{r_{14}}}{\dfrac{1}{p_4^2}-\dfrac{r_2}{r_{14}}}\,, & l_{36}= \dfrac{\dfrac{2 m^2 s+2 m^2 t-t p_4^2}{t}+r_1 r_6}{\dfrac{2 m^2 s+2 m^2 t-t p_4^2}{t}-r_1 r_6}\,, \\[2.5em]
 l_{37}= \dfrac{\dfrac{2 s+t-p_4^2}{t-p_4^2}+\dfrac{r_6}{r_1}}{\dfrac{2 s+t-p_4^2}{t-p_4^2}-\dfrac{r_6}{r_1}}\,, & l_{38}= \dfrac{1+\dfrac{r_{12}}{r_2 r_5}}{1-\dfrac{r_{12}}{r_2 r_5}}\,, \\[2.5em]
 l_{39}= \dfrac{1+\dfrac{r_{11}}{r_2 r_8}}{1-\dfrac{r_{11}}{r_2 r_8}}\,, & l_{40}= \dfrac{1+\dfrac{r_{11}}{r_3 r_7}}{1-\dfrac{r_{11}}{r_3 r_7}}\,,\quad\quad\quad\quad\quad\quad\quad\quad\quad\quad\quad\quad\quad\,\,
 \end{array}
 \end{align*}
 \begin{align*}
\begin{array}{ll}
 l_{41}= \dfrac{1+\dfrac{r_{13}}{r_3 r_5}}{1-\dfrac{r_{13}}{r_3 r_5}}\,, & l_{42}= \dfrac{\dfrac{1}{p_4^2}+\dfrac{r_5}{r_{15}}}{\dfrac{1}{p_4^2}-\dfrac{r_5}{r_{15}}} \,,\\[2.5em]
 l_{43}= \dfrac{1+\dfrac{r_2 r_6}{r_1 r_7}}{1-\dfrac{r_2 r_6}{r_1 r_7}}\,, & l_{44}= \dfrac{1+\dfrac{r_3 r_6}{r_1 r_8}}{1-\dfrac{r_3 r_6}{r_1 r_8}} \,,\\[2.5em]
 l_{45}= \dfrac{1+\dfrac{r_4 r_7}{r_2 r_9}}{1-\dfrac{r_4 r_7}{r_2 r_9}}\,, & l_{46}= \dfrac{1+\dfrac{r_{12}}{r_2 r_{10}}}{1-\dfrac{r_{12}}{r_2 r_{10}}} \,,\\[2.5em]
 l_{47}= \dfrac{\left(st-2 m^2 s-2 m^2 t\right)+r_2 r_3 r_{11}}{\left(st-2 m^2 s-2 m^2 t\right)-r_2 r_3 r_{11}}\,, & l_{48}= \dfrac{1+\dfrac{r_5 r_6}{r_1 r_{10}}}{1-\dfrac{r_5 r_6}{r_1 r_{10}}} \,,\\[2.5em]
 l_{49}= \dfrac{1+\dfrac{r_{12}}{r_5 r_7}}{1-\dfrac{r_{12}}{r_5 r_7}}\,, & l_{50}= \dfrac{1+\dfrac{r_4 r_{10}}{r_5 r_9}}{1-\dfrac{r_4 r_{10}}{r_5 r_9}} \,,\\[2.5em]
 l_{51}= \dfrac{-\dfrac{t+p_4^2}{t-p_4^2}+\dfrac{r_{12}}{r_2 r_5}}{-\dfrac{t+p_4^2}{t-p_4^2}-\dfrac{r_{12}}{r_2 r_5}}\,, & l_{52}= \dfrac{-\dfrac{s+p_4^2}{s-p_4^2}+\dfrac{r_{13}}{r_3 r_5}}{-\dfrac{s+p_4^2}{s-p_4^2}-\dfrac{r_{13}}{r_3 r_5}} \,,\\[2.5em]
 l_{53}= \dfrac{1+\dfrac{r_1 r_{11}}{r_2 r_3 r_6}}{1-\dfrac{r_1 r_{11}}{r_2 r_3 r_6}}\,, & l_{54}= \dfrac{-q_2+r_1 r_4 r_6 r_9}{-q_2-r_1 r_4 r_6 r_9} \,,\\[2.5em]
 l_{55}= \dfrac{\left(2 m^2 t{-}2 m^2 s{+}s t{+}t^2{-}t p_4^2\right)+r_3 r_{10} r_{13}}{\left(2 m^2 t{-}2 m^2 s{+}s t{+}t^2{-}t p_4^2\right)-r_3 r_{10} r_{13}}\,,\;\; & l_{56}= \dfrac{-q_4+r_5 r_8 r_{13}}{-q_4-r_5 r_8 r_{13}} \,,\\[2.5em]
 l_{57}= \dfrac{-q_8+r_1 r_2 r_3 r_6 r_{11}}{-q_8-r_1 r_2 r_3 r_6 r_{11}}\,, & l_{58}= \dfrac{q_7+r_1 r_2 r_6 r_{14}}{q_7-r_1 r_2 r_6 r_{14}} \,,\\[2.5em]
 l_{59}= \dfrac{-q_{13}+r_1 r_5 r_6 r_{15}}{-q_{13}-r_1 r_5 r_6 r_{15}}\,, & l_{60}= \dfrac{-q_{15}+r_1 r_3 r_5 r_6 r_{13}}{-q_{15}-r_1 r_3 r_5 r_6 r_{13}} \,,\\[2em]
 l_{61}= \dfrac{-q_{14}+r_1 r_2 r_5 r_6 r_{12}}{-q_{14}-r_1 r_2 r_5 r_6 r_{12}}\,, & l_{62}= \dfrac{-q_{16}+r_1 r_3 r_5 r_6 r_{13}}{-q_{16}-r_1 r_3 r_5 r_6 r_{13}} \,,\\[2em]
 l_{63}= \dfrac{\left(\dfrac{-q_3}{2}+r_3 r_{11} r_{14}\right) \left(\dfrac{-q_{17}}{p_4^2}+r_3 r_{11} r_{14}\right)}{\left(\dfrac{-q_3}{2}-r_3 r_{11} r_{14}\right) \left(\dfrac{-q_{17}}{p_4^2}-r_3 r_{11} r_{14}\right)}\,, & l_{64}= \dfrac{\left(-\dfrac{q_5}{p_4^2}+r_2 r_{12} r_{15}\right) \left(-\dfrac{q_9}{2}+r_2 r_{12} r_{15}\right)}{\left(-\dfrac{q_5}{p_4^2}-r_2 r_{12} r_{15}\right) \left(-\dfrac{q_9}{2}-r_2 r_{12} r_{15}\right)}\,,
  \end{array}
 \end{align*}
 \begin{align*}
\begin{array}{ll}
 l_{65}= \dfrac{\left(-\dfrac{q_6}{p_4^2}+r_3 r_{13} r_{15}\right) \left(-\dfrac{q_{10}}{2}+r_3 r_{13} r_{15}\right)}{\left(-\dfrac{q_6}{p_4^2}-r_3 r_{13} r_{15}\right) \left(-\dfrac{q_{10}}{2}-r_3 r_{13} r_{15}\right)}\,, & \\[3em]
 l_{66}= \dfrac{\dfrac{q_{24}}{\left(2 s+t-p_4^2\right) \left(t+p_4^2\right)}+r_1 r_2 r_5 r_6 r_{12}}{\dfrac{q_{24}}{\left(2 s+t-p_4^2\right) \left(t+p_4^2\right)}-r_1 r_2 r_5 r_6 r_{12}}\,,\\[3em]
 l_{67}= \dfrac{\left(\dfrac{q_{18}}{2}+r_5 r_{12} r_{14}\right) \left(-\dfrac{q_{19}}{p_4^2}+r_5 r_{12} r_{14}\right)}{\left(\dfrac{q_{18}}{2}-r_5 r_{12} r_{14}\right) \left(-\dfrac{q_{19}}{p_4^2}-r_5 r_{12} r_{14}\right)}\,, & \\
 l_{68}= \dfrac{\left(\dfrac{q_{20}}{2}+r_2 r_3 r_5 r_{11} r_{15}\right)\left(\dfrac{q_{22}}{p_4^2}+r_2 r_3 r_5 r_{11} r_{15}\right) }{\left(\dfrac{q_{20}}{2}-r_2 r_3 r_5 r_{11} r_{15}\right) \left(\dfrac{q_{22}}{p_4^2}-r_2 r_3 r_5 r_{11} r_{15}\right)} \,,\\[2em]
 l_{69}= \dfrac{ \left(\dfrac{q_{21}}{2}+r_2 r_3 r_5 r_{13} r_{14}\right)\left(\dfrac{q_{23}}{p_4^2}+r_2 r_3 r_5 r_{13} r_{14}\right)}{\left(\dfrac{q_{21}}{2}-r_2 r_3 r_5 r_{13} r_{14}\right) \left(\dfrac{q_{23}}{p_4^2}-r_2 r_3 r_5 r_{13} r_{14}\right)}\,. & \quad\quad\quad\quad\quad\quad\quad\quad\quad\quad\quad\quad\quad\quad\quad\quad\quad\quad\quad\quad\quad\quad\quad\quad\quad\quad
\end{array}
\end{align*}
The $q_i$ are given by the following polynomials,
\begin{align*}
 q_1&= -2 m^2 p_4^2+2 m^2 s+p_4^2 t\, \\
 q_2&= -4 m^2 p_4^2+2 m^2 t-p_4^2 t+p_4^4\, \\
 q_3&= -8 m^2 p_4^2 t+8 m^2 s t+8 m^2 t^2+p_4^4 (-s)-s t^2\, \\
 q_4&= -2 m^2 p_4^2+4 m^2 s+2 m^2 t+p_4^2 t-s t-t^2\, \\
 q_5&= 2 m^2 p_4^4 t+2 m^2 s^2 t-p_4^4 s^2-p_4^4 s t+p_4^6 s\, \\
 q_6&= 2 m^2 p_4^4 s+2 m^2 s t^2-p_4^4 s t-p_4^4 t^2+p_4^6 t\, \\
 q_7&= 2 m^2 p_4^2 s-2 m^2 p_4^2 t+2 m^2 s t+2 m^2 t^2+p_4^4 (-s)\, \\
 q_8&= -2 m^2 p_4^2 s+2 m^2 p_4^2 t+2 m^2 s^2+2 m^2 s t-p_4^2 s t\, \\
 q_9&= 8 m^2 s t+p_4^2 s^2-p_4^4 s-p_4^4 t+p_4^6-s^3-s^2 t\, \\
 q_{10}&= 8 m^2 s t-p_4^4 s+p_4^2 t^2-p_4^4 t+p_4^6-s t^2-t^3\, \\
 q_{11}&= -2 m^2 p_4^2 t+m^2 p_4^4+m^2 t^2+p_4^2 s^2+p_4^2 s t-p_4^4 s\, \\
 q_{12}&= -2 m^2 p_4^2 s+m^2 p_4^4+m^2 s^2+p_4^2 s t+p_4^2 t^2-p_4^4 t\, \\
 q_{13}&= 2 m^2 p_4^2 s+2 m^2 p_4^2 t-2 m^2 p_4^4+2 m^2 s t-p_4^4 s-p_4^4 t+p_4^6\, \\
 q_{14}&= -2 m^2 p_4^2 s+2 m^2 p_4^2 t+2 m^2 s^2+2 m^2 s t-p_4^2 s^2-p_4^2 s t+p_4^4 s\, \\
 q_{15}&= 2 m^2 p_4^2 s-2 m^2 p_4^2 t+2 m^2 s t+2 m^2 t^2-p_4^2 s t-p_4^2 t^2+p_4^4 t\, \\
 q_{16}&= 2 m^2 p_4^2 s+2 m^2 p_4^2 t-2 m^2 p_4^4+2 m^2 s t-p_4^2 s t-p_4^2 t^2+p_4^4 t\, \\
 q_{17}&= 2 m^2 p_4^4 s-2 m^2 p_4^2 t^2+2 m^2 p_4^4 t-2 m^2 p_4^6+2 m^2 s t^2+2 m^2 t^3-p_4^4 s t\, \\
 q_{18}&= -8 m^2 p_4^2 t+8 m^2 s t+8 m^2 t^2+2 p_4^2 s^2+2 p_4^2 s t-2 p_4^4 s-s^3-2 s^2 t-s t^2\, \\
 q_{19}&= -4 m^2 p_4^2 s t-4 m^2 p_4^2 t^2+4 m^2 p_4^4 t+2 m^2 s^2 t+4 m^2 s t^2+2 m^2 t^3-p_4^4 s^2-p_4^4 s t+p_4^6 s\, \\
 q_{20}&= -8 m^2 p_4^2 s t+8 m^2 s^2 t+8 m^2 s t^2-p_4^4 s^2-2 p_4^4 s t+2 p_4^6 s-p_4^4 t^2+2 p_4^6 t-p_4^8-s^2 t^2\, \\
 q_{21}&= -8 m^2 p_4^2 s t+8 m^2 s^2 t+8 m^2 s t^2-p_4^4 s^2+2 p_4^2 s t^2+2 p_4^2 t^3-p_4^4 t^2-s^2 t^2-2 s t^3-t^4\, \\
 q_{22}&= 2 m^2 p_4^4 s^2+4 m^2 p_4^4 s t-4 m^2 p_4^6 s+2 m^2 p_4^4 t^2-4 m^2 p_4^6 t+2 m^2 p_4^8+2 m^2 s^2 t^2-p_4^4 s^2 t\\&\quad-p_4^4 s t^2+p_4^6 s t\, \\
 q_{23}&= 2 m^2 p_4^4 s^2-4 m^2 p_4^2 s t^2-4 m^2 p_4^2 t^3+2 m^2 p_4^4 t^2+2 m^2 s^2 t^2+4 m^2 s t^3+2 m^2 t^4-p_4^4 s^2 t\\&\quad-p_4^4 s t^2+p_4^6 s t\, \\
 q_{24}&= 12 m^2 p_4^2 s^2 t+2 m^2 p_4^4 s^2+10 m^2 p_4^2 s t^2-10 m^2 p_4^4 s t-2 m^2 p_4^6 s+2 m^2 p_4^2 t^3-4 m^2 p_4^4 t^2\\&\quad+2 m^2 p_4^6 t+2 m^2 s^2 t^2+2 m^2 s t^3-2 p_4^2 s^4-4 p_4^2 s^3 t+4 p_4^4 s^3-3 p_4^2 s^2 t^2+4 p_4^4 s^2 t-3 p_4^6 s^2\\&\quad-p_4^2 s t^3+p_4^4 s t^2-p_4^6 s t+p_4^8 s\,.
\end{align*}
The terms labelled by $\{r_i\}$ are the square roots. The first 15 roots appear in the polylogarithmic sectors, while $r_{16}$ appears in the elliptic sectors (see next section),
\begin{align}
\label{eq:famFroots}
\def\arraystretch{1.3}
\begin{array}{ll}
 r_1= \sqrt{-p_4^2}\,, & r_2= \sqrt{-s}\,, \\
 r_3= \sqrt{-t}\,, & r_4= \sqrt{t-p_4^2}\,, \\
 r_5= \sqrt{s+t-p_4^2}\,, & r_6= \sqrt{4 m^2-p_4^2}\,, \\
 r_7= \sqrt{4 m^2-s}\,, & r_8= \sqrt{4 m^2-t}\,, \\
 r_9= \sqrt{4 m^2-p_4^2+t}\,, & r_{10}= \sqrt{4 m^2-p_4^2+s+t}\,, \\
 r_{11}= \sqrt{4 m^2 (p_4^2- s-t)+s t}\,, & r_{12}= \sqrt{4 m^2 t+s(p_4^2 -s- t)}\,, \\
 r_{13}= \sqrt{4 m^2 s+t(p_4^2 -s -t)}\,, & r_{14}= \sqrt{4 m^2 t(s+t -p_4^2)-\left(p_4^2\right)^2 s}\,, \\
 r_{15}= \sqrt{-4 m^2 s t+\left(p_4^2\right)^2 (s +t-p_4^2)}\,, & r_{16}= \sqrt{16 m^2 t+\left(p_4^2-t\right)^2}\,.
\end{array}
\end{align}
The labelling has been chosen so that the radicands of the roots are irreducible polynomials. In the polylogarithmic basis elements $B_1,\ldots,B_{65}$, the 15 roots only appear in 10 combinations,
\begin{align}
    \left\{r_1 r_6,r_2 r_7,r_3 r_8,r_4 r_9,r_5 r_{10},r_2 r_3 r_{11},r_2 r_5 r_{12},r_3 r_5 r_{13},r_2 r_{14},r_5 r_{15}\right\}\,.
\end{align}
It may also be verified that these same 10 combinations are sufficient to express all ratios of roots appearing in the letters. Hence, in principle is it possible to combine these and work with just 10 independent square roots for the polylogarithmic sectors. 

In the choice of basis for the elliptic sectors, roots $r_2$ and $r_{16}$ appear separately. Therefore, there are 12 independent combinations of roots in the full basis, given in Appendix \ref{app:CanonicalBasis}.

\section{Canonical basis and basis for elliptic sectors}
\label{app:CanonicalBasis}
\begin{figure}[!h]
\centering
\includegraphics[width=0.95\textwidth]{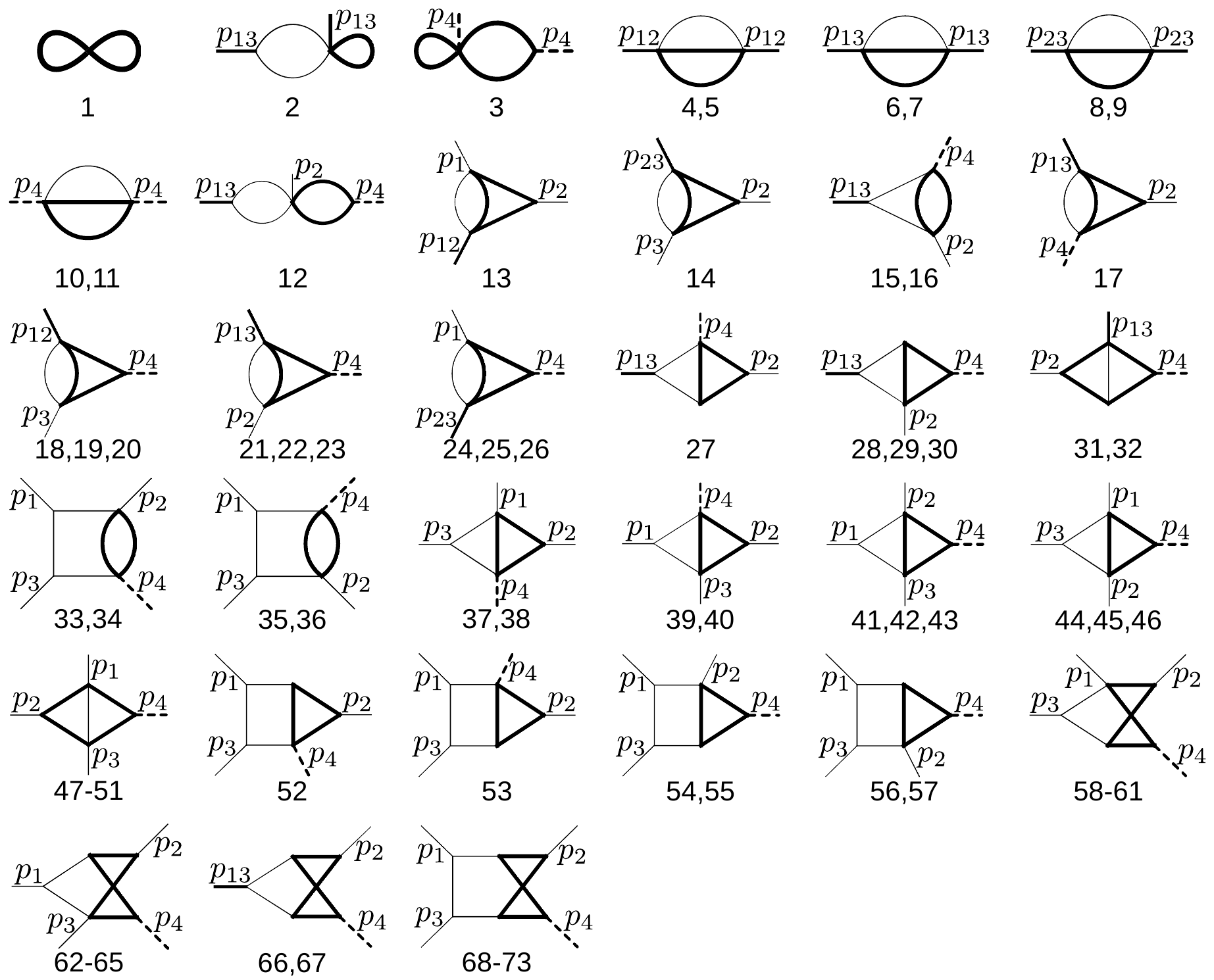}
\caption{The 73 master integrals. Shown on the figure is the sector, i.e. the set of propagators, to which the master integrals belong. Higher powers of propagators, numerators, or prefactors are not shown. External momenta are labelled using $p_{ij} = p_1{+}p_j$ and $p_4 = p_1{+}p_2{+}p_3$. Masses (internal as well as external) are indicated with a thicker line.}
\label{fig:masters}
\end{figure}

The set of 73 master integrals is represented in Figure \ref{fig:masters}.
The canonical basis of the polylogarithmic sectors is given by the following 65 integrals,
\begin{align}
 B_{1}&= \epsilon ^2 \text{I}_{0,0,0,0,2,0,2,0,0}\,,\nonumber\\   B_{2}&= t \epsilon ^2 \text{I}_{0,2,0,1,0,0,2,0,0}\,,\nonumber\\   B_{3}&= \epsilon ^2 r_1 r_6 \text{I}_{0,0,2,0,1,0,2,0,0}\,,\nonumber\\   B_{4}&= s
   \epsilon ^2 \text{I}_{1,0,2,0,0,2,0,0,0}\,,\nonumber\\   B_{5}&= \epsilon ^2 r_2 r_7 \left(\frac{1}{2} \text{I}_{1,0,2,0,0,2,0,0,0}+\text{I}_{2,0,2,0,0,1,0,0,0}\right)\,,\nonumber\\   B_{6}&= t \epsilon ^2
   \text{I}_{0,0,2,1,0,0,2,0,0}\,,\nonumber\\   B_{7}&= \epsilon ^2 r_3 r_8 \left(\frac{1}{2} \text{I}_{0,0,2,1,0,0,2,0,0}+\text{I}_{0,0,2,2,0,0,1,0,0}\right)\,,\nonumber\\   B_{8}&= \epsilon ^2
   \left(-s-t+p_4^2\right) \text{I}_{1,0,0,0,2,0,2,0,0}\,,\nonumber\\   B_{9}&= \epsilon ^2 r_5 r_{10} \left(\frac{1}{2} \text{I}_{1,0,0,0,2,0,2,0,0}+\text{I}_{2,0,0,0,1,0,2,0,0}\right)\,,\nonumber\\   B_{10}&=
   \epsilon ^2 p_4^2 \text{I}_{0,0,2,1,0,2,0,0,0}\,,\nonumber\\   B_{11}&= \epsilon ^2 r_1 r_6 \left(\frac{1}{2} \text{I}_{0,0,2,1,0,2,0,0,0}+\text{I}_{0,0,2,2,0,1,0,0,0}\right)\,,\nonumber\\   B_{12}&= t
   \epsilon ^2 r_1 r_6 \text{I}_{0,1,1,2,2,0,0,0,0}\,,\nonumber\\   B_{13}&= s \epsilon ^3 \text{I}_{1,0,2,0,0,1,1,0,0}\,,\nonumber\\   B_{14}&= \epsilon ^3 \left(-s-t+p_4^2\right)
   \text{I}_{1,0,0,0,2,1,1,0,0}\,,\nonumber\\   B_{15}&= \epsilon ^3 \left(p_4^2-t\right) \text{I}_{0,1,0,1,2,0,1,0,0}\,,\nonumber\\   B_{16}&= \epsilon ^2 r_4 r_9 \left(\frac{1}{2}
   \text{I}_{0,0,2,1,0,2,0,0,0}+\text{I}_{0,0,2,2,0,1,0,0,0}+t \text{I}_{0,2,0,1,2,0,1,0,0}\right)\,,\nonumber\\   B_{17}&= \epsilon ^3 \left(p_4^2-t\right) \text{I}_{0,0,2,1,0,1,1,0,0}\,,\nonumber\\   B_{18}&=
   \epsilon ^3 \left(p_4^2-s\right) \text{I}_{1,0,1,0,1,2,0,0,0}\,,\nonumber\\   B_{19}&= \epsilon ^2 m^2 \left(p_4^2-s\right) \text{I}_{1,0,1,0,1,3,0,0,0}\,,\nonumber\\   B_{20}&= \frac{\epsilon ^2
   r_1 r_6}{p_4^2-2 s} \left(\left(p_4^2-s\right) \left(-\frac{3\epsilon}{2}\text{I}_{1,0,1,0,1,2,0,0,0}+m^2
   \text{I}_{1,0,1,0,1,3,0,0,0}\right)\right. -\nonumber\\ &- \left. \frac{3}{4} s \text{I}_{1,0,2,0,0,2,0,0,0}+\left(s^2-p_4^2 s+m^2 p_4^2\right) \text{I}_{1,0,2,0,1,2,0,0,0}\right)\,,\nonumber\\   B_{21}&= \epsilon ^3
   \left(p_4^2-t\right) \text{I}_{0,0,1,1,1,0,2,0,0}\,,\nonumber\\   B_{22}&= \epsilon ^2 m^2 \left(p_4^2-t\right) \text{I}_{0,0,1,1,1,0,3,0,0}\,,\nonumber\\   B_{23}&= \frac{\epsilon ^2 r_1
   r_6}{p_4^2-2 t} \left(\left(p_4^2-t\right) \left(-\frac{3\epsilon}{2} \text{I}_{0,0,1,1,1,0,2,0,0}+m^2
   \text{I}_{0,0,1,1,1,0,3,0,0}\right)\right. -\nonumber\\ &- \left.\frac{3}{4} t \text{I}_{0,0,2,1,0,0,2,0,0}+\left(t^2-p_4^2 t+m^2 p_4^2\right) \text{I}_{0,0,2,1,1,0,2,0,0}\right)\,,\nonumber\\   B_{24}&= (s+t) \epsilon ^3
   \text{I}_{1,0,1,0,1,0,2,0,0}\,,\nonumber\\   B_{25}&= (s+t) \epsilon ^2 m^2 \text{I}_{1,0,1,0,1,0,3,0,0}\,,\nonumber\\   B_{26}&= \frac{\epsilon ^2 r_1 r_6}{-2 s-2 t+p_4^2}
   \left(\frac{3}{4} \left(-s-t+p_4^2\right) \text{I}_{1,0,0,0,2,0,2,0,0}+(s+t) \left(\frac{3 \epsilon }{2}\times \right. \right. \nonumber\\ &\times \left.
   \text{I}_{1,0,1,0,1,0,2,0,0}  -m^2 \text{I}_{1,0,1,0,1,0,3,0,0}\bigg)+\left(\left(s+t-m^2\right) p_4^2-(s+t)^2\right) \text{I}_{1,0,1,0,2,0,2,0,0}\right)\,,\nonumber\\   B_{27}&= \epsilon ^4
   \left(p_4^2-t\right) \text{I}_{0,1,0,1,1,1,1,0,0}\,,\nonumber\\   B_{28}&= \epsilon ^4 \left(p_4^2-t\right) \text{I}_{0,1,1,1,1,0,1,0,0}\,,\nonumber\\   B_{29}&= \epsilon ^3 \left(p_4^2-t\right) r_1 r_6
   \text{I}_{0,1,1,1,2,0,1,0,0}\,,\nonumber\\   B_{30}&= -2 p_4^2 \text{I}_{0,1,1,1,1,0,1,0,0} \epsilon ^4+\left(t+4 m^2-3 p_4^2\right) \text{I}_{0,1,0,1,1,0,2,0,0} \epsilon ^3+t m^2
   \text{I}_{0,1,1,1,1,0,2,0,0} \epsilon ^2+\nonumber\\ &+    \left(t+4 m^2-p_4^2\right) \left(-\frac{1}{2} \text{I}_{0,0,2,1,0,2,0,0,0}-\epsilon  \text{I}_{0,1,1,1,0,2,0,0,0}-t
   \text{I}_{0,1,1,2,0,2,0,0,0}\right) \epsilon ^2+ \nonumber\\ &+   \left(4 m^2-p_4^2\right) \left(-\frac{1}{2} \text{I}_{0,0,2,0,1,0,2,0,0}+\frac{1}{4} \text{I}_{0,1,0,0,2,0,2,0,0}+\frac{1}{2}
   \epsilon  \left(t+p_4^2\right) \text{I}_{0,1,1,1,2,0,1,0,0}+ \right.\nonumber\\ &+   t \text{I}_{0,2,2,1,1,0,0,0,0}\bigg) \epsilon ^2+\frac{\epsilon ^2}{p_4^2-2 t} \left(\frac{1}{2}
   \epsilon  \left(5 \left(p_4^2\right){}^2-7 t p_4^2-12 m^2 p_4^2+12 t m^2\right) \text{I}_{0,0,1,1,1,0,2,0,0}+ \right. \nonumber\\ &+\left. \left(4 m^2-p_4^2\right) \bigg(m^2 \left(p_4^2-t\right)
   \text{I}_{0,0,1,1,1,0,3,0,0}+\left(t^2-p_4^2 t+m^2 p_4^2\right) \text{I}_{0,0,2,1,1,0,2,0,0}- \right.\nonumber\\ &- \left. \left.\frac{3}{4} t \text{I}_{0,1,0,0,2,2,0,0,0}\right)\right)+\frac{\epsilon
   ^2}{p_4^2-t} \left(t \left(t+4 m^2-p_4^2\right) \text{I}_{0,0,2,2,0,1,0,0,0}+\frac{1}{2} \left(\left(p_4^2\right){}^2- \right. \right. \nonumber\\ &-  \left.  t p_4^2-4 m^2 p_4^2-4 t m^2\right)
   \text{I}_{0,2,0,0,1,0,2,0,0}\bigg)\,,\nonumber\\   B_{31}&= \epsilon ^4 \left(p_4^2-t\right) \text{I}_{0,0,1,1,1,1,1,0,0}\,,\nonumber\\   B_{32}&= \epsilon ^3 \left(p_4^2-t\right) r_1 r_6
   \text{I}_{0,0,2,1,1,1,1,0,0}\,,\nonumber\\   B_{33}&= t \epsilon ^3 r_2 r_7 \text{I}_{1,1,1,1,0,2,0,0,0}\,,\nonumber\\   B_{34}&= t \epsilon ^2 \left(\epsilon  \left((-1+2 \epsilon )
   \text{I}_{1,1,1,1,0,1,0,0,0}+\left(-4 m^2+s\right) \text{I}_{1,1,1,1,0,2,0,0,0}\right)-\text{I}_{0,1,2,2,0,0,0,0,0}\right)\,,\nonumber\\   B_{35}&= t \epsilon ^3 r_5 r_{10}
   \text{I}_{1,1,0,1,2,0,1,0,0}\,,\nonumber\\   B_{36}&= t \epsilon ^2 \left(\epsilon  \left((-1+2 \epsilon ) \text{I}_{1,1,0,1,1,0,1,0,0}+\left(-4 m^2+p_4^2-s-t\right)
   \text{I}_{1,1,0,1,2,0,1,0,0}\right)-\text{I}_{0,2,0,1,0,0,2,0,0}\right)\,,\nonumber\\   B_{37}&= \epsilon ^4 \left(-s-t+p_4^2\right) \text{I}_{1,0,1,1,0,1,1,0,0}\,,\nonumber\\   B_{38}&= \epsilon ^3 r_2 r_3
   r_{11} \text{I}_{1,0,2,1,0,1,1,0,0}\,,\nonumber\\   B_{39}&= s \epsilon ^4 \text{I}_{1,1,0,0,1,1,1,0,0}\,,\nonumber\\   B_{40}&= \epsilon ^3 r_3 r_5 r_{13} \text{I}_{1,1,0,0,2,1,1,0,0}\,,\nonumber\\   B_{41}&= (s+t) \epsilon
   ^4 \text{I}_{1,1,1,0,1,1,0,0,0}\,,\nonumber\\   B_{42}&= \epsilon ^3 r_2 r_3 r_{11} \text{I}_{1,1,1,0,1,2,0,0,0}\,,\nonumber\\   B_{43}&= \left(\frac{1}{2} \left(-s t+2 m^2 t+2 s m^2\right)
   \text{I}_{1,1,1,0,1,2,0,0,0}+\frac{m^2 s^2+2 t m^2 s-t p_4^2 s+t^2 m^2}{s}\right. \times \nonumber\\ &\times \text{I}_{1,1,1,0,2,1,0,0,0}\bigg) \epsilon ^3+\left(2 s m^2+2 t m^2-s
   p_4^2\right) \left(\frac{1}{s \left(p_4^2-2 t\right)} \left(\frac{3}{4} t \text{I}_{0,1,0,0,2,2,0,0,0}+ \right.\right.\nonumber\\ &+ \left.\left.\left(p_4^2-t\right) \left(\frac{3}{2} \epsilon 
   \text{I}_{0,1,1,0,1,2,0,0,0}-m^2 \text{I}_{0,1,1,0,1,3,0,0,0}\right)+\left(-t^2+p_4^2 t-m^2 p_4^2\right) \text{I}_{0,1,1,0,2,2,0,0,0}\right)+\right.\nonumber\\ &+\left. \frac{1}{p_4^2-2
   s} \left(\frac{1}{s} \left(\left(p_4^2-s\right) \left(m^2 \text{I}_{1,0,1,0,1,3,0,0,0}-\frac{3}{2} \epsilon 
   \text{I}_{1,0,1,0,1,2,0,0,0}\right)+ \right.\right.\right.\nonumber\\ &+\left. \left(s^2-p_4^2 s+ m^2 p_4^2\right) \text{I}_{1,0,2,0,1,2,0,0,0}\bigg)-\frac{3}{4} \text{I}_{1,0,2,0,0,2,0,0,0}\right)\Bigg) \epsilon
   ^2\,,\nonumber\\   B_{44}&= \epsilon ^4 \left(p_4^2-s\right) \text{I}_{1,0,1,1,1,0,1,0,0}\,,\nonumber\\   B_{45}&= \epsilon ^3 r_3 r_5 r_{13} \text{I}_{1,0,1,1,1,0,2,0,0}\,,\nonumber\\   B_{46}&= \epsilon ^2
   \left(\frac{-2 s m^2+2 p_4^2 m^2-t p_4^2}{p_4^2-2 t} \left(\frac{1}{t}\left(\left(p_4^2-t\right) \left(m^2
   \text{I}_{0,0,1,1,1,0,3,0,0}-\frac{3}{2} \epsilon  \text{I}_{0,0,1,1,1,0,2,0,0}\right)+ \right.\right.\right.\nonumber\\&+\left.\left. \left(t^2-p_4^2 t+m^2 p_4^2\right) \text{I}_{0,0,2,1,1,0,2,0,0}\bigg)-\frac{3}{4}
   \text{I}_{0,0,2,1,0,0,2,0,0}\right)+\frac{-2 s m^2+2 p_4^2 m^2-t p_4^2}{t \left(2 s+2 t-p_4^2\right)}\times \right. \nonumber\\&\times \left. \left(\frac{3}{4} \left(-s-t+p_4^2\right)
   \text{I}_{1,0,0,0,2,0,2,0,0}+(s+t) \left(\frac{3}{2} \epsilon  \text{I}_{1,0,1,0,1,0,2,0,0}-m^2 \text{I}_{1,0,1,0,1,0,3,0,0}\right)+ \right.\right.\nonumber\\ &+ \left.\left(-s^2-2 t s+p_4^2 s-t^2+t p_4^2-m^2
   p_4^2\right) \text{I}_{1,0,1,0,2,0,2,0,0}\bigg)+\frac{1}{2} \epsilon  \left(t^2+s t-p_4^2 t-2 s m^2+\right.\right. \nonumber\\ &+ \left.\left. 2 m^2 p_4^2\right) \text{I}_{1,0,1,1,1,0,2,0,0}+\epsilon 
   \frac{m^2 s^2+t p_4^2 s-2 m^2 p_4^2 s-t \left(p_4^2\right){}^2+m^2 \left(p_4^2\right){}^2+t^2 p_4^2}{t} \times \right.\nonumber\\&\times \text{I}_{1,0,1,1,2,0,1,0,0}\bigg)\,,\nonumber\\   B_{47}&= \epsilon
   ^4 \left(p_4^2-t\right) \text{I}_{1,0,1,0,1,1,1,0,0}\,,\nonumber\\   B_{48}&= \epsilon ^3 r_1 r_6 \left(\left(s+t-p_4^2\right) \text{I}_{1,0,1,0,2,1,1,0,0}+s
   \text{I}_{1,0,2,0,1,1,1,0,0}\right)\,,\nonumber\\   B_{49}&= \epsilon ^3 r_2 r_5 r_{12} \text{I}_{2,0,1,0,1,1,1,0,0}\,,\nonumber\\   B_{50}&= \frac{1}{2} \epsilon ^3 \left(\left(-p_4^4+\left(2 m^2+s+t\right)
   p_4^2-2 m^2 t\right) \text{I}_{1,0,1,0,2,1,1,0,0}+\left(\left(2 m^2-s\right) p_4^2-2 m^2 t\right) \text{I}_{1,0,2,0,1,1,1,0,0}\right)\,,\nonumber\\   B_{51}&= \epsilon ^2 \left(2
   \left(s+p_4^2\right) \left(-\epsilon  \text{I}_{1,0,1,0,1,0,2,0,0}+m^2 \text{I}_{1,0,1,0,1,0,3,0,0}\right)+\epsilon  \left(p_4^2-t\right)
   \text{I}_{2,0,1,0,1,1,1,-1,0}\right)\,,\nonumber\\   B_{52}&= s t \epsilon ^4 \text{I}_{1,1,1,1,0,1,1,0,0}\,,\nonumber\\   B_{53}&= t \epsilon ^4 \left(-s-t+p_4^2\right) \text{I}_{1,1,0,1,1,1,1,0,0}\,,\nonumber\\   B_{54}&= t
   \epsilon ^4 \left(p_4^2-s\right) \text{I}_{1,1,1,1,1,1,0,0,0}\,,\nonumber\\   B_{55}&= t \epsilon ^3 r_1 r_6 \left(-\text{I}_{1,1,1,0,1,2,0,0,0}+2
   \text{I}_{1,1,1,1,0,2,0,0,0}+\left(p_4^2-s\right) \text{I}_{1,1,2,1,1,1,0,0,0}\right)\,,\nonumber\\   B_{56}&= t (s+t) \epsilon ^4 \text{I}_{1,1,1,1,1,0,1,0,0}\,,\nonumber\\   B_{57}&= t \epsilon ^3 r_1 r_6
   \left(-\text{I}_{1,0,1,1,1,0,2,0,0}+2 \text{I}_{1,1,0,1,1,0,2,0,0}+(s+t) \text{I}_{1,1,1,1,2,0,1,0,0}\right)\,,\nonumber\\   B_{58}&= \epsilon ^4 r_5 r_{15}
   \text{I}_{1,0,1,1,1,1,1,0,0}\,,\nonumber\\   B_{59}&= \epsilon ^4 \left(p_4^2-t\right) \left(\text{I}_{1,0,1,1,0,1,1,0,0}-\text{I}_{1,0,1,1,1,1,1,-1,0}\right)\,,\nonumber\\   B_{60}&=
   \frac{s^2-p_4^2 s+t^2-t p_4^2}{p_4^2-s} \text{I}_{1,0,1,0,1,1,1,0,0} \epsilon ^4+\left(-p_4^2+s+t\right) \left(\text{I}_{1,-1,1,1,1,1,1,0,0}+ \right.\nonumber\\ &+ \left.t
   \text{I}_{1,0,1,1,1,1,1,0,0}\right) \epsilon ^4+\frac{t}{p_4^2-s} \left(\frac{1}{4} (B_{6}+B_{10})+\frac{1}{2}
   (B_{8}-B_{13}-B_{14}+\right.\nonumber\\ &+ B_{18}+B_{21})-B_{22}-B_{44}+B_{46}+B_{50}-B_{59}\bigg)\,,\nonumber\\   B_{61}&= \epsilon ^3 r_1 r_6 \left(\left(-s-t+p_4^2\right) \left((-2 \epsilon )
   \text{I}_{1,0,1,1,1,1,1,0,0}-\text{I}_{1,0,1,1,1,0,2,0,0}\right)+s \text{I}_{1,0,2,1,0,1,1,0,0}+ \right.\nonumber\\ &+ \left.\left(t-p_4^2\right) \text{I}_{1,0,2,1,1,1,1,-1,0}\right)\,,\nonumber\\   B_{62}&= \epsilon ^4
   r_2 r_{14} \text{I}_{1,1,1,0,1,1,1,0,0}\,,\nonumber\\   B_{63}&= \epsilon ^4 \left(p_4^2-t\right) \left(\text{I}_{1,1,1,0,1,1,0,0,0}-\text{I}_{1,1,1,0,1,1,1,0,-1}\right)\,,\nonumber\\   B_{64}&= s
   \text{I}_{1,1,1,-1,1,1,1,0,0} \epsilon ^4+(s t) \text{I}_{1,1,1,0,1,1,1,0,0} \epsilon ^4+\frac{t}{s+t} \left(\frac{1}{4} (-B_{6}-B_{10})+B_{22}+ \right. \nonumber\\ &+ \left. \frac{1}{2}
   (-B_{4}+B_{13}+B_{14}-B_{21}-B_{24})-B_{31}+B_{41}-B_{43}-B_{50}\right)+ \nonumber\\ &+\frac{1}{s+t} \left(\left(-s^2-t s-2 t^2+2 t p_4^2\right) \text{I}_{1,0,1,0,1,1,1,0,0}
   \epsilon ^4+s B_{63}\right)\,,\nonumber\\   B_{65}&= r_1 r_6 \left(s \left(\text{I}_{1,1,0,0,2,1,1,0,0}+2 \epsilon 
   \text{I}_{1,1,1,0,1,1,1,0,0}+\text{I}_{1,1,1,0,1,2,0,0,0}\right)+\left(t-p_4^2\right)
   \left(\text{I}_{1,0,1,0,2,1,1,0,0}- \right.\right.\nonumber\\ &-\left.\left. \text{I}_{1,1,1,0,2,1,0,0,0}+\text{I}_{1,1,1,0,2,1,1,0,-1}\right)\right) \epsilon ^3+\frac{\epsilon ^2 \left(p_4^2-t\right)
   \left(2 s m^2+2 t m^2-s p_4^2\right) r_1 r_6}{\left(p_4^2-2 t\right) \left(m^2 s^2+2 t m^2 s-t p_4^2 s+t^2 m^2\right)}\times \nonumber\\ &\times  \left(\frac{3}{4} t
   \left(\text{I}_{0,1,0,0,2,2,0,0,0}-\text{I}_{0,0,2,1,0,0,2,0,0}\right)+\left(p_4^2-t\right) \left(-\frac{3}{2} \epsilon  \text{I}_{0,0,1,1,1,0,2,0,0}+m^2
   \text{I}_{0,0,1,1,1,0,3,0,0}+ \right. \right.\nonumber\\ &+ \left.\left. \frac{3}{2} \epsilon  \text{I}_{0,1,1,0,1,2,0,0,0}-m^2 \text{I}_{0,1,1,0,1,3,0,0,0}\right)+\left(t^2-p_4^2 t+m^2 p_4^2\right)
   \left(\text{I}_{0,0,2,1,1,0,2,0,0}-\text{I}_{0,1,1,0,2,2,0,0,0}\right)\right)\,.
\end{align}
In addition we consider the following choice of basis for the elliptic sectors,
\begin{align}
   B_{66}&= s \epsilon ^4 r_2 \text{I}_{0,1,1,1,1,1,1,0,0}\,,\nonumber\\   B_{67}&= \epsilon ^4 r_2
   \text{I}_{-2,1,1,1,1,1,1,0,0}\,,\nonumber\\   B_{68}&= t \epsilon ^4 \left(p_4^2-t\right) \left(\text{I}_{1,1,1,1,1,1,1,-1,0}-\text{I}_{1,1,1,1,1,1,1,0,-1}\right)\,,\nonumber\\   B_{69}&= t \epsilon ^4
   \left(\text{I}_{1,1,1,1,1,1,1,-2,0}-\text{I}_{1,1,1,1,1,1,1,0,-2}+s \left(\text{I}_{1,1,1,1,1,1,1,-1,0}-\text{I}_{1,1,1,1,1,1,1,0,-1}\right)\right)\,,\nonumber\\   B_{70}&= t \epsilon ^4 r_{16}
   \left(\text{I}_{1,1,1,1,1,1,1,-1,0}+\text{I}_{1,1,1,1,1,1,1,0,-1}\right)\,,\nonumber\\   B_{71}&= \frac{t \epsilon ^4 \left(p_4^2-t\right){}^2}{\left(2 s+t-p_4^2\right)
   r_{16}} \text{I}_{1,1,1,1,1,1,1,-1,-1}\,,\nonumber\\   B_{72}&= t \epsilon ^4 r_2 r_5 r_{12} \text{I}_{1,1,1,1,1,1,1,0,0}\,,\nonumber\\   B_{73}&= t \epsilon ^4
   \left(\text{I}_{1,1,1,1,1,1,1,-2,0}+\frac{4 s}{-p_4^2+2 s+t} \text{I}_{1,1,1,1,1,1,1,-1,-1}+\text{I}_{1,1,1,1,1,1,1,0,-2}+ \right.\nonumber\\ &+ \left. \frac{1}{4} \left(4
   s+t-p_4^2\right) \left(\text{I}_{1,1,1,1,1,1,1,-1,0}+\text{I}_{1,1,1,1,1,1,1,0,-1}\right)\right)
 \end{align}
 \FloatBarrier
 
\section{Maximal cut of the elliptic sectors}
\label{app:ellipticcurveappendix}
Of the discussed integral families, only two are genuinely elliptic as we have seen. These are the six-propagator sector 126 consisting of integrals $I_{66}$ and $I_{67}$, and the seven-propagator sector 127 consisting of the six integrals $I_{68}$-$I_{73}$. Representing these sectors by their most basic members $I_{011111100}$ and $I_{111111100}$ we may derive their maximal cut using the loop-by-loop Baikov representation \cite{Frellesvig:2017aai}, in both cases yielding a one-fold integral representation. In $d=4$ these cut integrals become
\begin{align}
I_{011111100} &\rightarrow \int \frac{d z}{(p_4^2 {-} t) \sqrt{ z \, (z {+} p_4^2 {-} t) \, \big(z^2 {+} (p_4^2 {-} t)z {-} 4 m^2 t \big) }} \\
I_{111111100} &\rightarrow \int \frac{d z}{t \, (z{+}s) \sqrt{ z \, (z {+} p_4^2 {-} t) \, \big(z^2 {+} (p_4^2 {-} t)z {-} 4 m^2 t \big) }}
\end{align}
of which the first has been analyzed in refs. \cite{Primo:2016ebd, Hidding:2017jkk}, and the second discussed in ref. \cite{Frellesvig:2019kgj}. These two integrals both evaluate to elliptic integrals, and can be expressed as integrals over an elliptic curve, proving that genuinely elliptic structures have to be present in representations of the integrals in these sectors.

We see that this elliptic curve,
\begin{align}
y^2 = z \, (z {+} p_4^2 {-} t) \, \big(z^2 {+} (p_4^2 {-} t)z {-} 4 m^2 t \big)
\end{align}
is in common between the two sectors, implying that it might be possible to express all integrals in the integral family considered in this paper, in terms of polylogarithmic functions defined on that elliptic curve in the spirit of refs. \cite{BrownLevin, Broedel:2017kkb}.

\section{Polylogarithmic sectors up to weight 2}
\label{app:PolylogsWeight2}
We provide below the results for the canonical basis integrals up to weight 2 in terms of logarithms and dilogarithms, provided in the region $\mathcal{R}$ defined in Eq. (\ref{eq:regionR}), where the result is manifestly real-valued. Let $\vec{B} = \sum_{k=0}^\infty \vec{B}^{(k) }\epsilon^k$. At weight 0 we have
\begin{align}
    B_1^{(0)} = 1\,,\quad B_2^{(0)} = 1\,,\quad B_i^{(0)} = 0\quad\text{for }i = 3,\ldots,65.
\end{align}
At weight 1 we have
\begin{align}
\begin{array}{lll}
 B_1^{(1)}=-2 \log \left(l_1\right) \,,& B_2^{(1)}=-\log \left(l_1\right)-\log \left(-l_4\right) \,,& B_3^{(1)}=\log \left(-l_{27}\right) \,,\\
 B_5^{(1)}=-\log \left(-l_{25}\right) \,,& B_7^{(1)}=-\log \left(-l_{26}\right) \,,& B_9^{(1)}=-\log \left(l_{28}\right) \,,\\
 B_{11}^{(1)}=-\log \left(-l_{27}\right) \,,& B_{12}^{(1)}=\log \left(-l_{27}\right) \,,& B_{33}^{(1)}=-\log \left(-l_{25}\right) \,,\\
 B_{35}^{(1)}=-\log \left(l_{28}\right) \,,& B_{55}^{(1)}=-2 \log \left(-l_{27}\right) \,,& B_{57}^{(1)}=-2 \log \left(-l_{27}\right) \,,\\
\end{array}
\end{align}
and $B_i^{(1)} = 0$ for all other $i\leq 65$. Finally, at weight 2 we have
\begin{align}
 B_1^{(2)}&=\zeta _2+2 \log ^2\left(l_1\right) \nonumber\\
 B_2^{(2)}&=\frac{1}{2} \log ^2\left(l_1\right)+\frac{1}{2} \log ^2\left(-l_4\right)+\log \left(-l_4\right) \log \left(l_1\right) \nonumber\\
 B_3^{(2)}&=\zeta _2+2 \text{Li}_2\left(l_{27}^{-1}\right)+\frac{1}{2} \log ^2\left(-l_{27}\right)-\log \left(l_1\right) \log \left(-l_{27}\right)-\log \left(-l_6\right) \log \left(-l_{27}\right) \nonumber\\
 B_4^{(2)}&=-\log ^2\left(-l_{25}\right) \nonumber\\
 B_5^{(2)}&=-\zeta _2-6 \text{Li}_2\left(l_{25}^{-1}\right)-2 \text{Li}_2\left(-l_{25}^{-1}\right)-2 \log ^2\left(-l_{25}\right)-2 \log \left(l_1\right) \log \left(-l_{25}\right)\nonumber\\&\quad+\log \left(-l_3\right) \log \left(-l_{25}\right)+3 \log \left(l_{10}\right) \log \left(-l_{25}\right) \nonumber\\
 B_6^{(2)}&=-\log ^2\left(-l_{26}\right) \nonumber\\
 B_7^{(2)}&=-\zeta _2-6 \text{Li}_2\left(l_{26}^{-1}\right)-2 \text{Li}_2\left(-l_{26}^{-1}\right)-2 \log ^2\left(-l_{26}\right)-2 \log \left(l_1\right) \log \left(-l_{26}\right)\nonumber\\&\quad+\log \left(-l_4\right) \log \left(-l_{26}\right)+3 \log \left(l_{11}\right) \log \left(-l_{26}\right) \nonumber\\
 B_8^{(2)}&=-\log ^2\left(l_{28}\right) \nonumber\\
 B_9^{(2)}&=-\zeta _2-2 \text{Li}_2\left(l_{28}^{-1}\right)-6 \text{Li}_2\left(-l_{28}^{-1}\right)-2 \log ^2\left(l_{28}\right)-2 \log \left(l_1\right) \log \left(l_{28}\right)\nonumber\\&\quad+\log \left(l_9\right) \log \left(l_{28}\right)+3 \log \left(l_{13}\right) \log \left(l_{28}\right) \nonumber\\
 B_{10}^{(2)}&=-\log ^2\left(-l_{27}\right) \nonumber\\
 B_{11}^{(2)}&=-\zeta _2-2 \text{Li}_2\left(-l_{27}^{-1}\right)-6 \text{Li}_2\left(l_{27}^{-1}\right)-2 \log ^2\left(-l_{27}\right)-2 \log \left(l_1\right) \log \left(-l_{27}\right)\nonumber\\&\quad+\log \left(-l_2\right) \log \left(-l_{27}\right)+3 \log \left(-l_6\right) \log \left(-l_{27}\right) \nonumber\\
 B_{12}^{(2)}&=\zeta _2+2 \text{Li}_2\left(l_{27}^{-1}\right)+\frac{1}{2} \log ^2\left(-l_{27}\right)-\log \left(-l_4\right) \log \left(-l_{27}\right)-\log \left(-l_6\right) \log \left(-l_{27}\right) \nonumber\\
 B_{13}^{(2)}&=0 \nonumber\\
 B_{14}^{(2)}&=0 \nonumber\\
 B_{15}^{(2)}&=0 \nonumber\\
 B_{16}^{(2)}&=\text{Li}_2\left(l_{29} l_{27}^{-1}\right)-\text{Li}_2\left(l_{27}^{-1} l_{29}^{-1}\right)-\log \left(-l_{27}\right) \log \left(-l_{29}\right)-\frac{1}{2} \log \left(-l_{27}\right) \log \left(l_{54}\right) \nonumber\\
 B_{17}^{(2)}&=0 \nonumber\\
 B_{18}^{(2)}&=0 \nonumber\\
 B_{19}^{(2)}&=\frac{1}{4} \log ^2\left(-l_{25}\right)-\frac{1}{4} \log ^2\left(-l_{27}\right) \nonumber\\
 B_{20}^{(2)}&=\zeta _2+\text{Li}_2\left(-l_{27}^{-1}\right)-\text{Li}_2\left(l_{25} l_{27}^{-1}\right)-\text{Li}_2\left(l_{25}^{-1} l_{27}^{-1}\right)-\frac{1}{2} \log ^2\left(-l_{25}\right)\nonumber\\&\quad-\frac{1}{4} \log ^2\left(-l_{27}\right)+\log \left(l_{43}\right) \log \left(-l_{25}\right)-\frac{1}{2} \log \left(l_1\right) \log \left(-l_{27}\right)-\frac{1}{2} \log \left(-l_2\right) \log \left(-l_{27}\right)\nonumber\\&\quad+\log \left(-l_7\right) \log \left(-l_{27}\right) \nonumber\\
 B_{21}^{(2)}&=0 \nonumber\\
 B_{22}^{(2)}&=\frac{1}{4} \log ^2\left(-l_{26}\right)-\frac{1}{4} \log ^2\left(-l_{27}\right) \nonumber\\
 B_{23}^{(2)}&=\zeta _2+\text{Li}_2\left(-l_{27}^{-1}\right)-\text{Li}_2\left(l_{26} l_{27}^{-1}\right)-\text{Li}_2\left(l_{26}^{-1} l_{27}^{-1}\right)-\frac{1}{2} \log ^2\left(-l_{26}\right)-\frac{1}{4} \log ^2\left(-l_{27}\right)\nonumber\\&\quad+\log \left(l_{44}\right) \log \left(-l_{26}\right)-\frac{1}{2} \log \left(l_1\right) \log \left(-l_{27}\right)-\frac{1}{2} \log \left(-l_2\right) \log \left(-l_{27}\right)+\log \left(-l_8\right) \log \left(-l_{27}\right) \nonumber\\
 B_{24}^{(2)}&=0 \nonumber\\
 B_{25}^{(2)}&=\frac{1}{4} \log ^2\left(l_{28}\right)-\frac{1}{4} \log ^2\left(-l_{27}\right) \nonumber\\
 B_{26}^{(2)}&=\zeta _2+\text{Li}_2\left(-l_{27}^{-1}\right)-\text{Li}_2\left(-l_{28} l_{27}^{-1}\right)-\text{Li}_2\left(-l_{27}^{-1} l_{28}^{-1}\right)-\frac{1}{4} \log ^2\left(-l_{27}\right)-\frac{1}{2} \log ^2\left(l_{28}\right)\nonumber\\&\quad-\frac{1}{2} \log \left(l_1\right) \log \left(-l_{27}\right)-\frac{1}{2} \log \left(-l_2\right) \log \left(-l_{27}\right)+\log \left(-l_5\right) \log \left(-l_{27}\right)+\log \left(l_{28}\right) \log \left(l_{48}\right) \nonumber\\
 B_{27}^{(2)}&=0 \nonumber\\
 B_{28}^{(2)}&=0 \nonumber\\
 B_{29}^{(2)}&=0 \nonumber\\
 B_{30}^{(2)}&=\frac{1}{2} \log ^2\left(-l_{26}\right) \nonumber\\
 B_{31}^{(2)}&=0 \nonumber\\
 B_{32}^{(2)}&=0 \nonumber\\
 B_{33}^{(2)}&=-\zeta _2-6 \text{Li}_2\left(l_{25}^{-1}\right)-2 \text{Li}_2\left(-l_{25}^{-1}\right)-\text{Li}_2\left(l_{25} l_{27}^{-1}\right)+\text{Li}_2\left(l_{25}^{-1} l_{27}^{-1}\right)-2 \log ^2\left(-l_{25}\right)\nonumber\\&\quad-2 \log \left(l_1\right) \log \left(-l_{25}\right)+\log \left(-l_3\right) \log \left(-l_{25}\right)+\log \left(-l_4\right) \log \left(-l_{25}\right)-\log \left(-l_7\right) \log \left(-l_{25}\right)\nonumber\\&\quad+3 \log \left(l_{10}\right) \log \left(-l_{25}\right)+\log \left(-l_{27}\right) \log \left(-l_{25}\right)-\log \left(-l_{27}\right) \log \left(l_{43}\right) \nonumber\\
 B_{34}^{(2)}&=\frac{1}{2} \log ^2\left(-l_{27}\right)-\log ^2\left(-l_{25}\right) \nonumber\\
 B_{35}^{(2)}&=-\zeta _2-\text{Li}_2\left(-l_{28} l_{27}^{-1}\right)-2 \text{Li}_2\left(l_{28}^{-1}\right)-6 \text{Li}_2\left(-l_{28}^{-1}\right)+\text{Li}_2\left(-l_{27}^{-1} l_{28}^{-1}\right)-2 \log ^2\left(l_{28}\right)\nonumber\\&\quad-2 \log \left(l_1\right) \log \left(l_{28}\right)+\log \left(-l_4\right) \log \left(l_{28}\right)-\log \left(-l_5\right) \log \left(l_{28}\right)+\log \left(l_9\right) \log \left(l_{28}\right)\nonumber\\&\quad+3 \log \left(l_{13}\right) \log \left(l_{28}\right)+\log \left(-l_{27}\right) \log \left(l_{28}\right)-\log \left(-l_{27}\right) \log \left(l_{48}\right) \nonumber\\
 B_{36}^{(2)}&=\frac{1}{2} \log ^2\left(-l_{27}\right)-\log ^2\left(l_{28}\right) \nonumber\\
 B_{37}^{(2)}&=0 \nonumber\\
 B_{38}^{(2)}&=-4 \text{Li}_2\left(-l_{33}^{-1}\right)+2 \text{Li}_2\left(l_{25} l_{33}^{-1}\right)+2 \text{Li}_2\left(l_{26} l_{33}^{-1}\right)-2 \text{Li}_2\left(l_{27} l_{33}^{-1}\right)+2 \text{Li}_2\left(l_{25}^{-1} l_{33}^{-1}\right)\nonumber\\&\quad+2 \text{Li}_2\left(l_{26}^{-1} l_{33}^{-1}\right)-2 \text{Li}_2\left(l_{27}^{-1} l_{33}^{-1}\right)+\log ^2\left(-l_{25}\right)+\log ^2\left(-l_{26}\right)-\log ^2\left(-l_{27}\right)\nonumber\\&\quad-2 \log \left(l_{40}\right) \log \left(-l_{25}\right)-2 \log \left(-l_{26}\right) \log \left(l_{39}\right)+2 \log \left(-l_{27}\right) \log \left(l_{53}\right) \nonumber\\
 B_{39}^{(2)}&=0 \nonumber\\
 B_{40}^{(2)}&=-4 \text{Li}_2\left(-l_{41}^{-1}\right)+2 \text{Li}_2\left(l_{26} l_{41}^{-1}\right)-2 \text{Li}_2\left(l_{27} l_{41}^{-1}\right)+2 \text{Li}_2\left(-l_{28} l_{41}^{-1}\right)+2 \text{Li}_2\left(l_{26}^{-1} l_{41}^{-1}\right)\nonumber\\&\quad-2 \text{Li}_2\left(l_{27}^{-1} l_{41}^{-1}\right)+2 \text{Li}_2\left(-l_{28}^{-1} l_{41}^{-1}\right)+\log ^2\left(-l_{26}\right)-\log ^2\left(-l_{27}\right)+\log ^2\left(l_{28}\right)+\log \left(l_{56}\right) \log \left(-l_{26}\right)\nonumber\\&\quad+\log \left(l_{28}\right) \log \left(l_{55}\right)+\log \left(-l_{27}\right) \log \left(l_{60}\right) \nonumber\\
 B_{41}^{(2)}&=0 \nonumber\\
 B_{42}^{(2)}&=-4 \text{Li}_2\left(-l_{33}^{-1}\right)+2 \text{Li}_2\left(l_{25} l_{33}^{-1}\right)+2 \text{Li}_2\left(l_{26} l_{33}^{-1}\right)-2 \text{Li}_2\left(l_{27} l_{33}^{-1}\right)+2 \text{Li}_2\left(l_{25}^{-1} l_{33}^{-1}\right)\nonumber\\&\quad+2 \text{Li}_2\left(l_{26}^{-1} l_{33}^{-1}\right)-2 \text{Li}_2\left(l_{27}^{-1} l_{33}^{-1}\right)+\log ^2\left(-l_{25}\right)+\log ^2\left(-l_{26}\right)-\log ^2\left(-l_{27}\right)\nonumber\\&\quad-2 \log \left(l_{40}\right) \log \left(-l_{25}\right)-2 \log \left(-l_{26}\right) \log \left(l_{39}\right)+2 \log \left(-l_{27}\right) \log \left(l_{53}\right) \nonumber\\
 B_{43}^{(2)}&=\frac{1}{2} \log ^2\left(-l_{25}\right)+\frac{1}{2} \log ^2\left(-l_{26}\right) \nonumber\\
 B_{44}^{(2)}&=0 \nonumber\\
 B_{45}^{(2)}&=-4 \text{Li}_2\left(-l_{41}^{-1}\right)+2 \text{Li}_2\left(l_{26} l_{41}^{-1}\right)-2 \text{Li}_2\left(l_{27} l_{41}^{-1}\right)+2 \text{Li}_2\left(-l_{28} l_{41}^{-1}\right)+2 \text{Li}_2\left(l_{26}^{-1} l_{41}^{-1}\right)\nonumber\\&\quad-2 \text{Li}_2\left(l_{27}^{-1} l_{41}^{-1}\right)+2 \text{Li}_2\left(-l_{28}^{-1} l_{41}^{-1}\right)+\log ^2\left(-l_{26}\right)-\log ^2\left(-l_{27}\right)+\log ^2\left(l_{28}\right)\nonumber\\&\quad+\log \left(l_{56}\right) \log \left(-l_{26}\right)+\log \left(l_{28}\right) \log \left(l_{55}\right)+\log \left(-l_{27}\right) \log \left(l_{60}\right) \nonumber\\
 B_{46}^{(2)}&=\frac{1}{2} \log ^2\left(-l_{26}\right)+\frac{1}{2} \log ^2\left(l_{28}\right) \nonumber\\
 B_{47}^{(2)}&=0 \nonumber\\
 B_{48}^{(2)}&=0 \nonumber\\
 B_{49}^{(2)}&=-4 \text{Li}_2\left(-l_{38}^{-1}\right)+2 \text{Li}_2\left(l_{25} l_{38}^{-1}\right)-2 \text{Li}_2\left(l_{27} l_{38}^{-1}\right)+2 \text{Li}_2\left(-l_{28} l_{38}^{-1}\right)+2 \text{Li}_2\left(l_{25}^{-1} l_{38}^{-1}\right)\nonumber\\&\quad-2 \text{Li}_2\left(l_{27}^{-1} l_{38}^{-1}\right)+2 \text{Li}_2\left(-l_{28}^{-1} l_{38}^{-1}\right)+\log ^2\left(-l_{25}\right)-\log ^2\left(-l_{27}\right)+\log ^2\left(l_{28}\right)\nonumber\\&\quad-2 \log \left(l_{49}\right) \log \left(-l_{25}\right)-2 \log \left(l_{28}\right) \log \left(l_{46}\right)+\log \left(-l_{27}\right) \log \left(l_{61}\right) \nonumber\\
 B_{50}^{(2)}&=0 \nonumber\\
 B_{51}^{(2)}&=\frac{1}{2} \log ^2\left(l_{28}\right)-\frac{1}{2} \log ^2\left(-l_{27}\right) \nonumber\\
 B_{52}^{(2)}&=\frac{1}{2} \log ^2\left(-l_{25}\right) \nonumber\\
 B_{53}^{(2)}&=\frac{1}{2} \log ^2\left(l_{28}\right) \nonumber\\
 B_{54}^{(2)}&=\frac{1}{2} \log ^2\left(-l_{27}\right)-\frac{1}{2} \log ^2\left(-l_{25}\right) \nonumber\\
 B_{55}^{(2)}&=4 \text{Li}_2\left(-l_{27}^{-1}\right)-4 \text{Li}_2\left(l_{25} l_{27}^{-1}\right)+2 \text{Li}_2\left(l_{26} l_{27}^{-1}\right)-4 \text{Li}_2\left(l_{25}^{-1} l_{27}^{-1}\right)\nonumber\\&\quad+2 \text{Li}_2\left(l_{26}^{-1} l_{27}^{-1}\right)-2 \log ^2\left(-l_{25}\right)+\log ^2\left(-l_{26}\right)+4 \log \left(l_{43}\right) \log \left(-l_{25}\right)\nonumber\\&\quad+2 \log \left(l_1\right) \log \left(-l_{27}\right)-2 \log \left(-l_2\right) \log \left(-l_{27}\right)+2 \log \left(-l_4\right) \log \left(-l_{27}\right)\nonumber\\&\quad+4 \log \left(-l_7\right) \log \left(-l_{27}\right)-2 \log \left(-l_8\right) \log \left(-l_{27}\right)-2 \log \left(-l_{26}\right) \log \left(l_{44}\right) \nonumber\\
 B_{56}^{(2)}&=\frac{1}{2} \log ^2\left(-l_{27}\right)-\frac{1}{2} \log ^2\left(l_{28}\right) \nonumber\\
 B_{57}^{(2)}&=4 \text{Li}_2\left(-l_{27}^{-1}\right)+2 \text{Li}_2\left(l_{26} l_{27}^{-1}\right)-4 \text{Li}_2\left(-l_{28} l_{27}^{-1}\right)+2 \text{Li}_2\left(l_{26}^{-1} l_{27}^{-1}\right)-4 \text{Li}_2\left(-l_{27}^{-1} l_{28}^{-1}\right)\nonumber\\&\quad+\log ^2\left(-l_{26}\right)-2 \log ^2\left(l_{28}\right)-2 \log \left(l_{44}\right) \log \left(-l_{26}\right)+2 \log \left(l_1\right) \log \left(-l_{27}\right)\nonumber\\&\quad-2 \log \left(-l_2\right) \log \left(-l_{27}\right)+2 \log \left(-l_4\right) \log \left(-l_{27}\right)+4 \log \left(-l_5\right) \log \left(-l_{27}\right)\nonumber\\&\quad-2 \log \left(-l_8\right) \log \left(-l_{27}\right)+4 \log \left(l_{28}\right) \log \left(l_{48}\right) \nonumber\\
 B_{58}^{(2)}&=0 \nonumber\\
 B_{59}^{(2)}&=0 \nonumber\\
 B_{60}^{(2)}&=0 \nonumber\\
 B_{61}^{(2)}&=2 \zeta _2+4 \text{Li}_2\left(-l_{27}^{-1}\right)+4 \text{Li}_2\left(l_{27}^{-1}\right)-2 \text{Li}_2\left(l_{25} l_{27}^{-1}\right)-2 \text{Li}_2\left(l_{26} l_{27}^{-1}\right)+2 \text{Li}_2\left(-l_{28} l_{27}^{-1}\right)\nonumber\\&\quad-2 \text{Li}_2\left(l_{25}^{-1} l_{27}^{-1}\right)-2 \text{Li}_2\left(l_{26}^{-1} l_{27}^{-1}\right)+2 \text{Li}_2\left(-l_{27}^{-1} l_{28}^{-1}\right)-\log ^2\left(-l_{25}\right)-\log ^2\left(-l_{26}\right)+\log ^2\left(-l_{27}\right)\nonumber\\&\quad+\log ^2\left(l_{28}\right)+2 \log \left(l_{43}\right) \log \left(-l_{25}\right)+2 \log \left(l_1\right) \log \left(-l_{27}\right)-2 \log \left(-l_2\right) \log \left(-l_{27}\right)\nonumber\\&\quad-2 \log \left(-l_5\right) \log \left(-l_{27}\right)-2 \log \left(-l_6\right) \log \left(-l_{27}\right)+2 \log \left(-l_7\right) \log \left(-l_{27}\right)\nonumber\\&\quad+2 \log \left(-l_8\right) \log \left(-l_{27}\right)+2 \log \left(-l_{26}\right) \log \left(l_{44}\right)-2 \log \left(l_{28}\right) \log \left(l_{48}\right) \nonumber\\
 B_{62}^{(2)}&=0 \nonumber\\
 B_{63}^{(2)}&=0 \nonumber\\
 B_{64}^{(2)}&=0 \nonumber\\
 B_{65}^{(2)}&=-2 \zeta _2-4 \text{Li}_2\left(-l_{27}^{-1}\right)-4 \text{Li}_2\left(l_{27}^{-1}\right)-2 \text{Li}_2\left(l_{25} l_{27}^{-1}\right)+2 \text{Li}_2\left(l_{26} l_{27}^{-1}\right)+2 \text{Li}_2\left(-l_{28} l_{27}^{-1}\right)\nonumber\\&\quad-2 \text{Li}_2\left(l_{25}^{-1} l_{27}^{-1}\right)+2 \text{Li}_2\left(l_{26}^{-1} l_{27}^{-1}\right)+2 \text{Li}_2\left(-l_{27}^{-1} l_{28}^{-1}\right)-\log ^2\left(-l_{25}\right)+\log ^2\left(-l_{26}\right)-\log ^2\left(-l_{27}\right)\nonumber\\&\quad+\log ^2\left(l_{28}\right)+2 \log \left(l_{43}\right) \log \left(-l_{25}\right)-2 \log \left(l_1\right) \log \left(-l_{27}\right)+2 \log \left(-l_2\right) \log \left(-l_{27}\right)\nonumber\\&\quad-2 \log \left(-l_5\right) \log \left(-l_{27}\right)+2 \log \left(-l_6\right) \log \left(-l_{27}\right)+2 \log \left(-l_7\right) \log \left(-l_{27}\right)\nonumber\\&\quad-2 \log \left(-l_8\right) \log \left(-l_{27}\right)-2 \log \left(-l_{26}\right) \log \left(l_{44}\right)-2 \log \left(l_{28}\right) \log \left(l_{48}\right)
\end{align}

\section{Plots of basis integrals}
\label{app:plotsappendix}
In this appendix we provide plots of all the integrals of our basis along the line going from the kinematic point $\textrm{P}_{\textrm{below}} = (s=2, t=-1, p_4^2=13/25)$  to $\textrm{P}_{\textrm{above}}=(6,-1,13/25)$, which are obtained using the series solution strategy. The blue and orange lines are the real and imaginary parts and the dots are values obtained with FIESTA for comparison. Note that the only noticeable mismatch is for $B_{68}^{(4)}$ and $B_{69}^{(4)}$, but it is well inside the uncertainty produced by FIESTA (not reported here).
 
{ \centering
\begin{table}[h!]
	\begin{tabular}{ p{4.75cm}  p{4.75cm}  p{4.75cm}}
\img{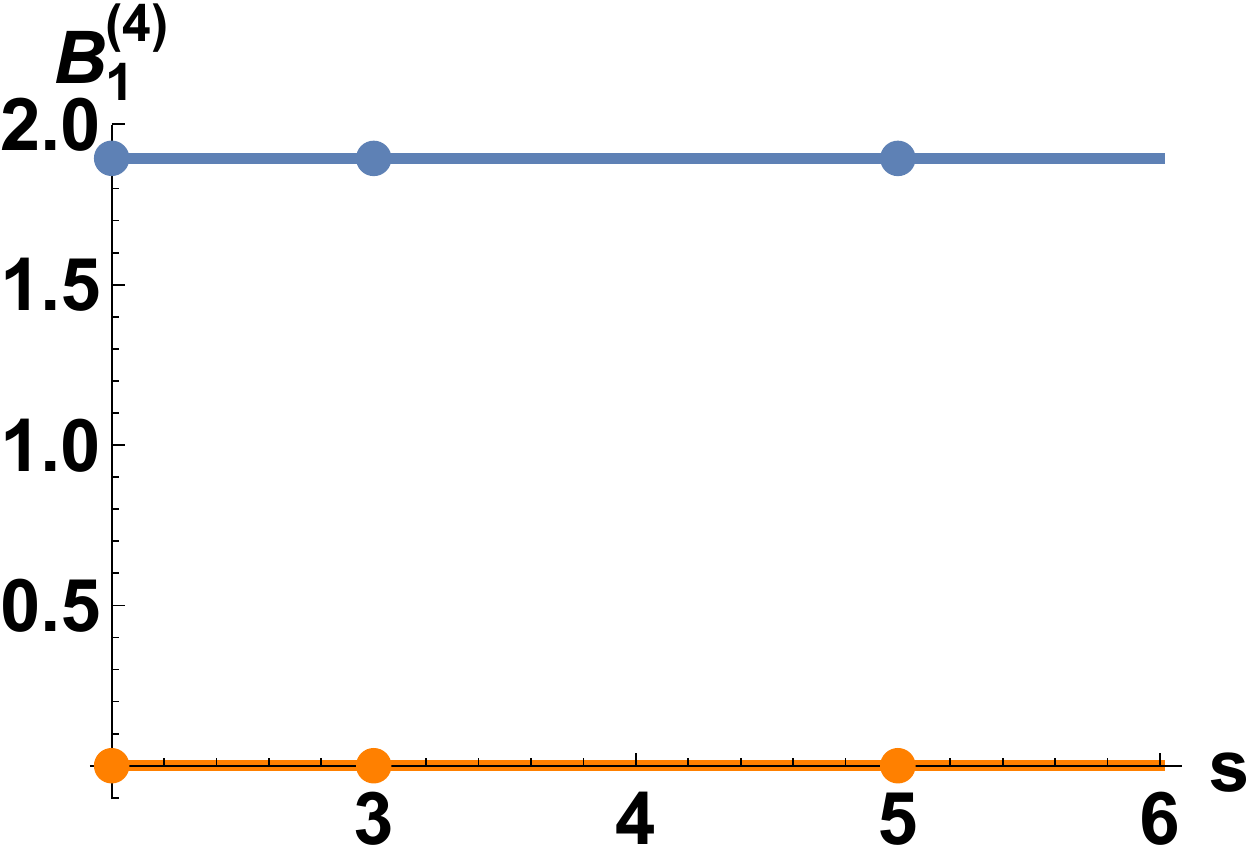} & \img{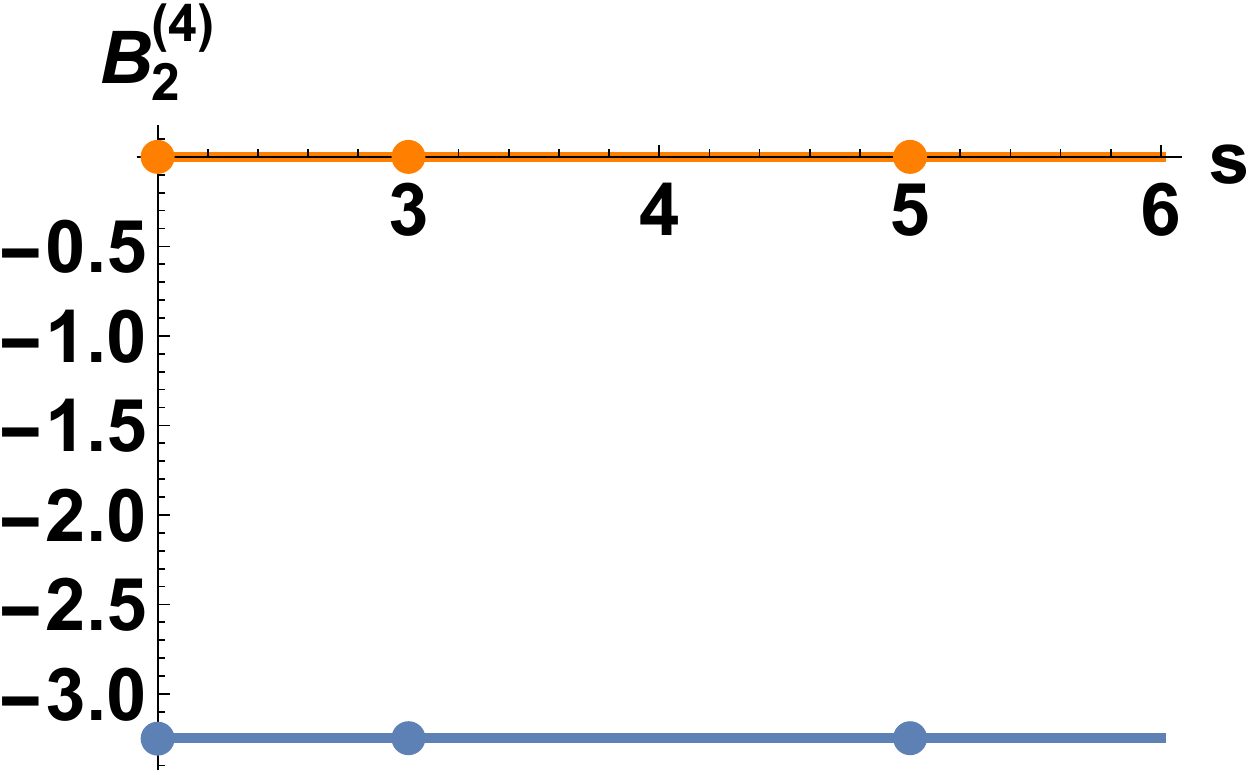} & \img{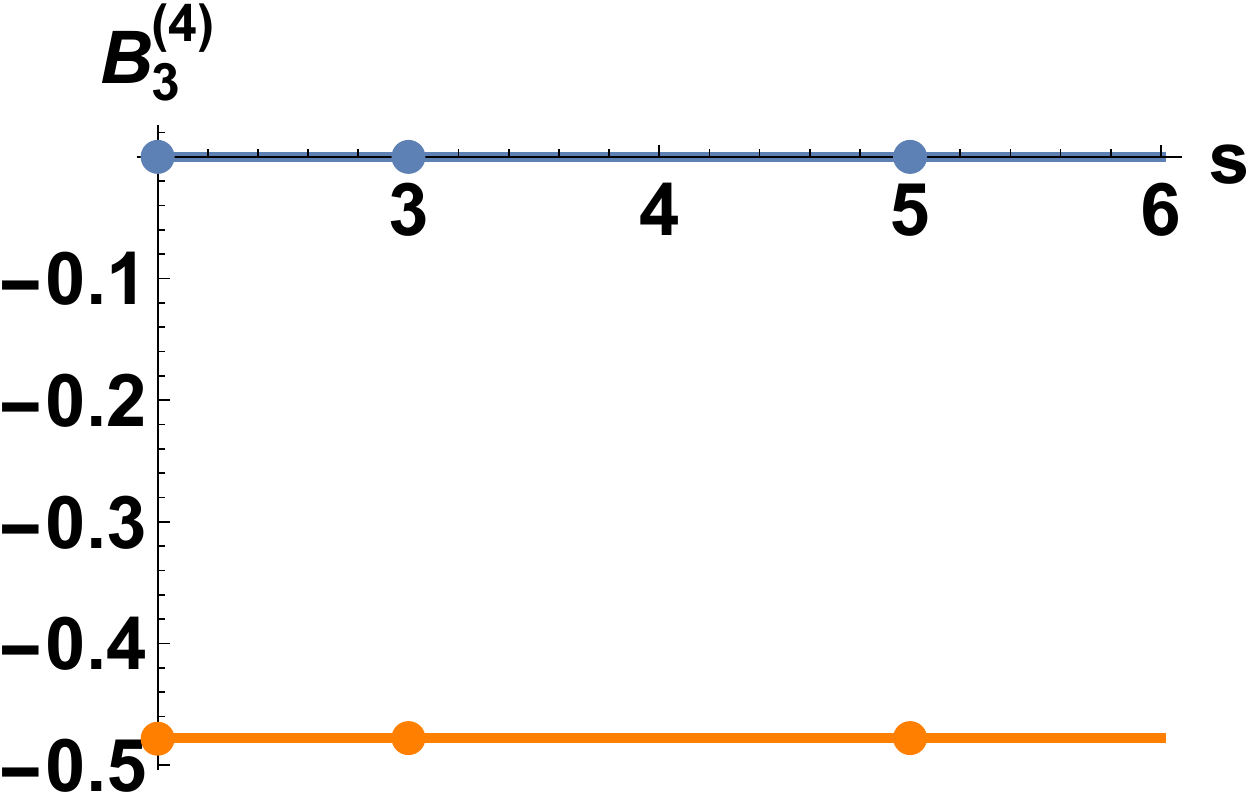} \\
\img{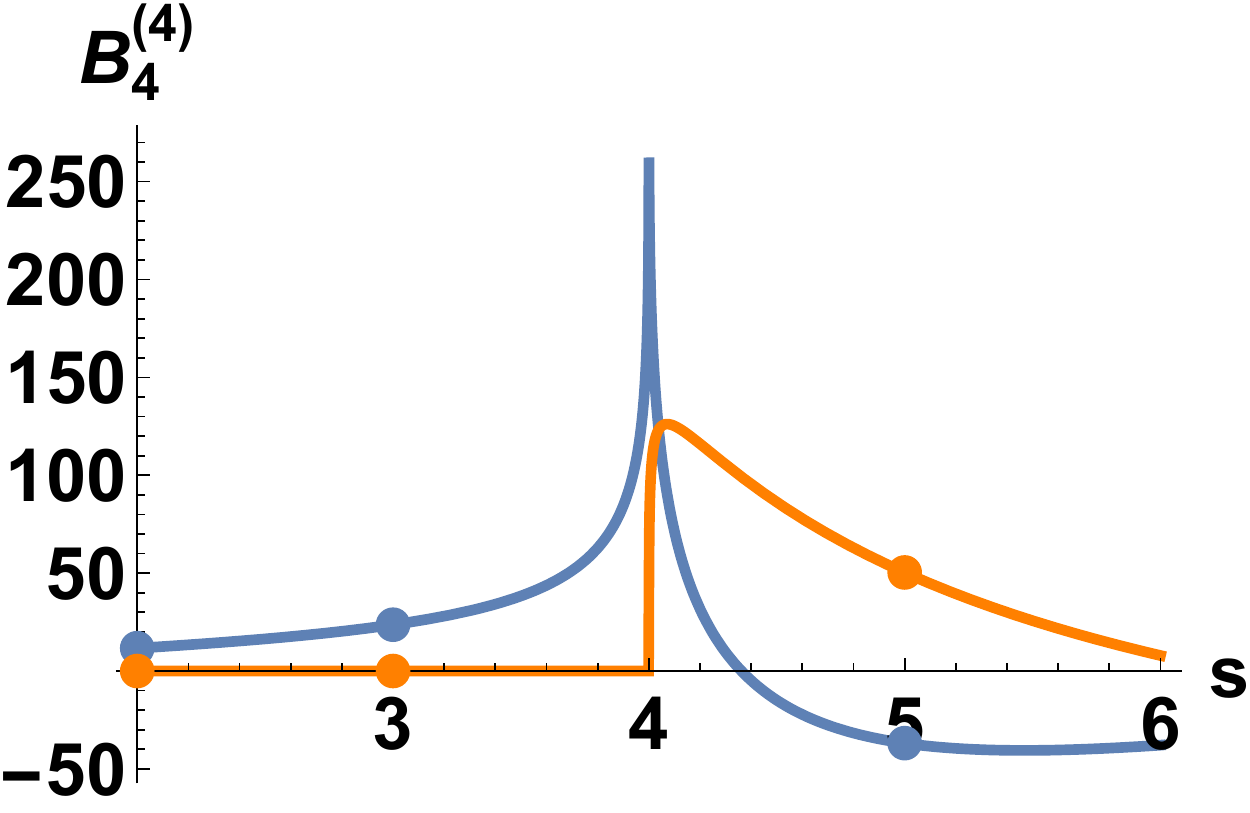} & \img{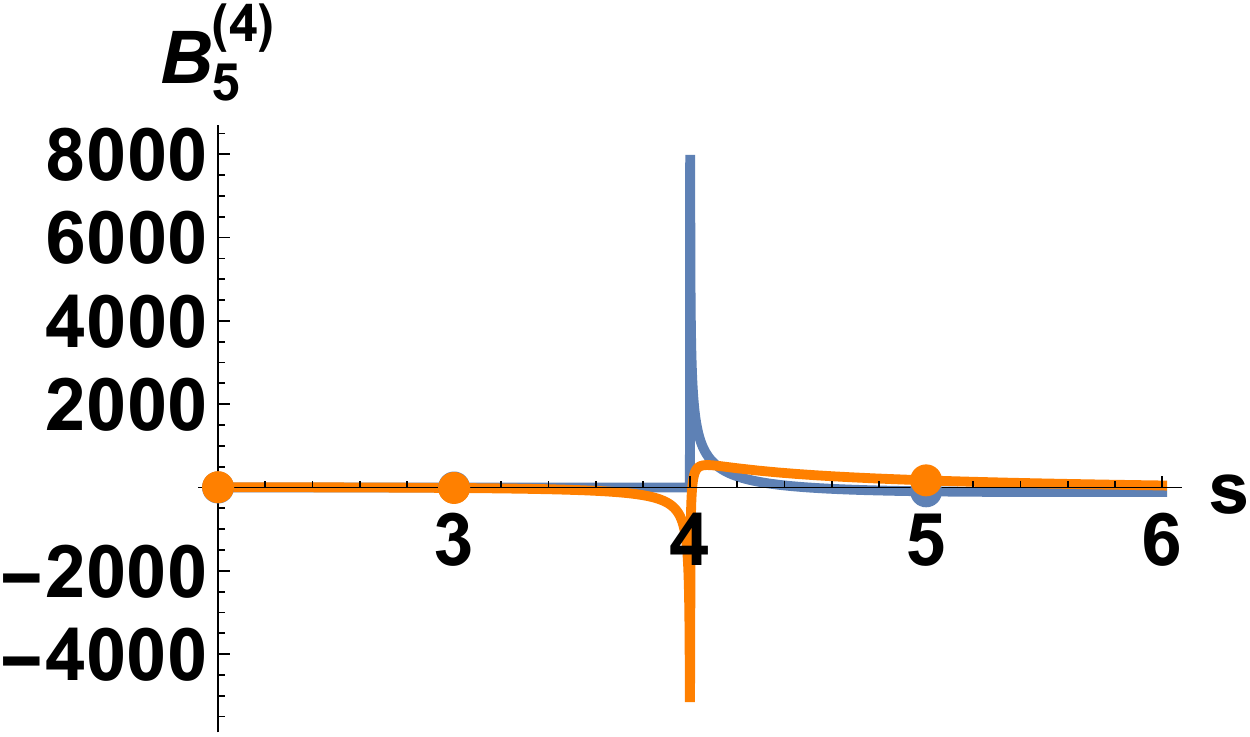} & \img{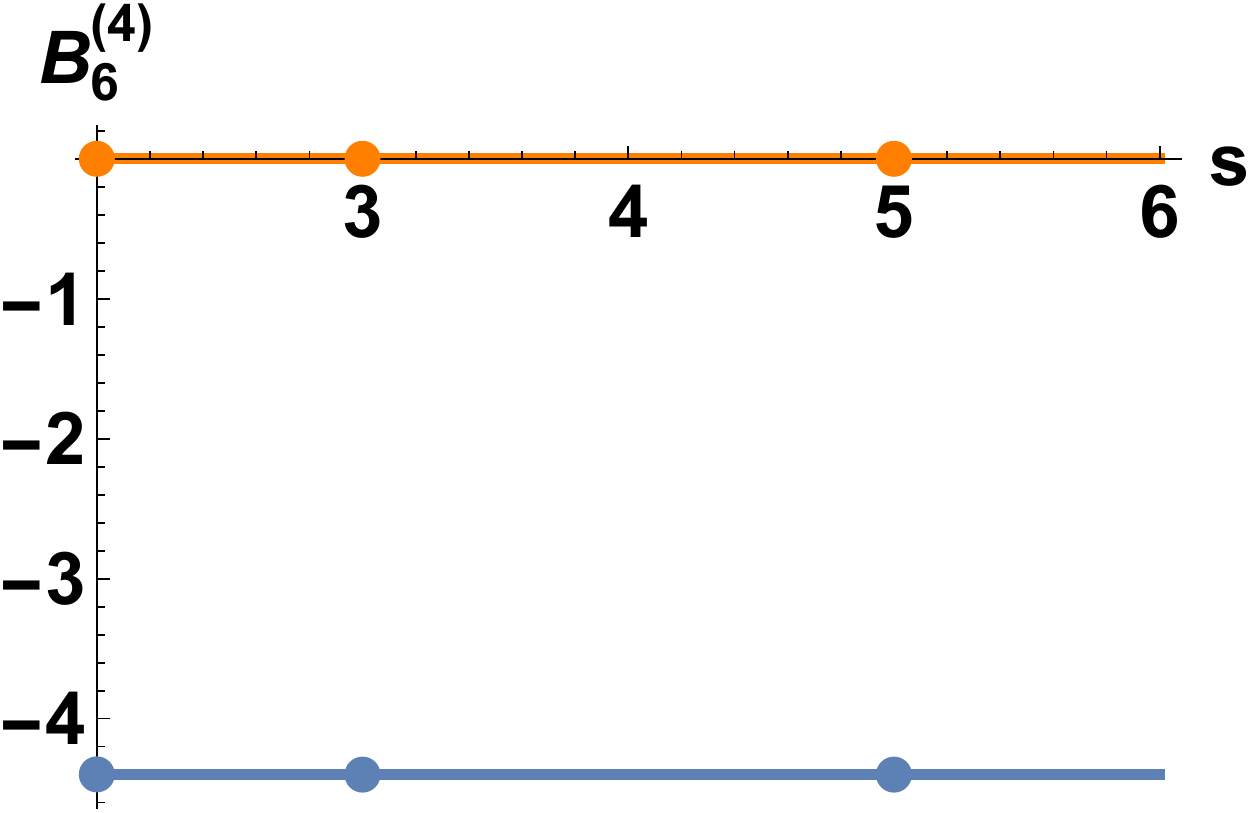} \\
\img{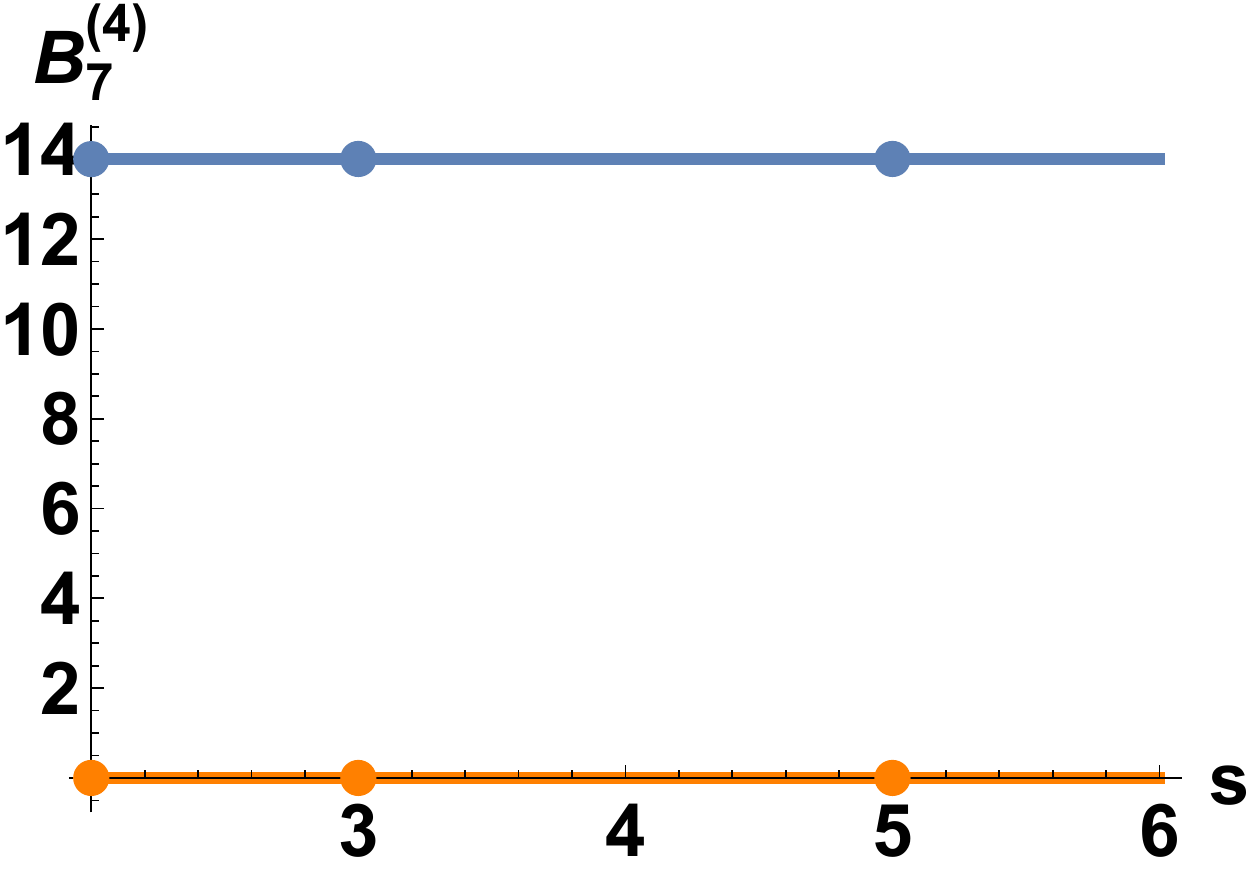} & \img{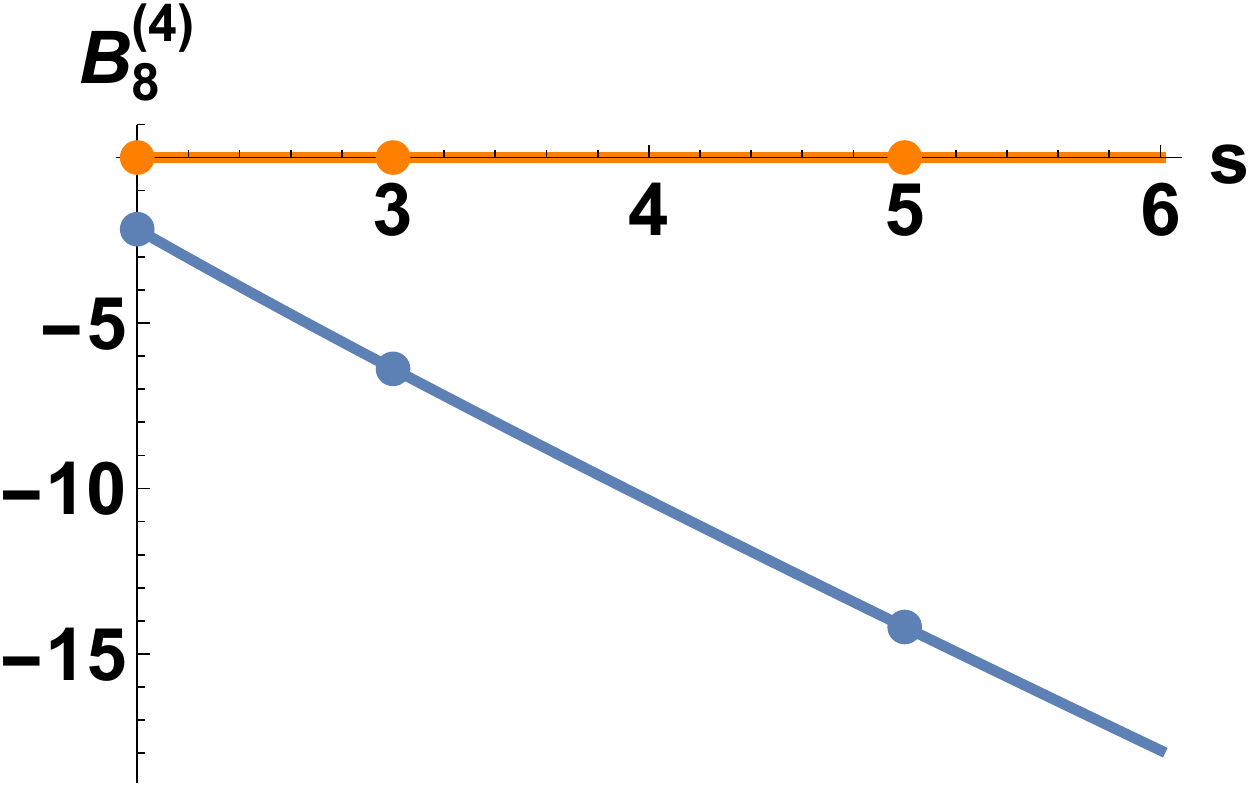} & \img{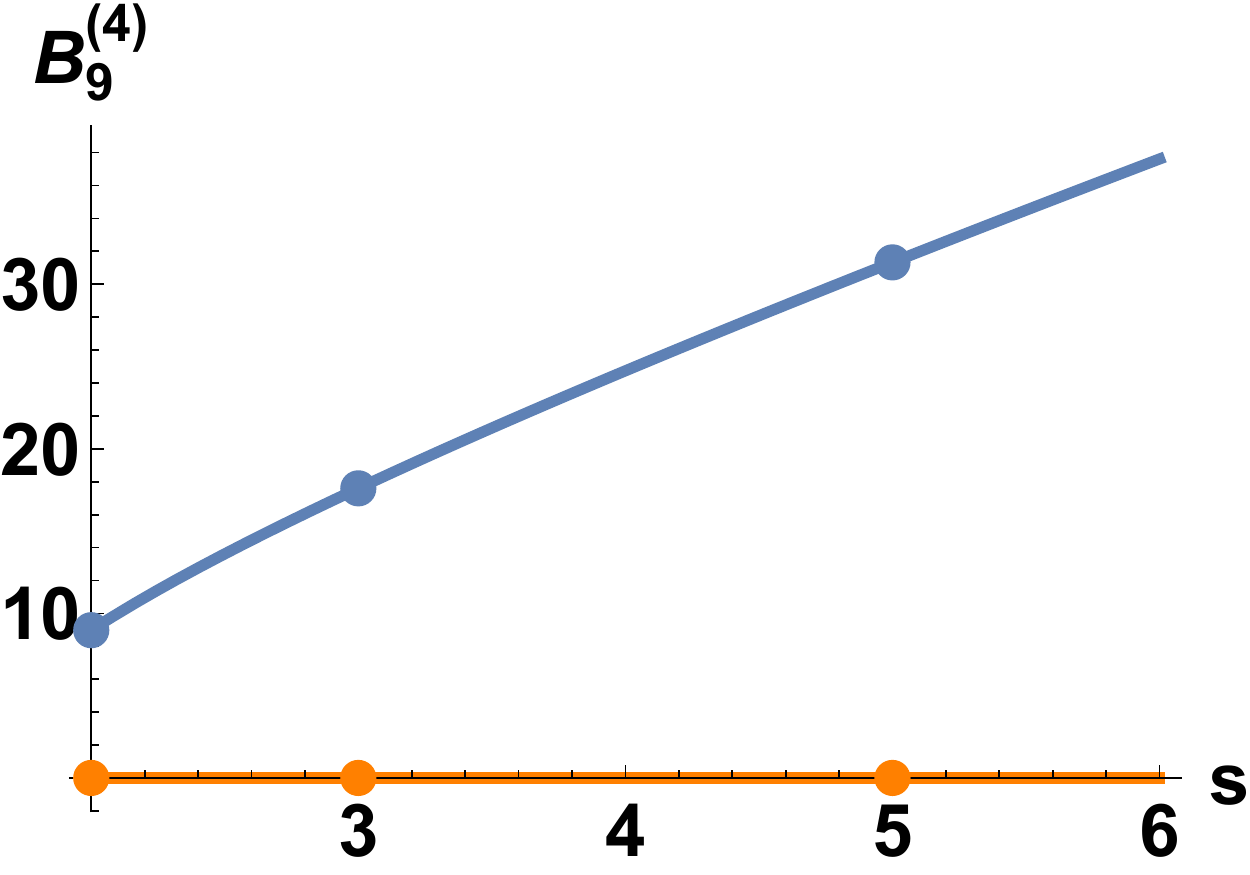} \\
\img{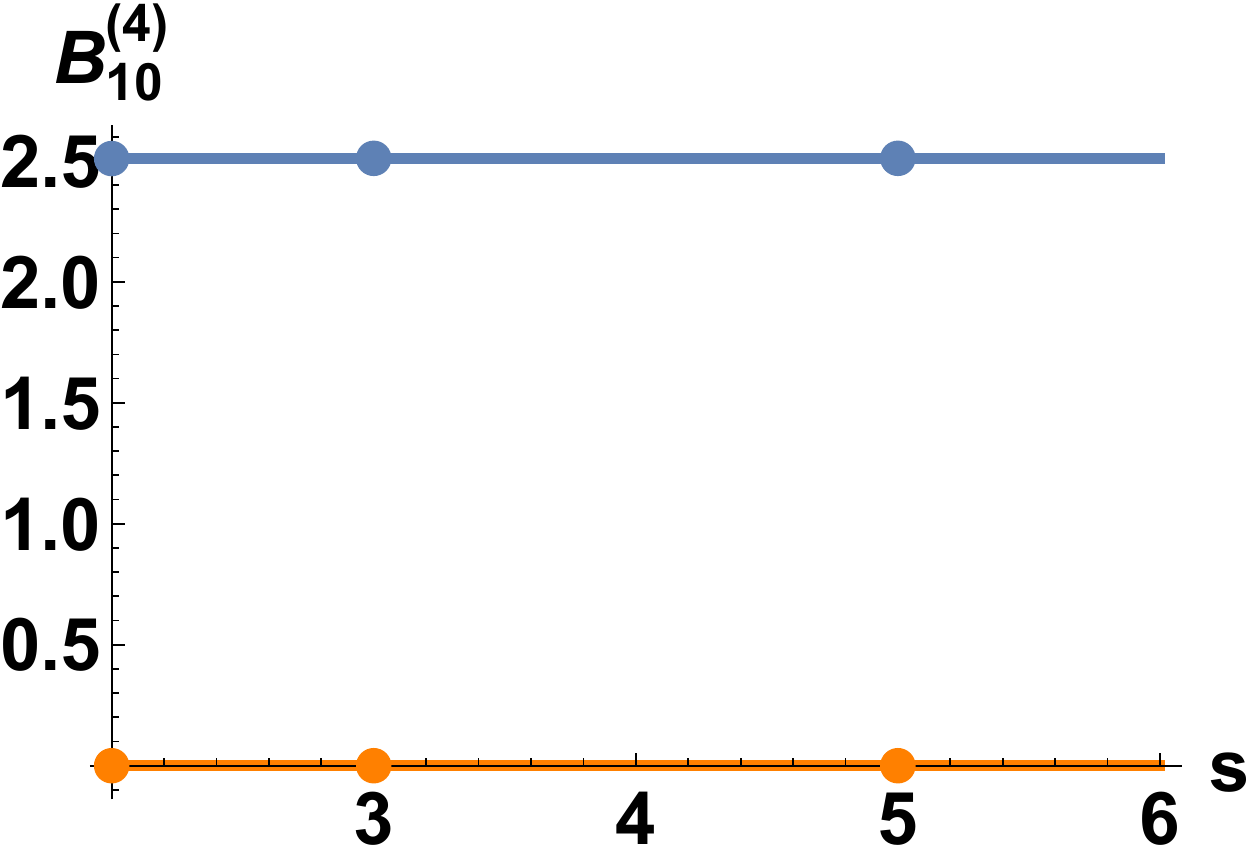} & \img{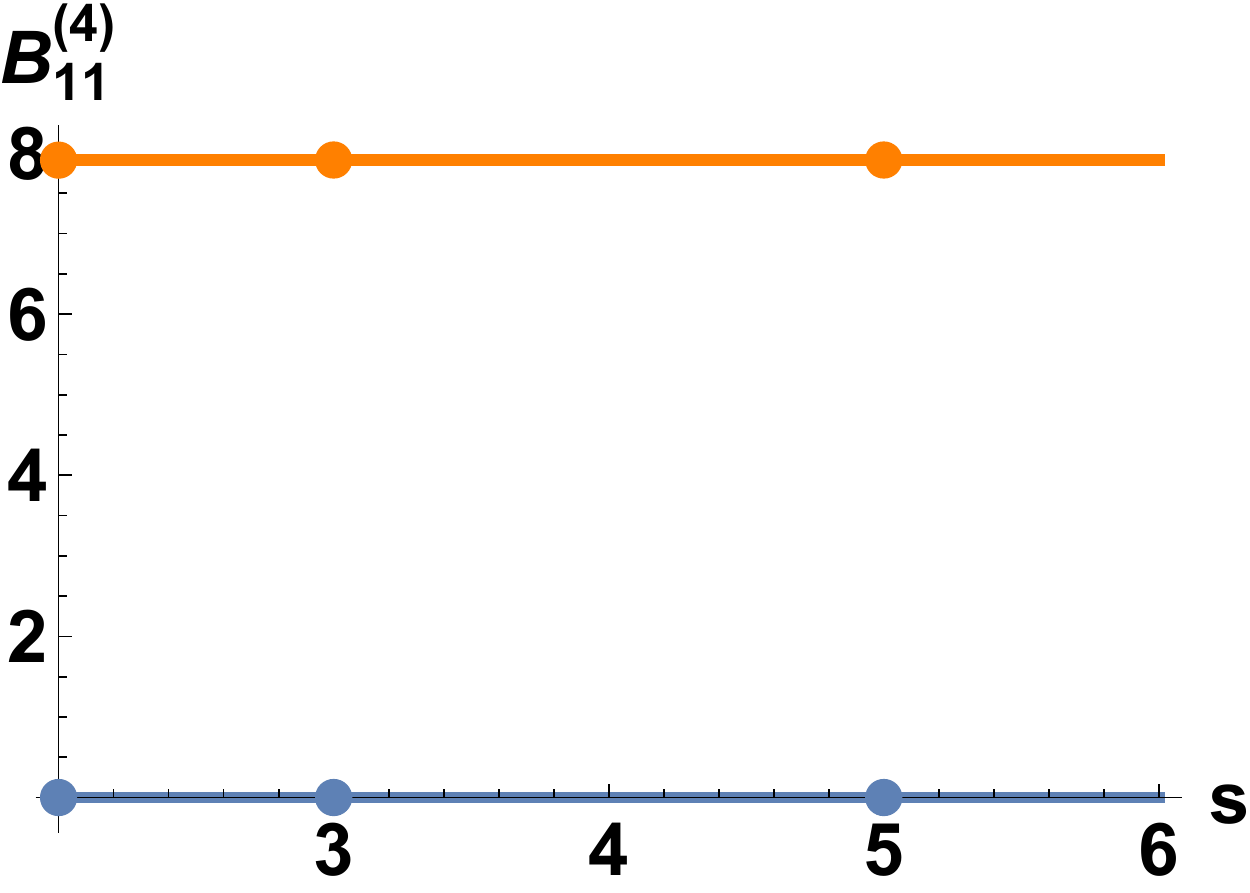} & \img{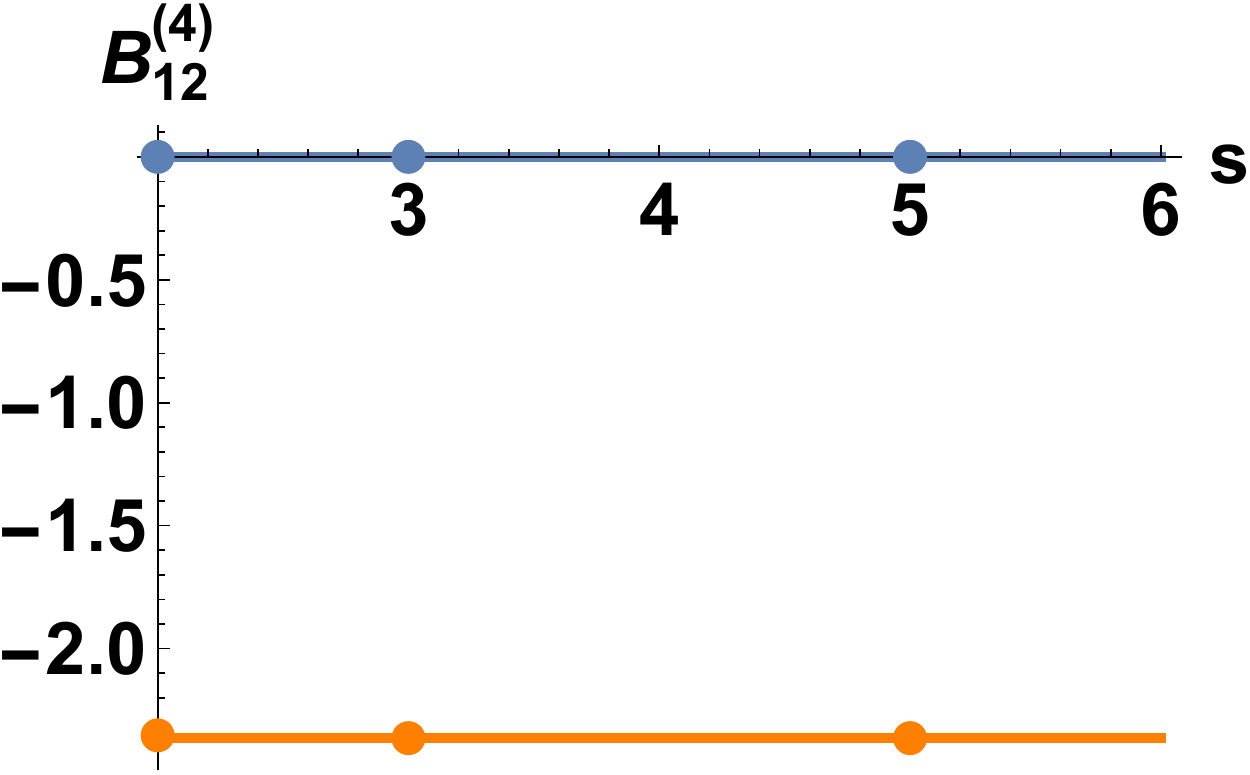} \\
\img{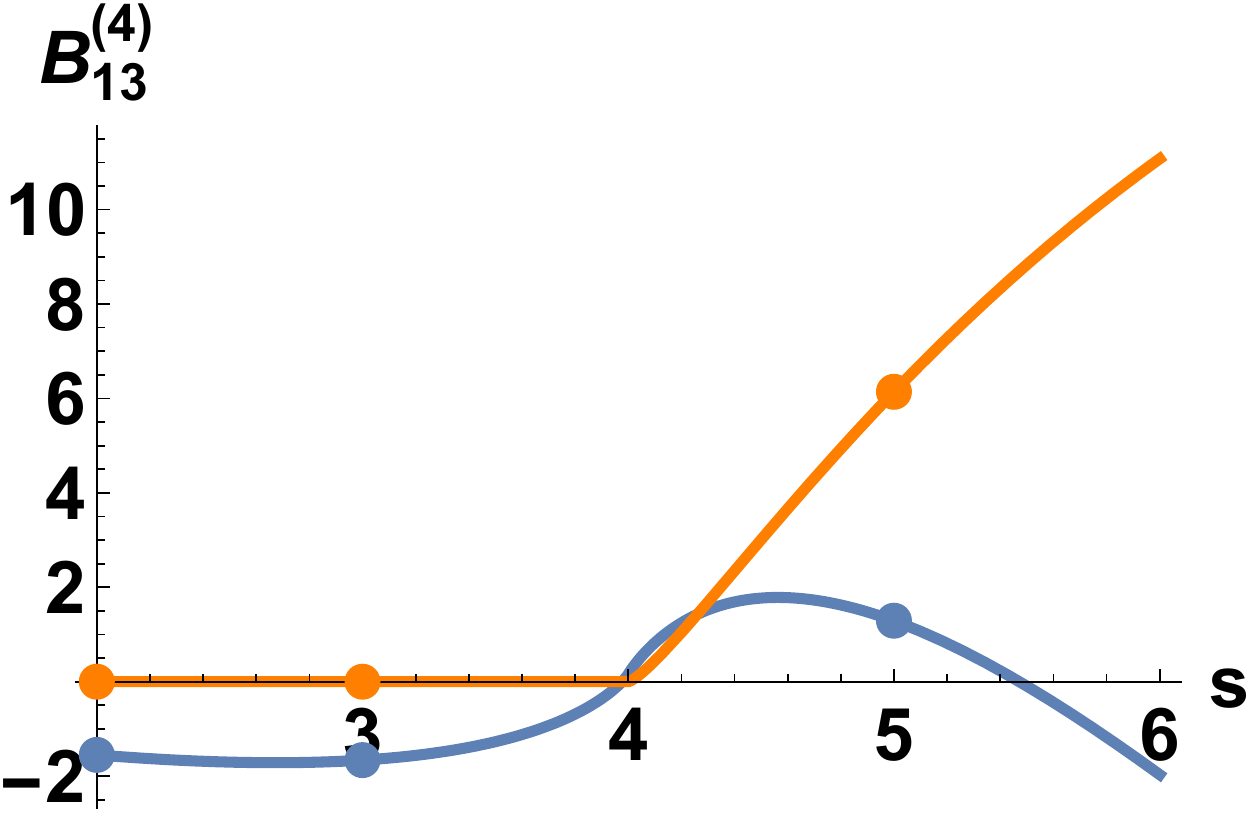} & \img{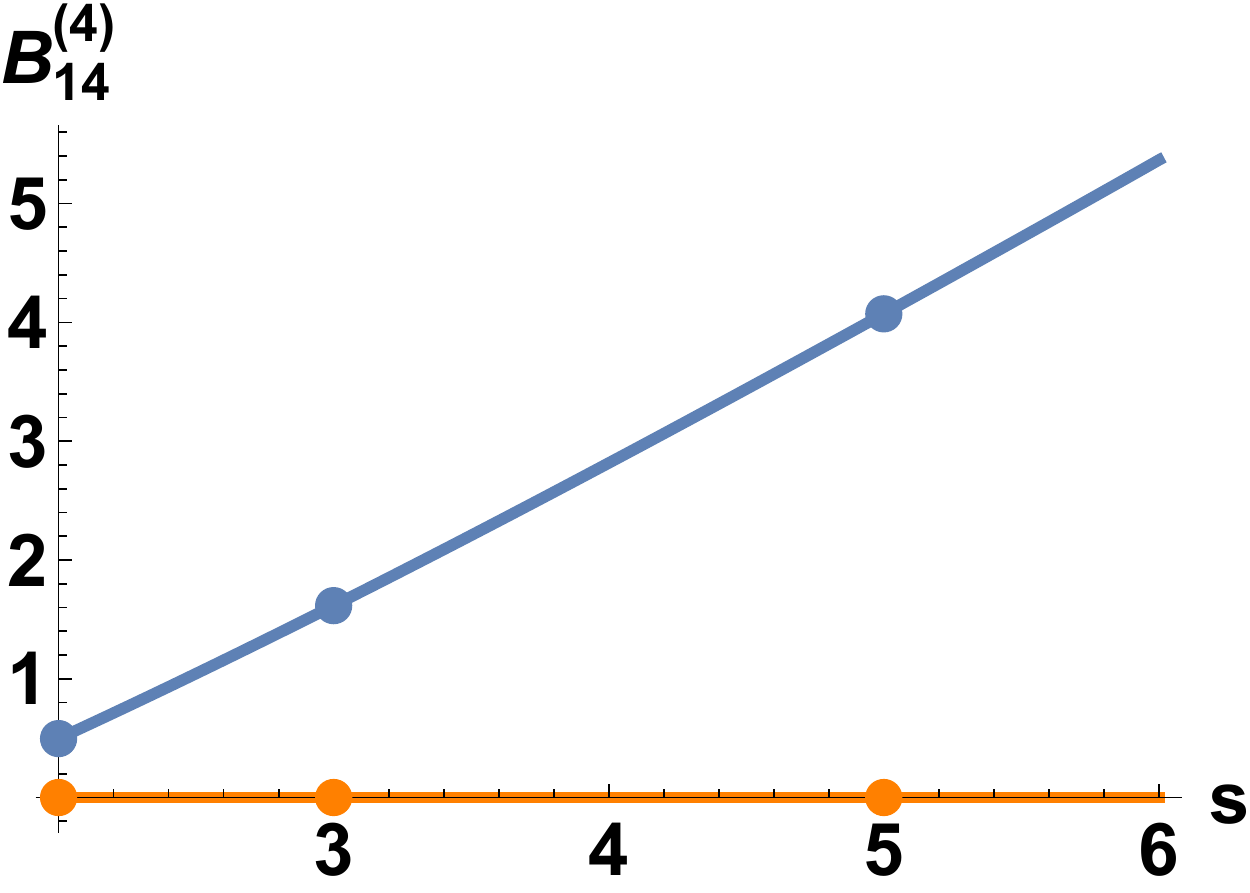} & \img{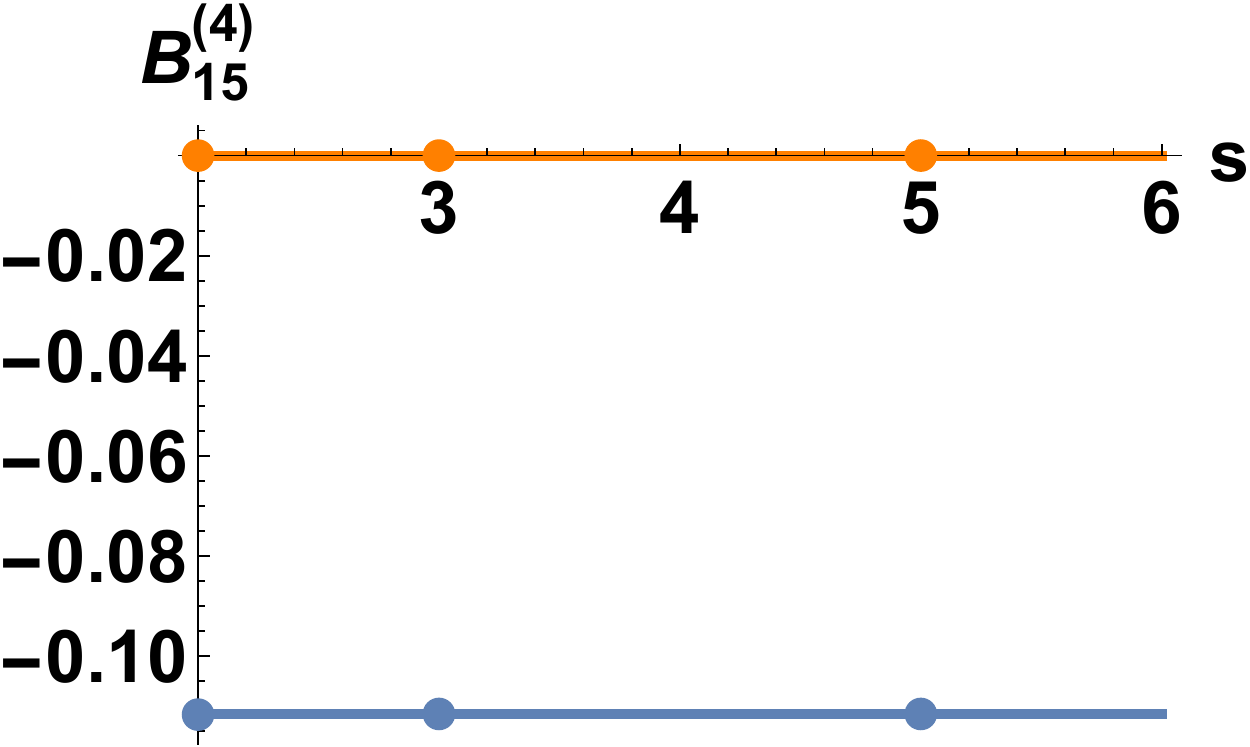} \\
\img{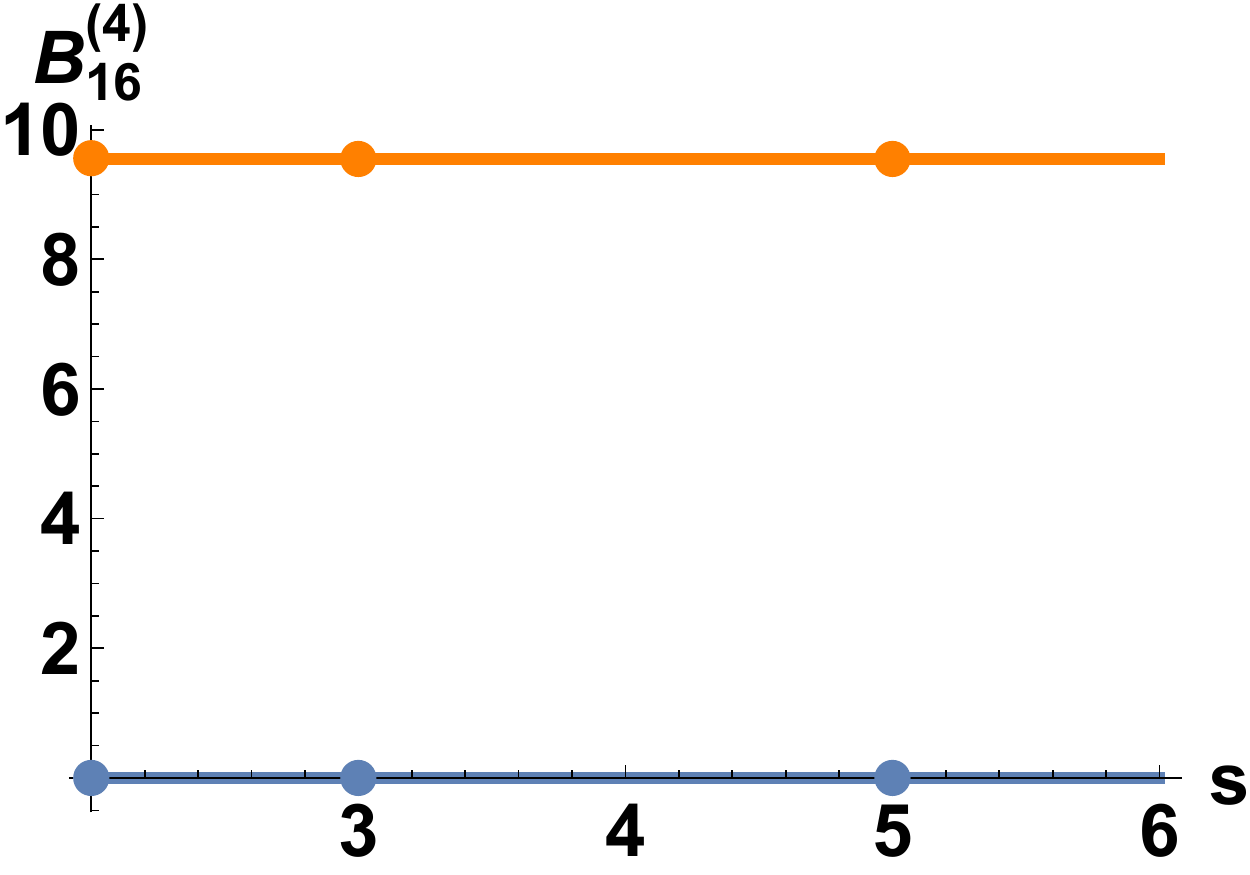} & \img{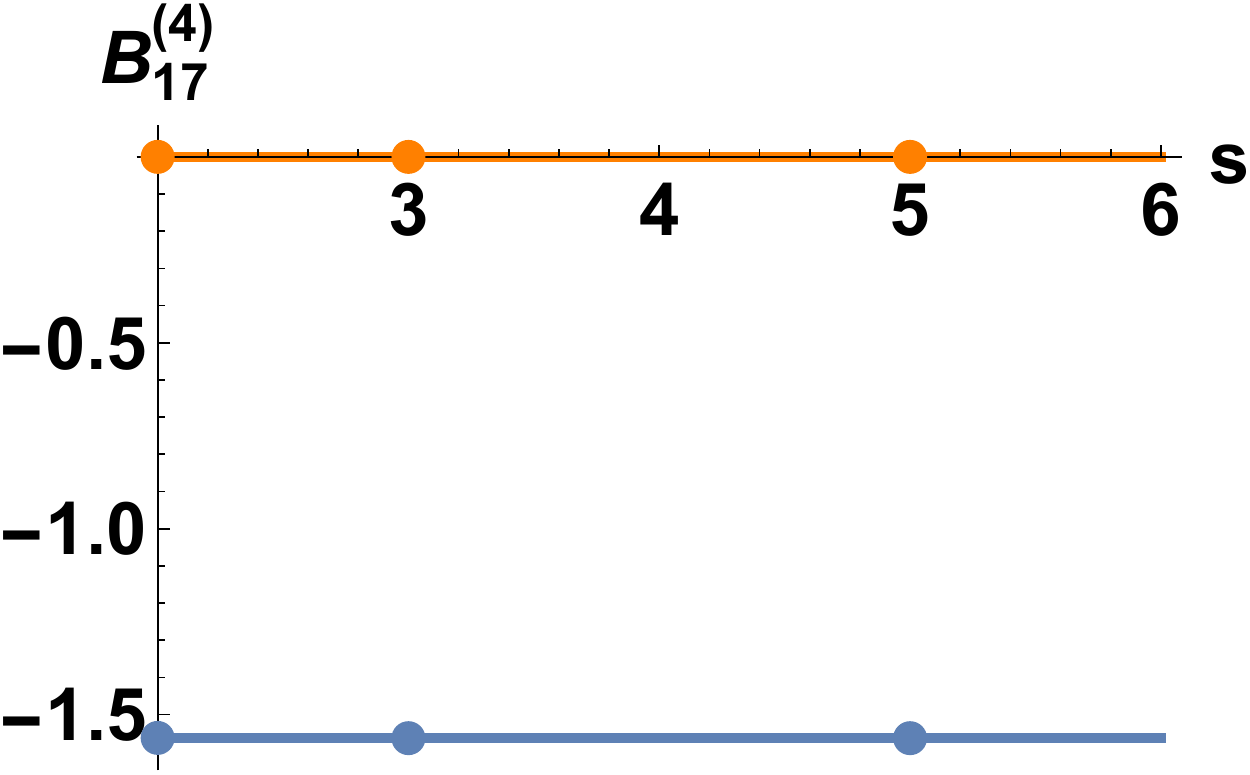} & \img{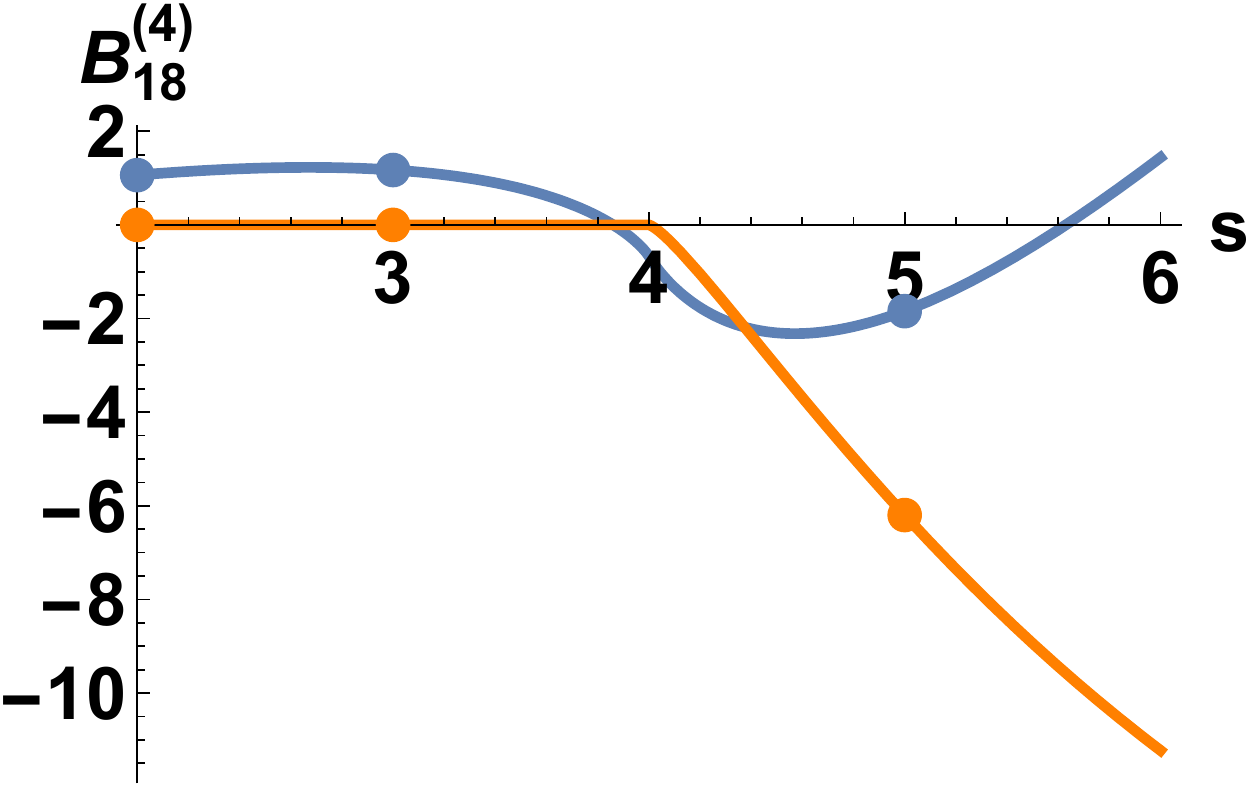} \\
\end{tabular}
\end{table}
\begin{table}[h!]
	\begin{tabular}{ p{4.75cm}  p{4.75cm}  p{4.75cm}}
\img{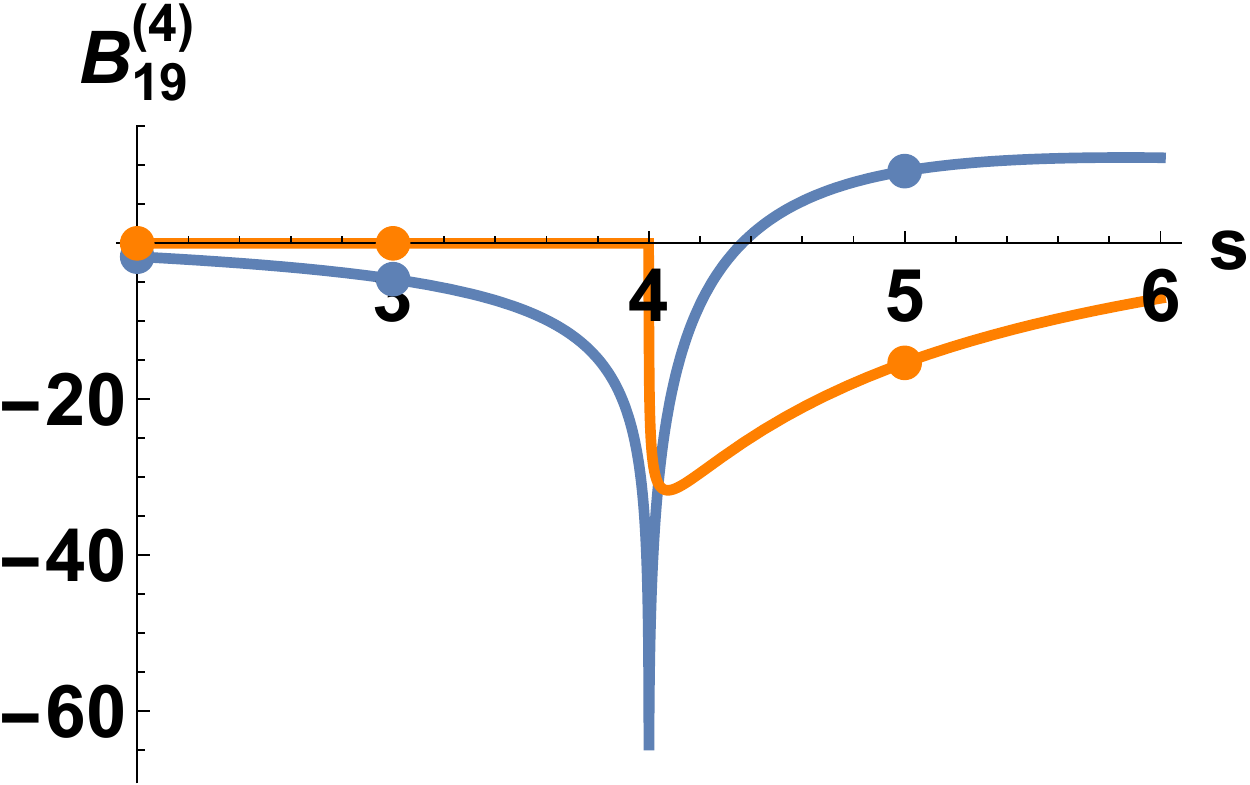} & \img{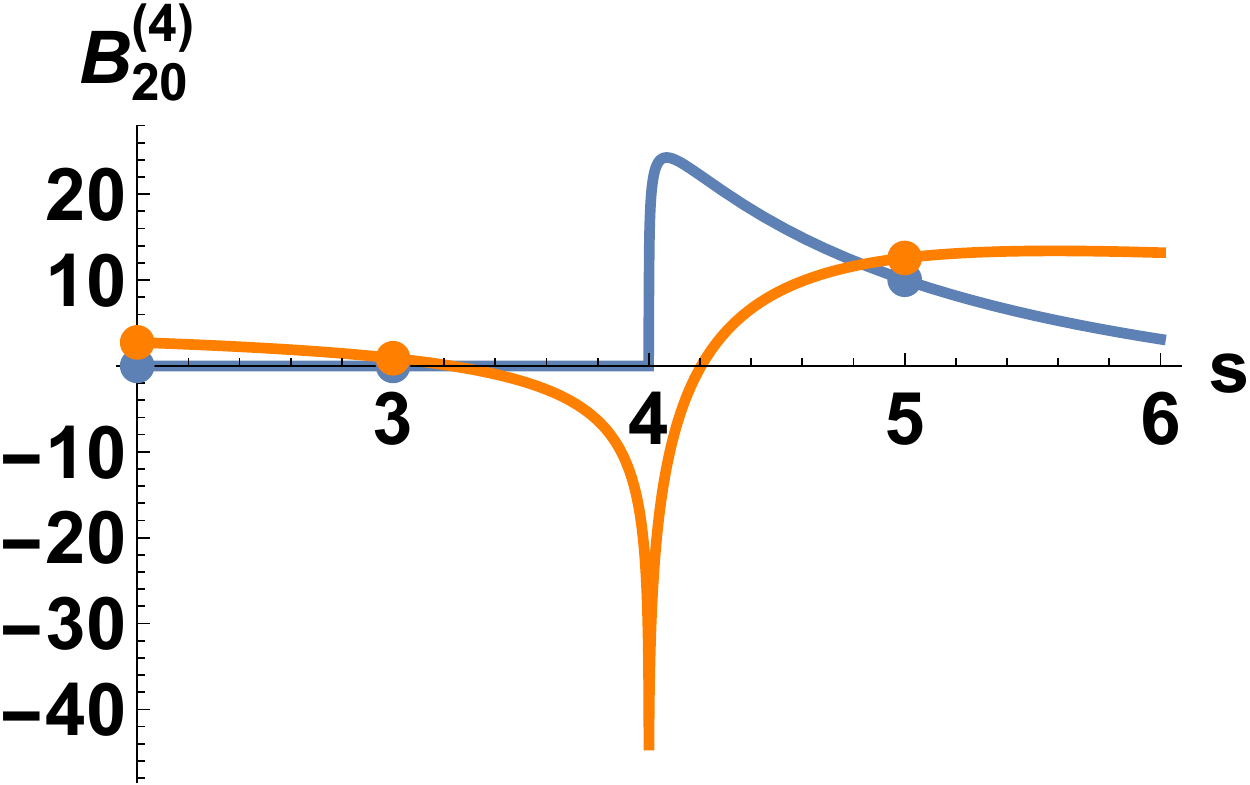} & \img{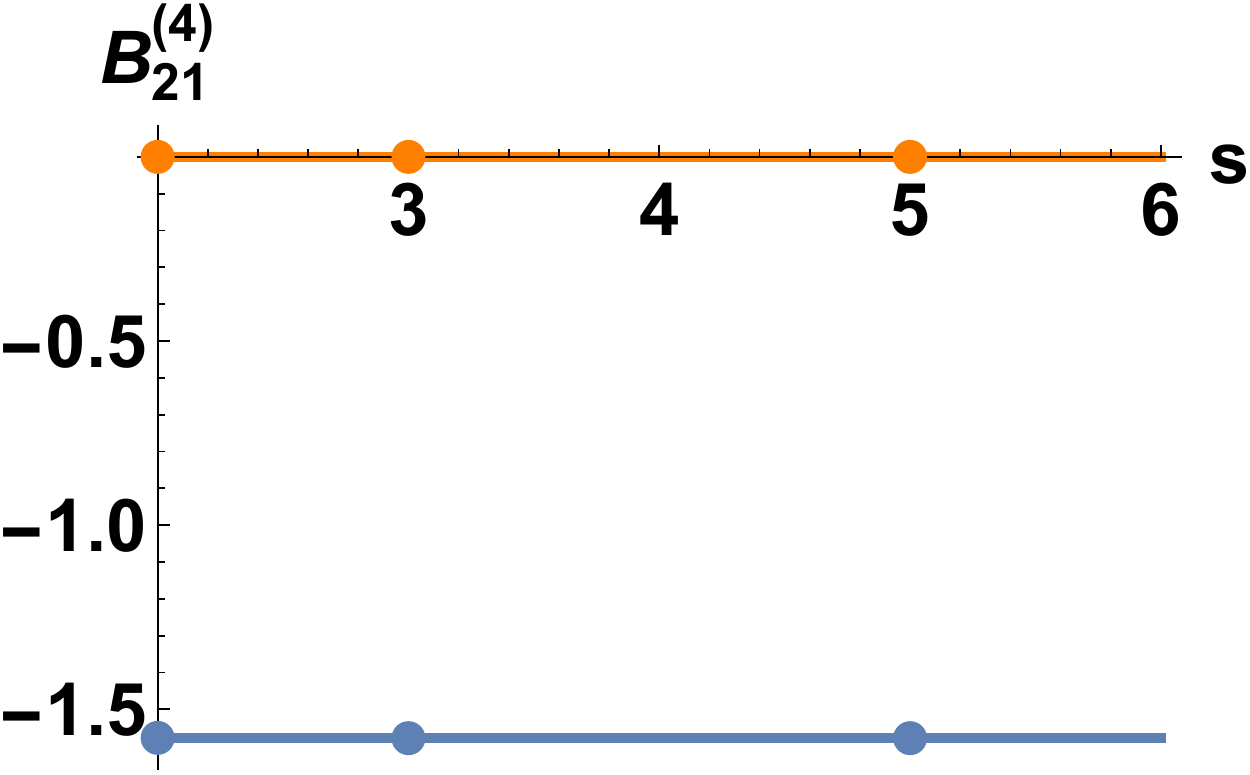} \\
\img{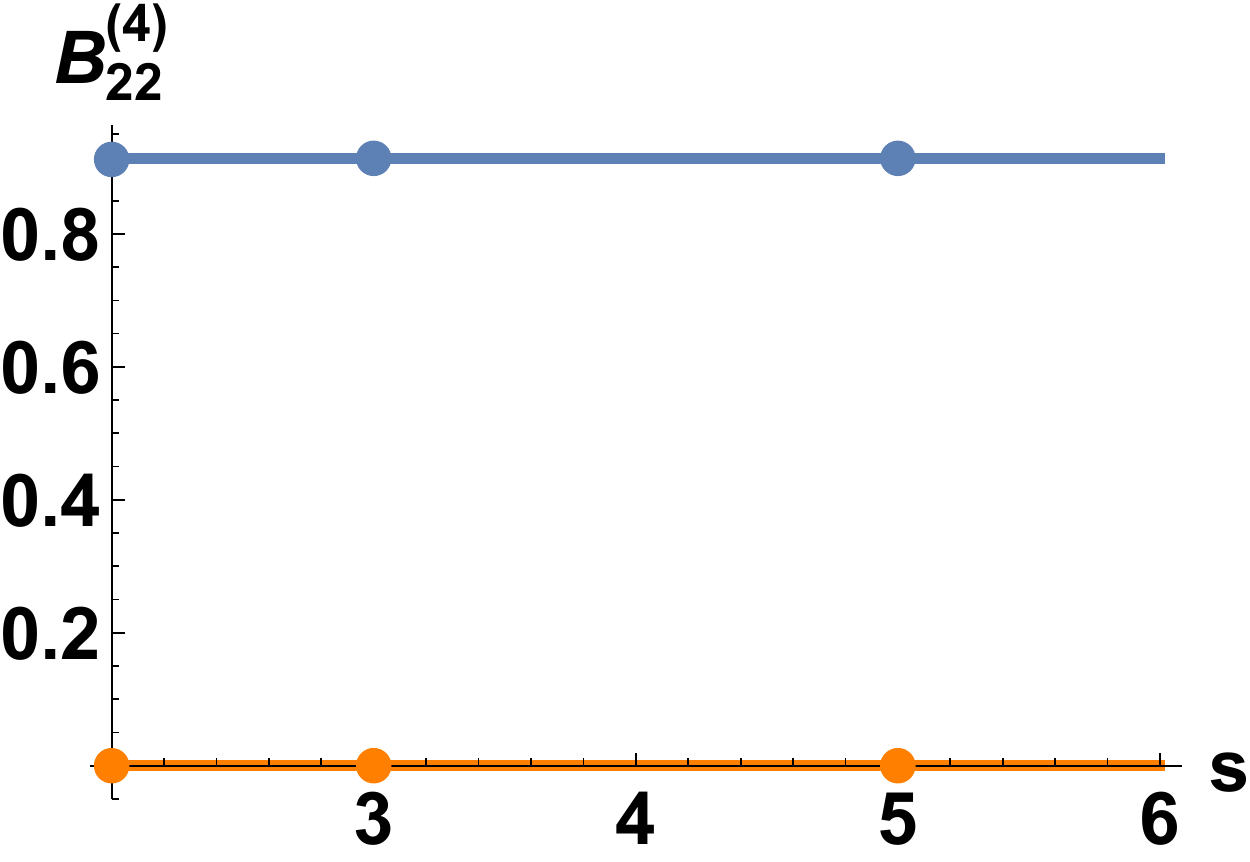} & \img{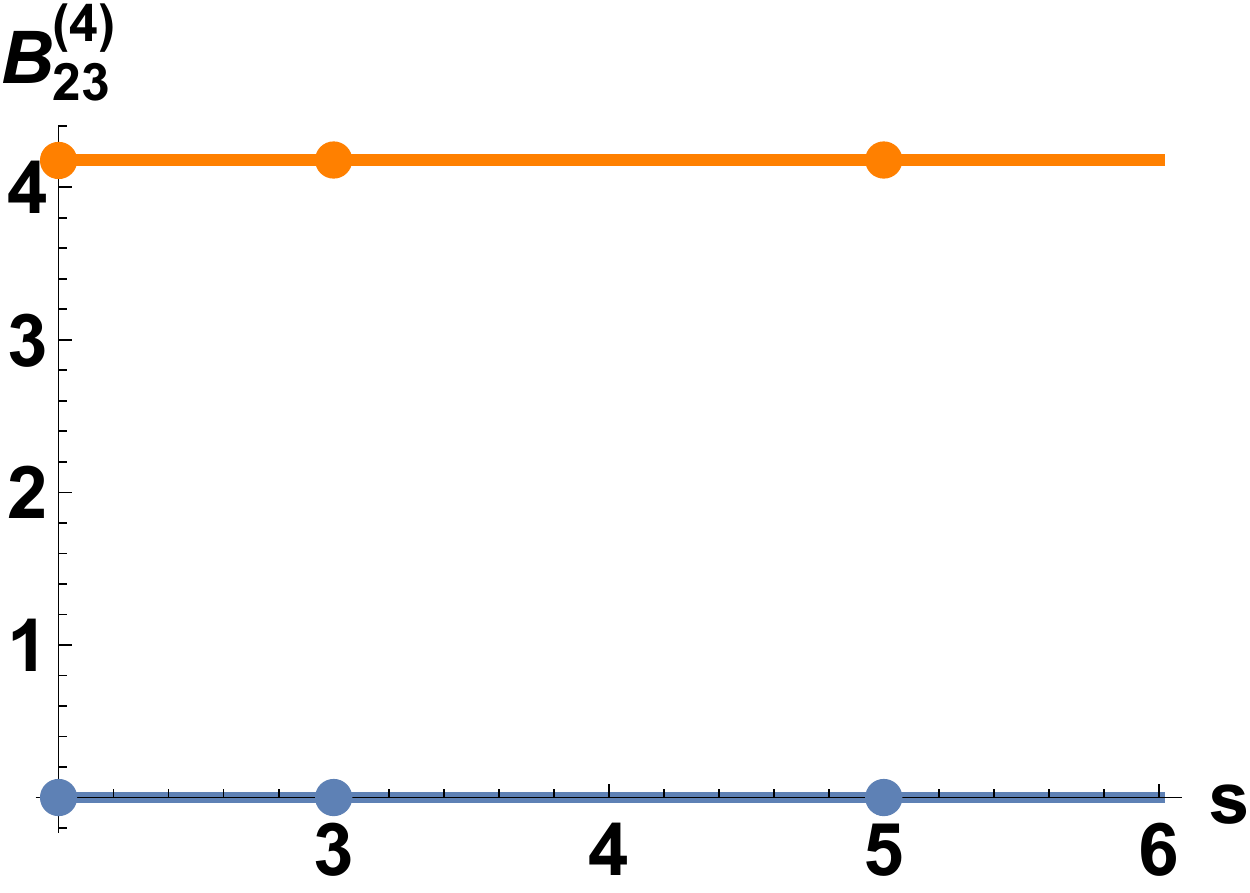} & \img{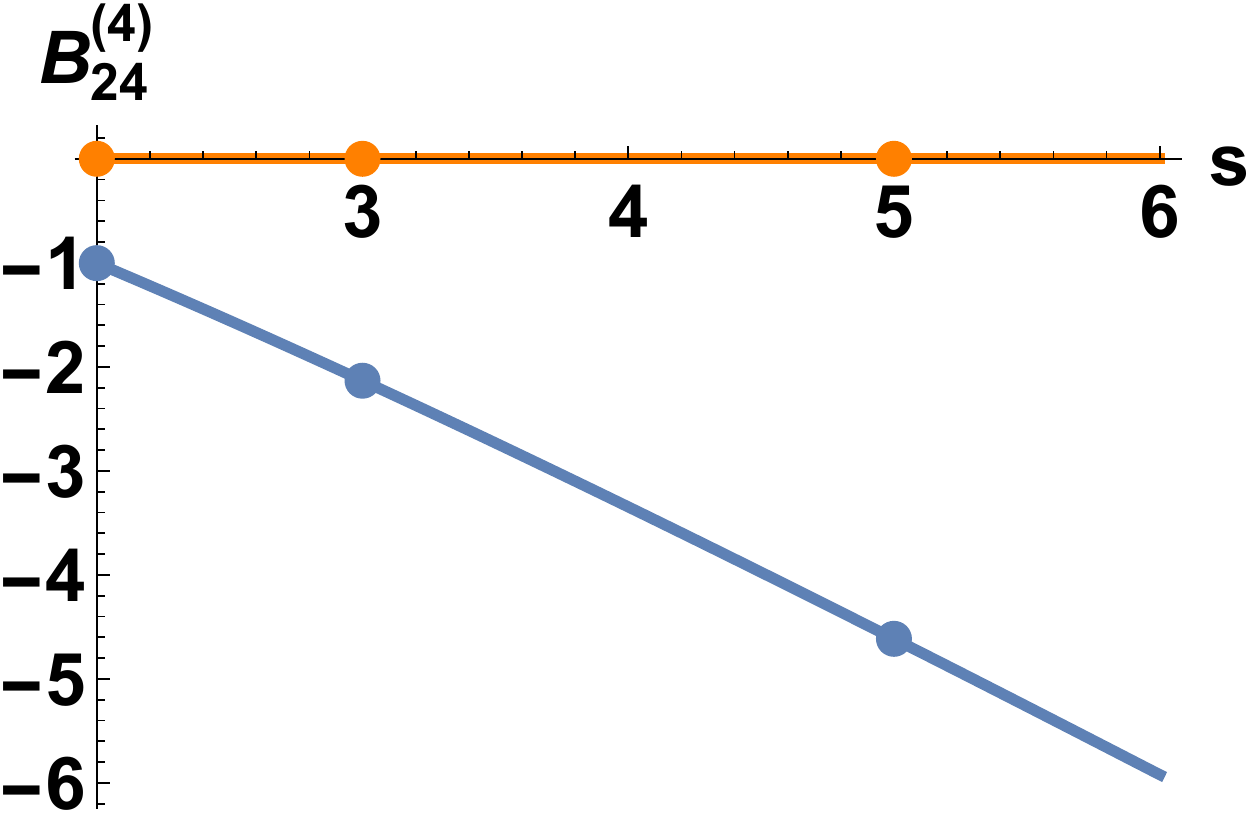} \\
\img{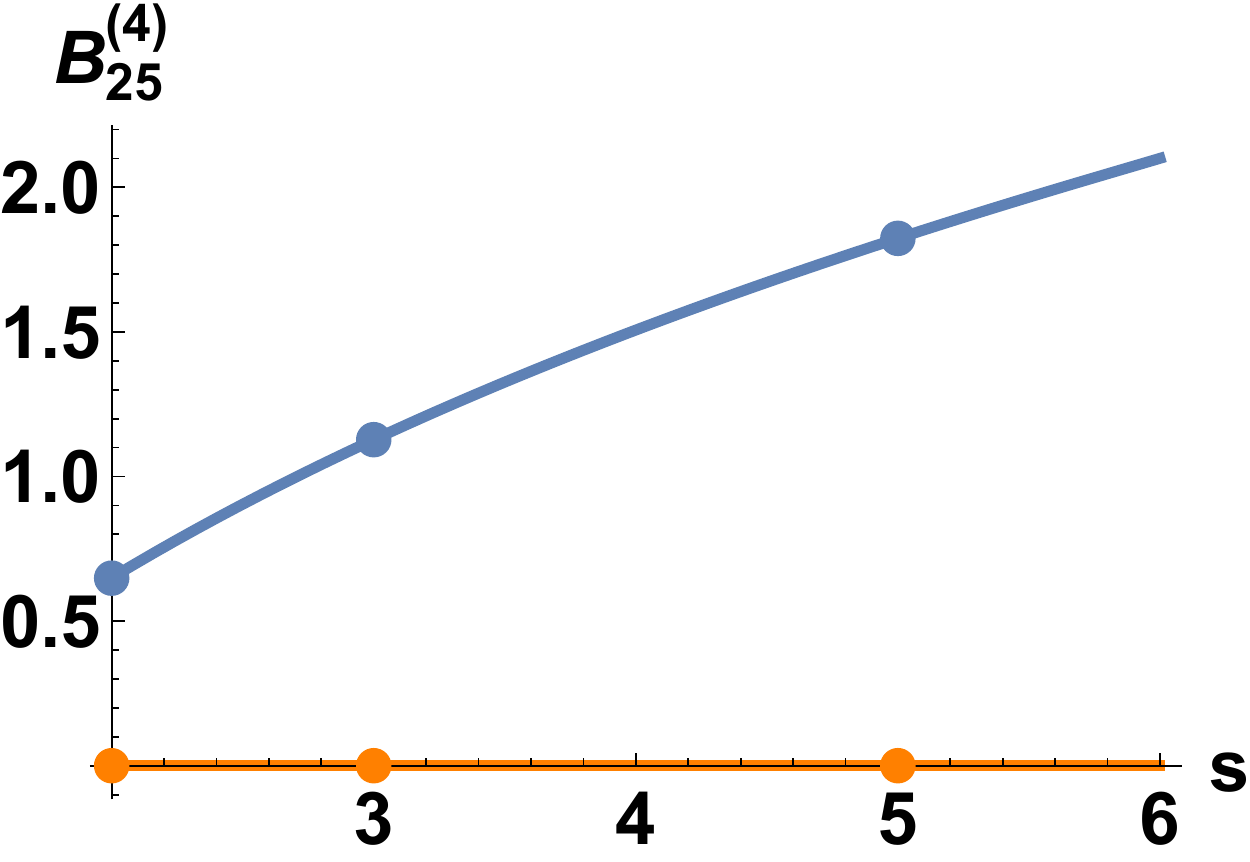} & \img{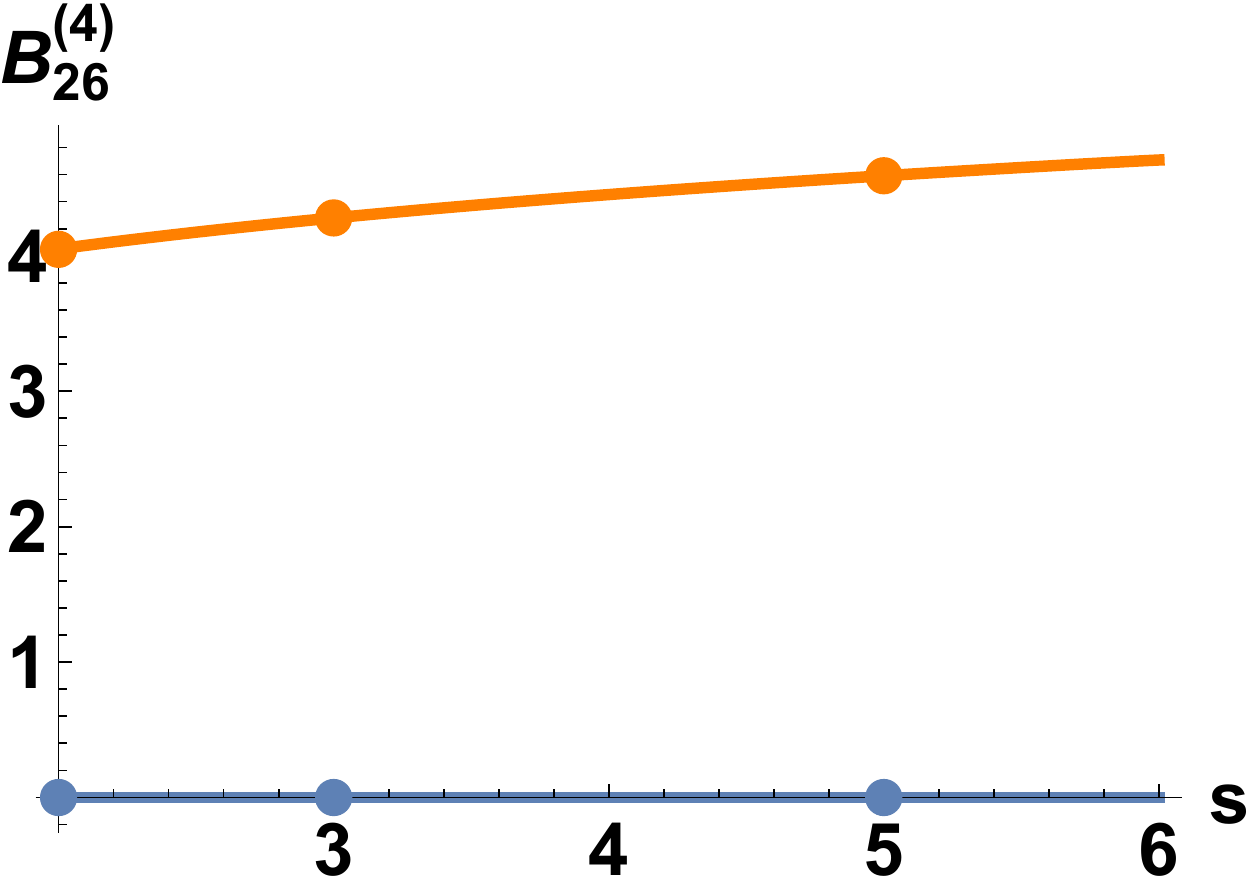} & \img{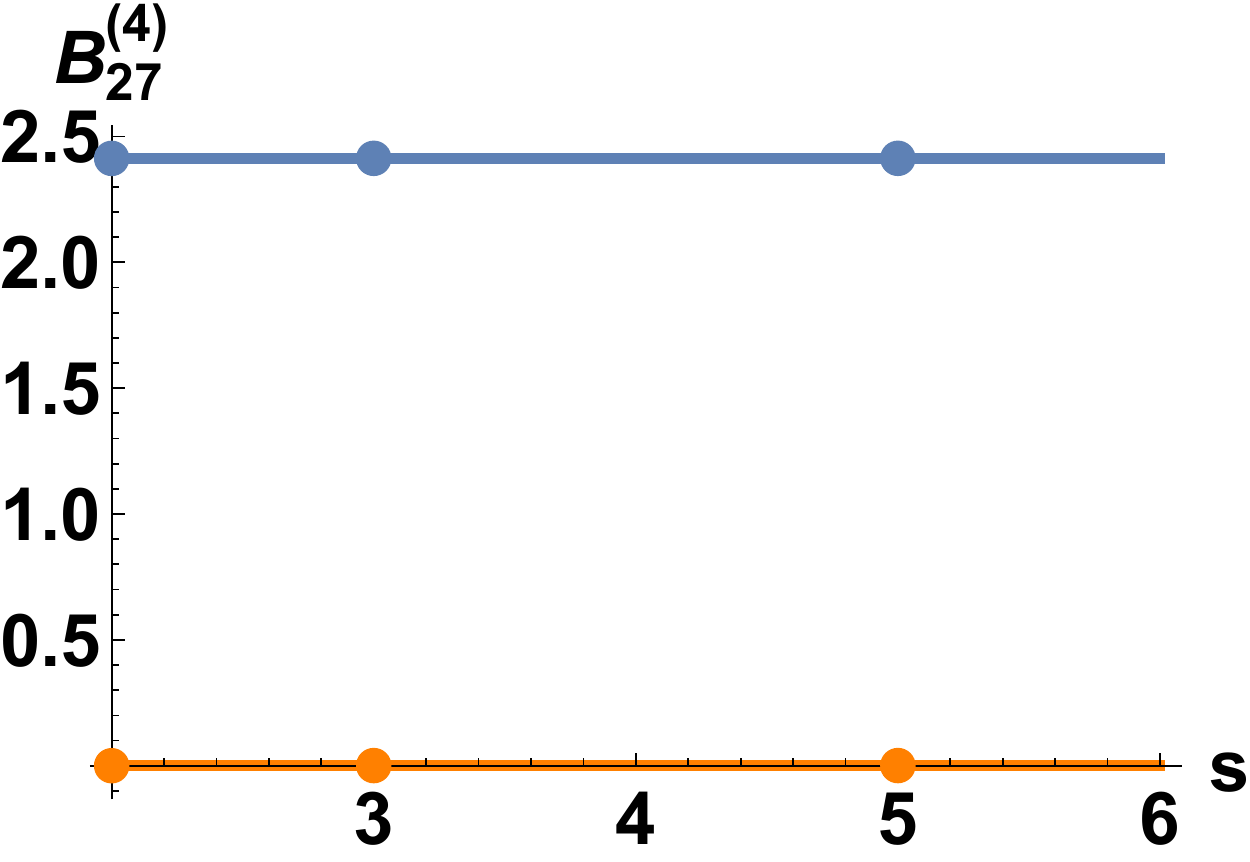} \\
\img{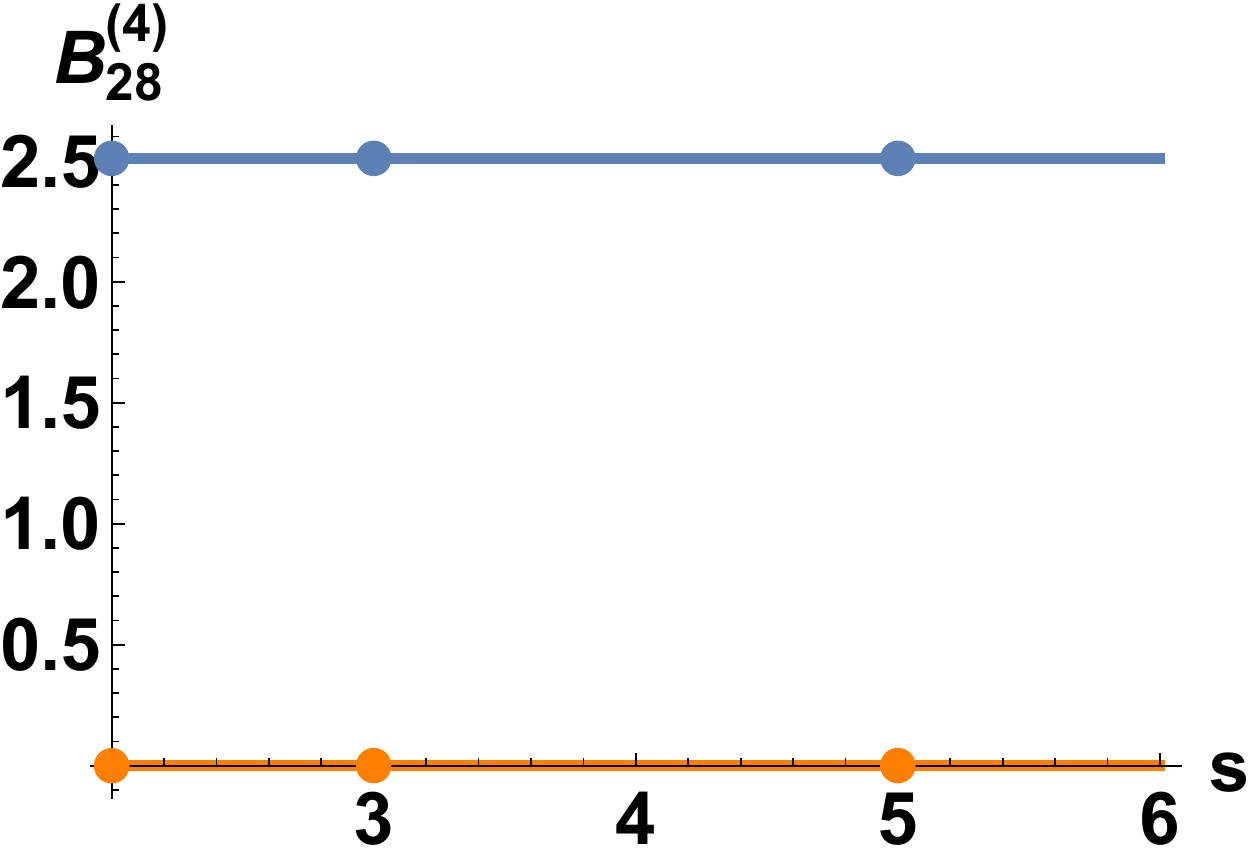} & \img{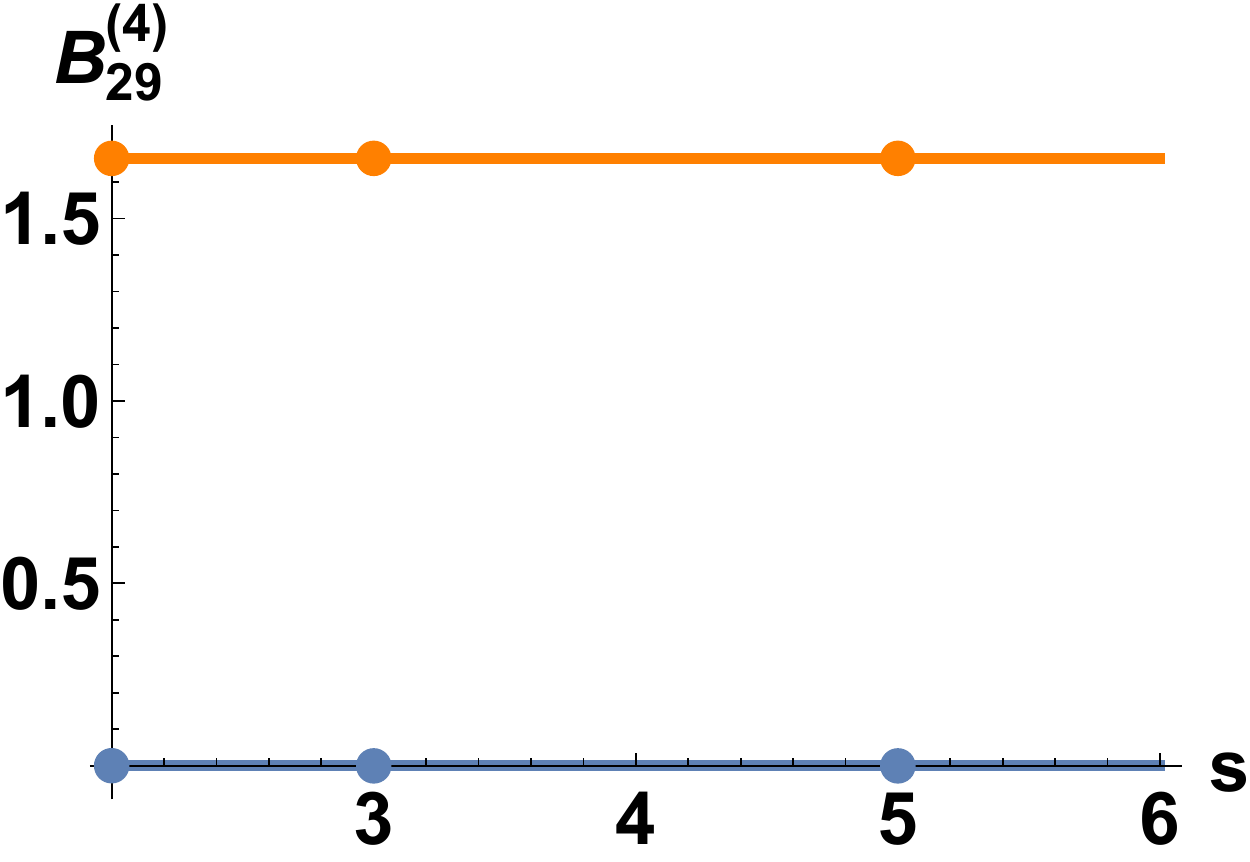} & \img{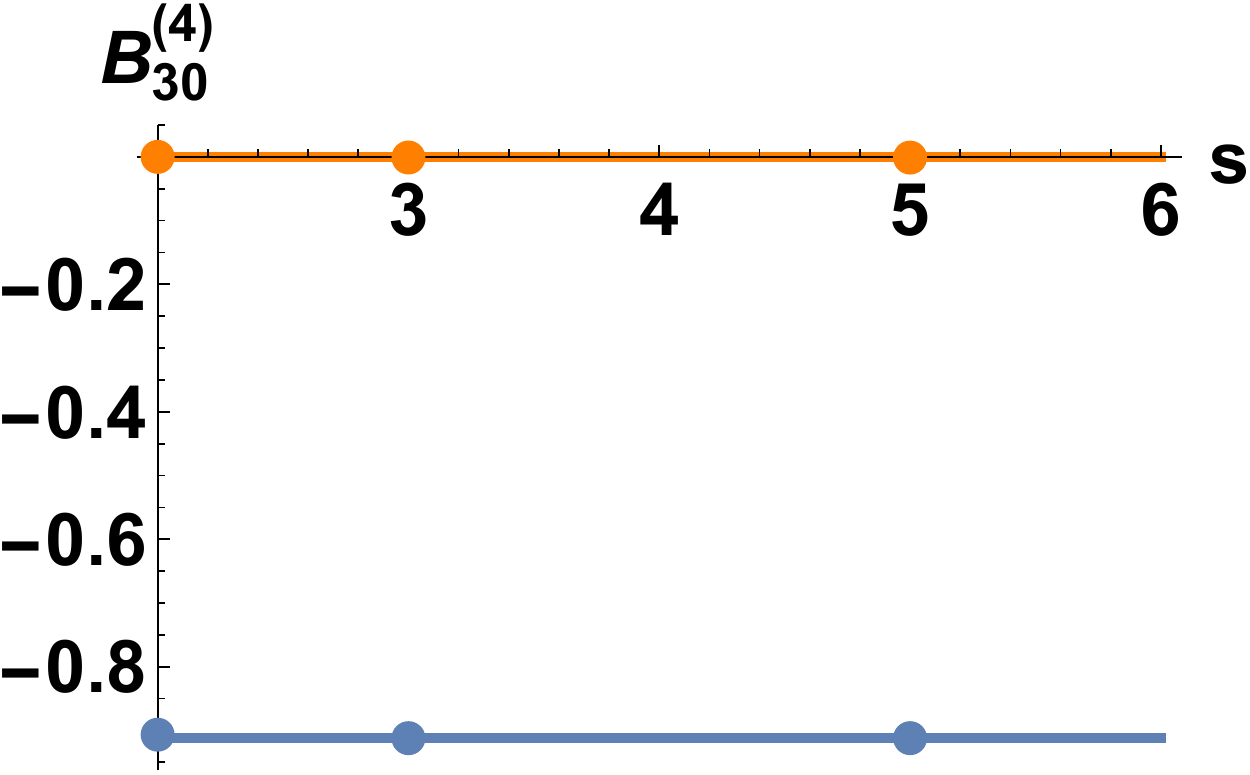} \\
\img{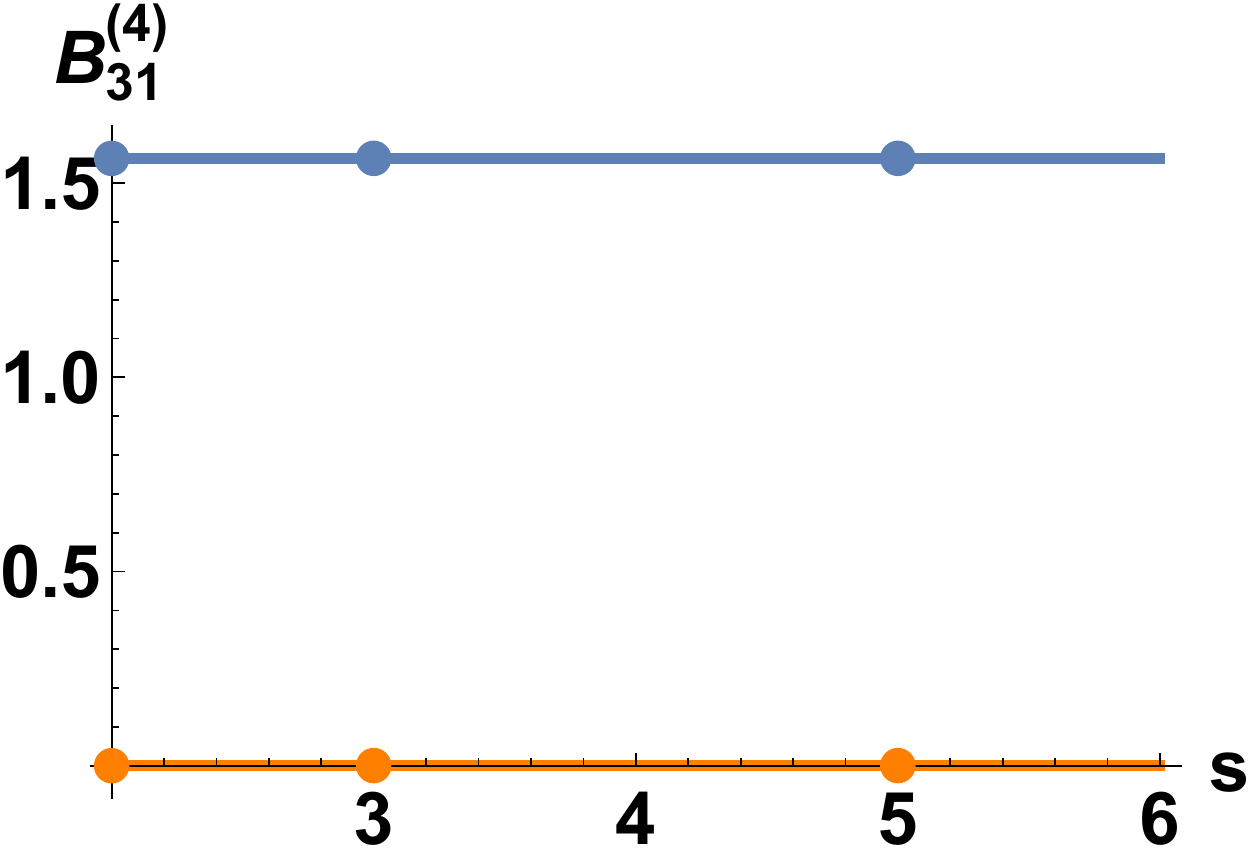} & \img{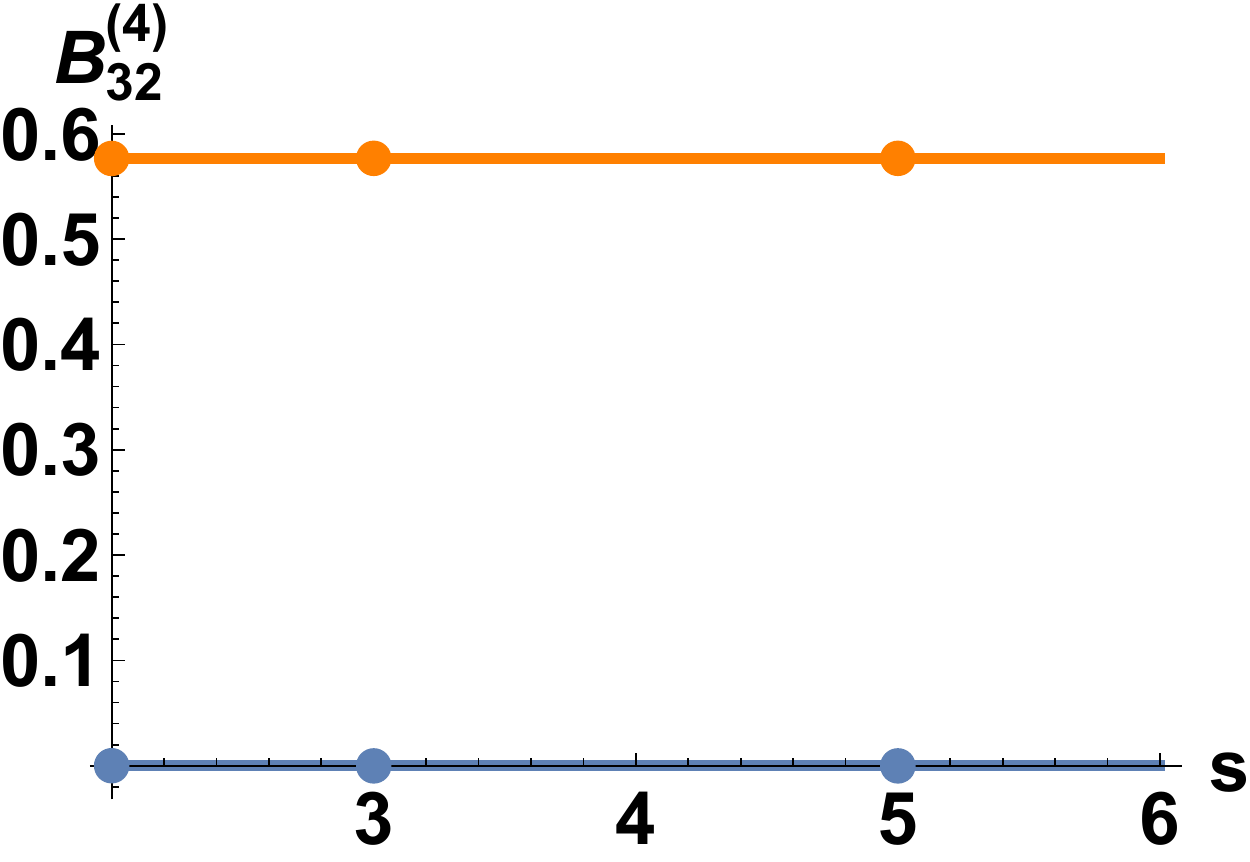} & \img{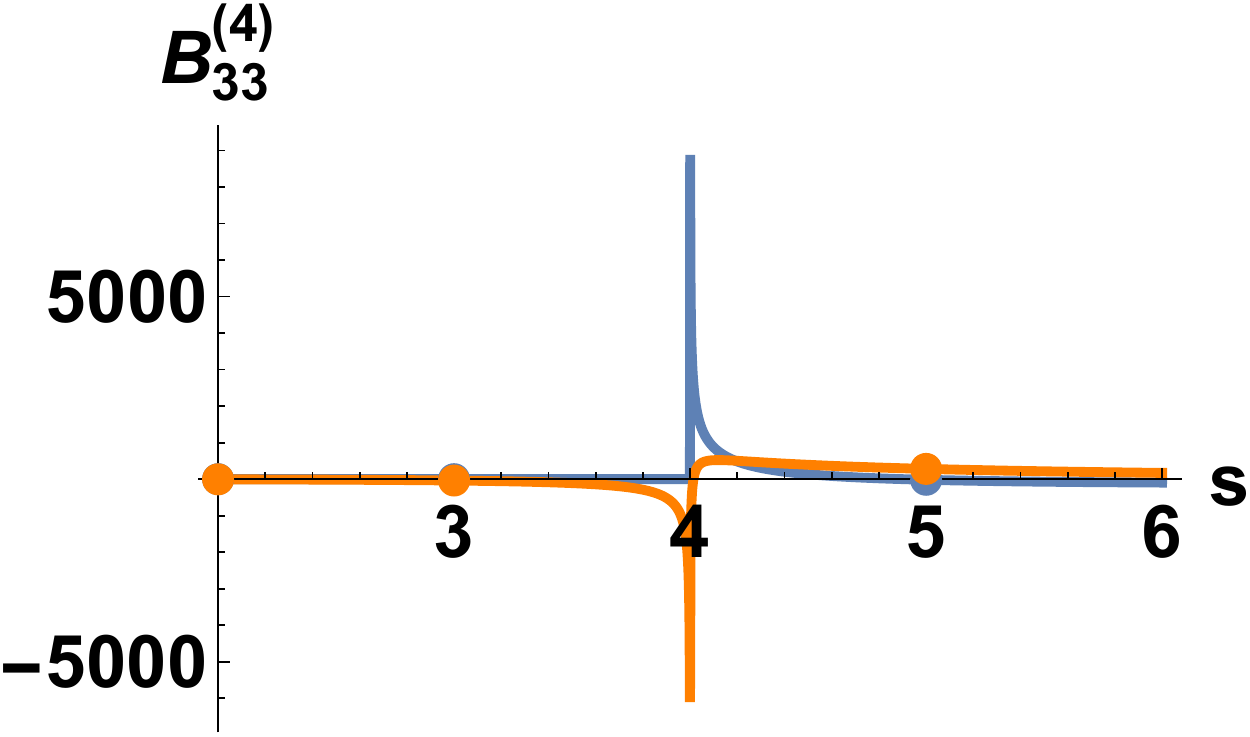} \\
\img{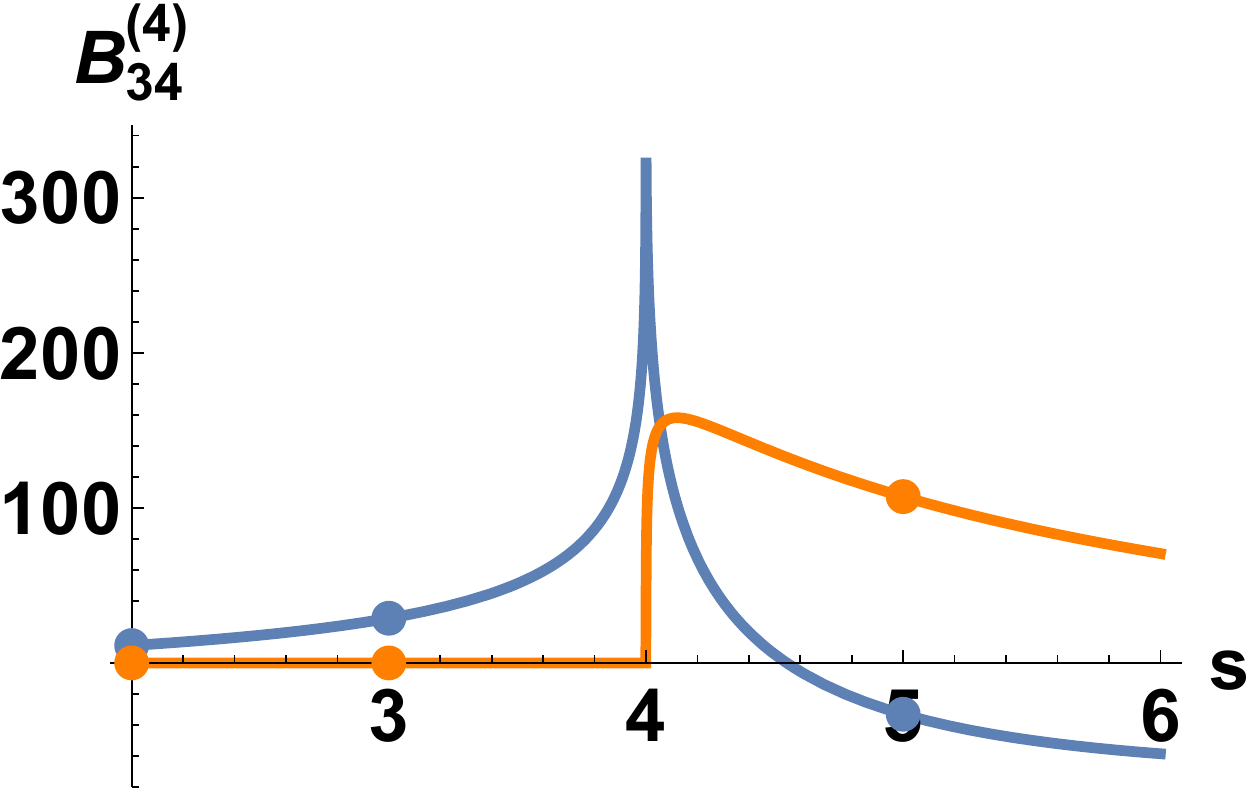} & \img{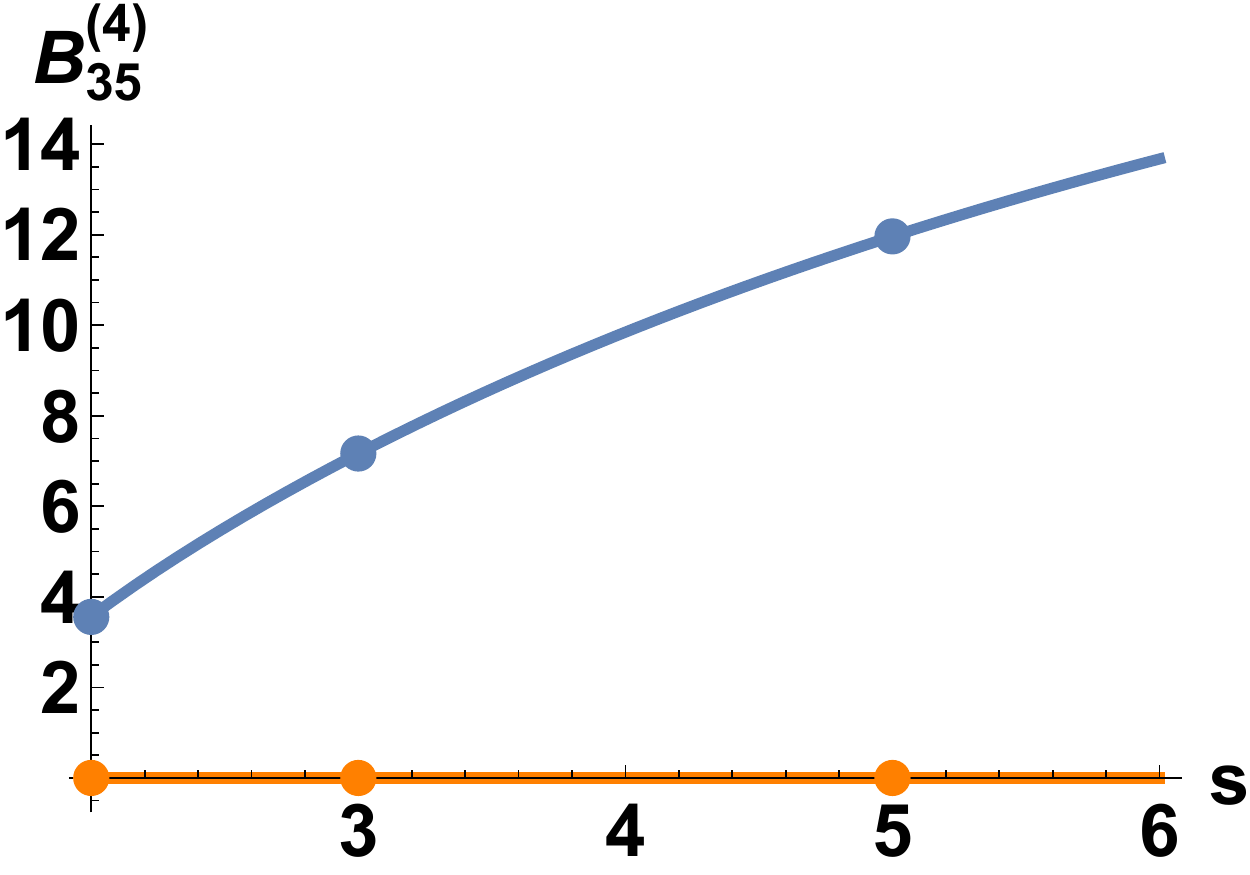} & \img{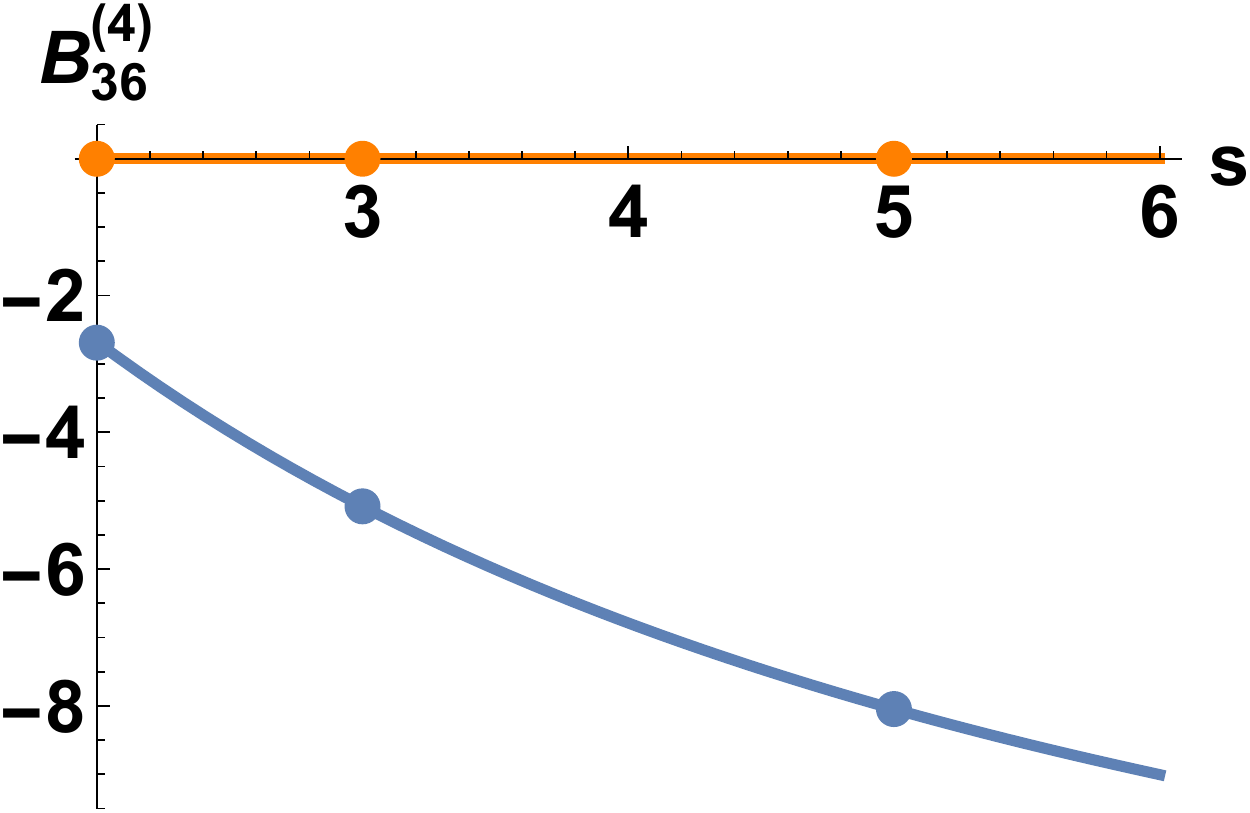} \\
\img{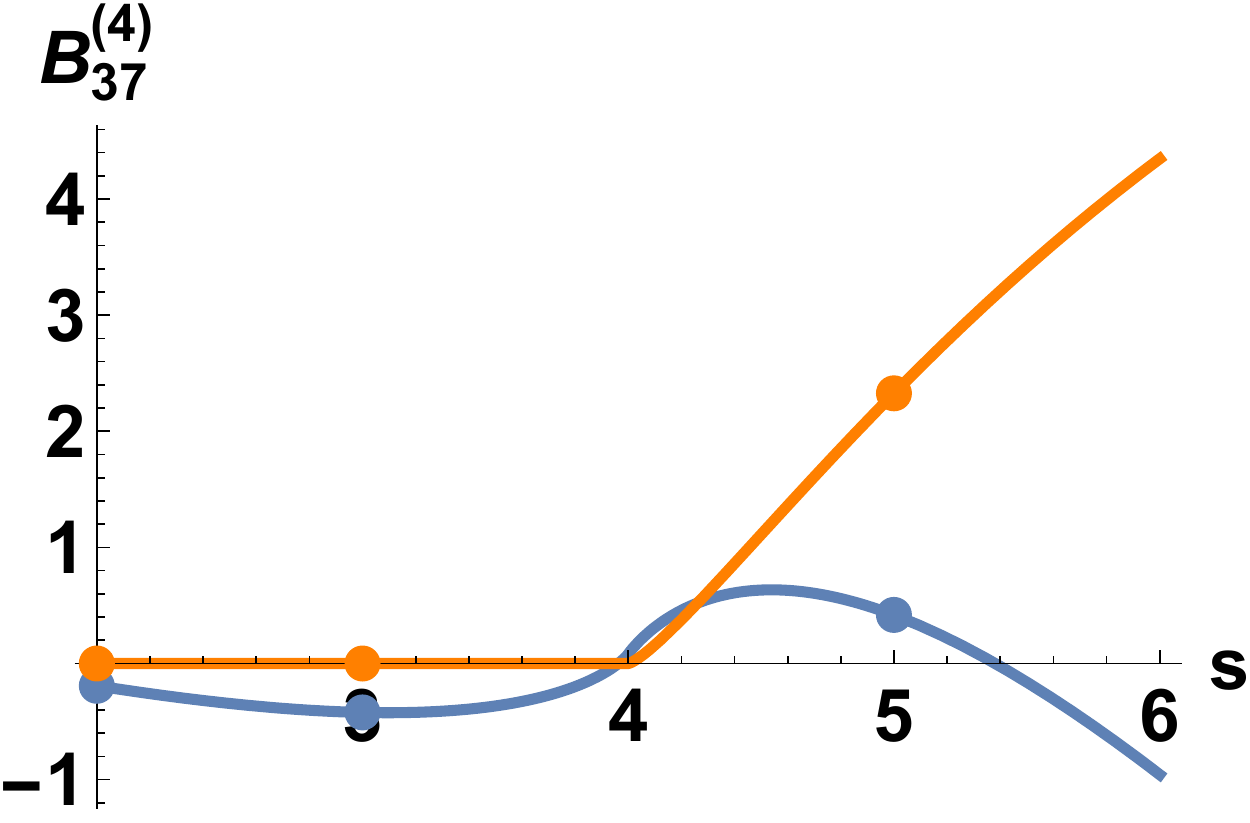} & \img{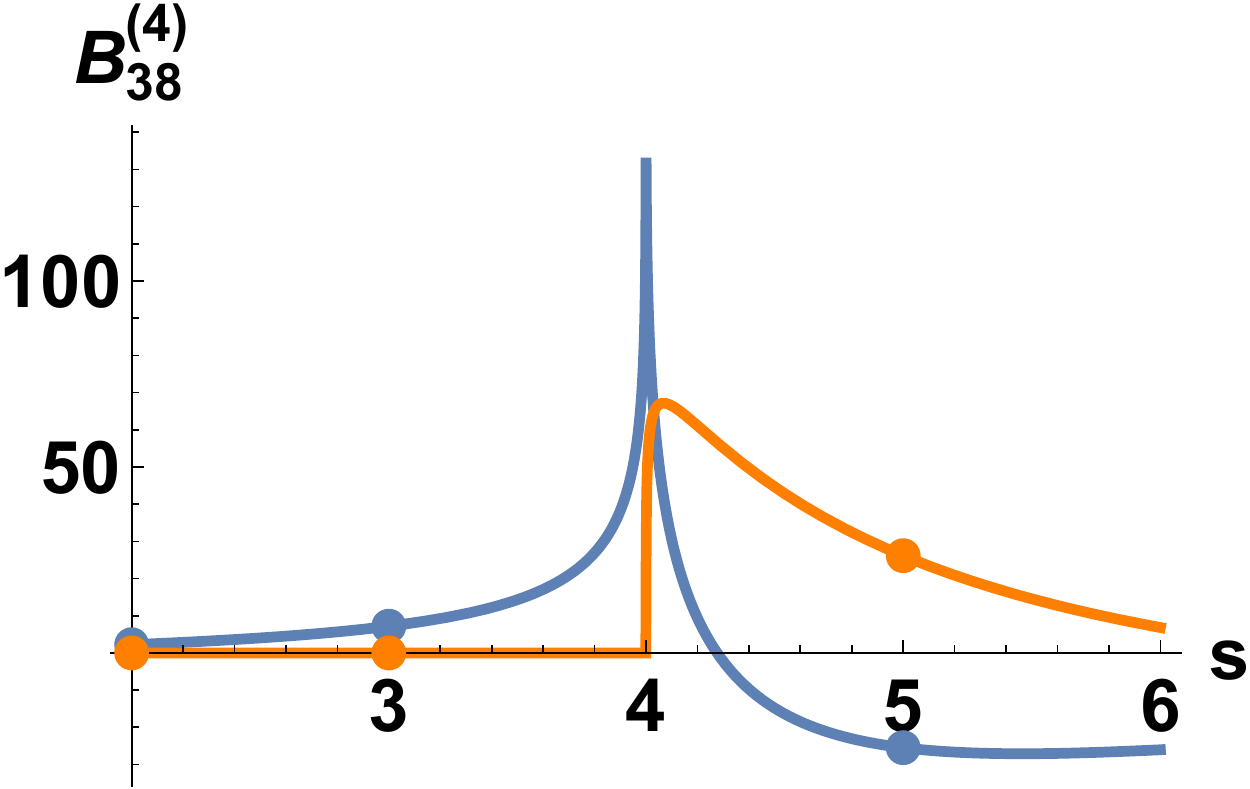} & \img{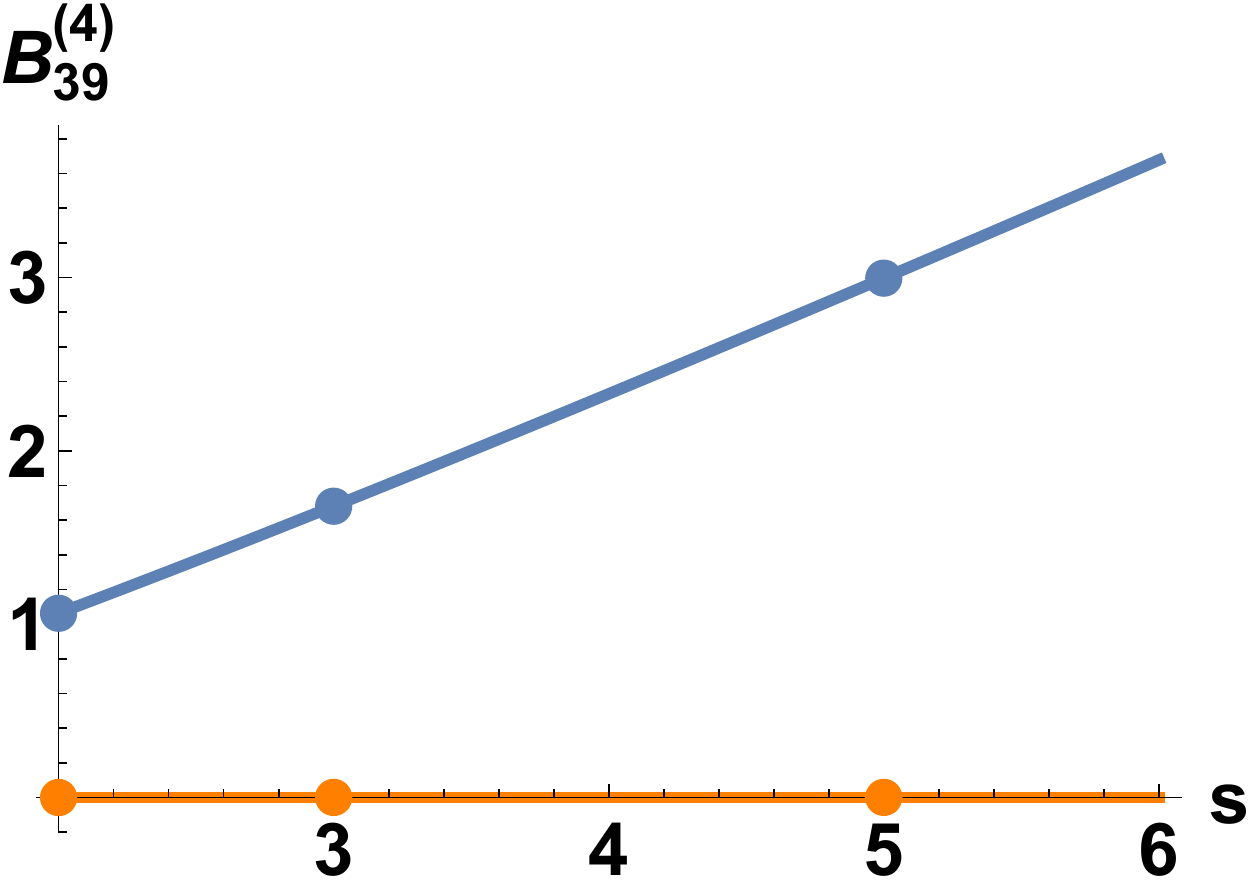} \\
\img{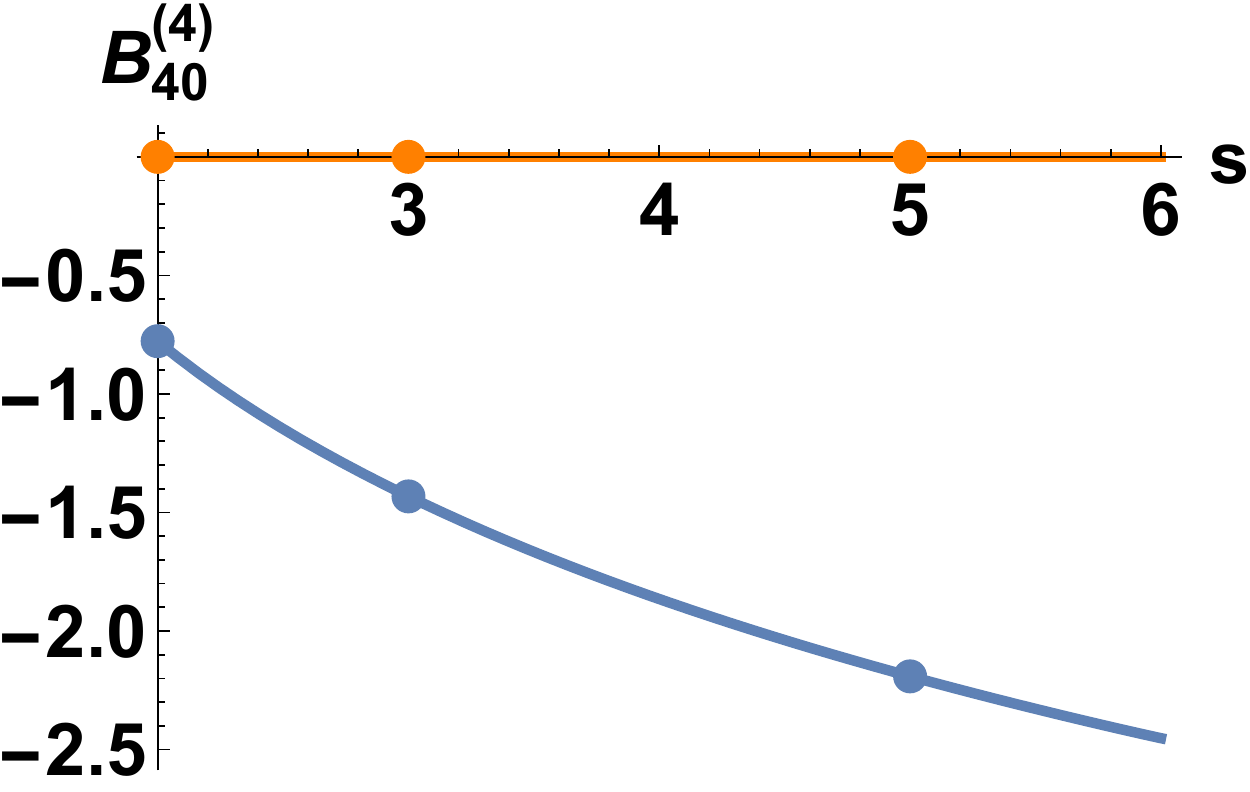} & \img{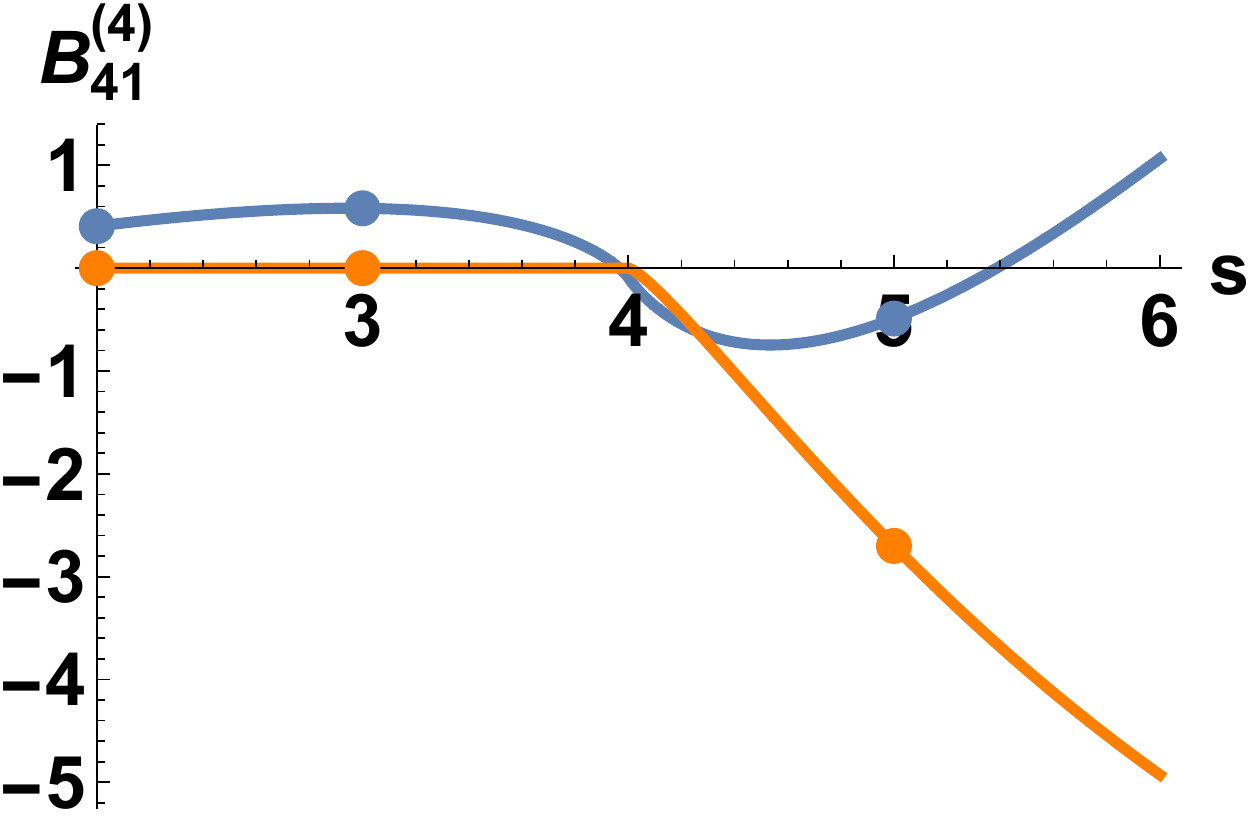} & \img{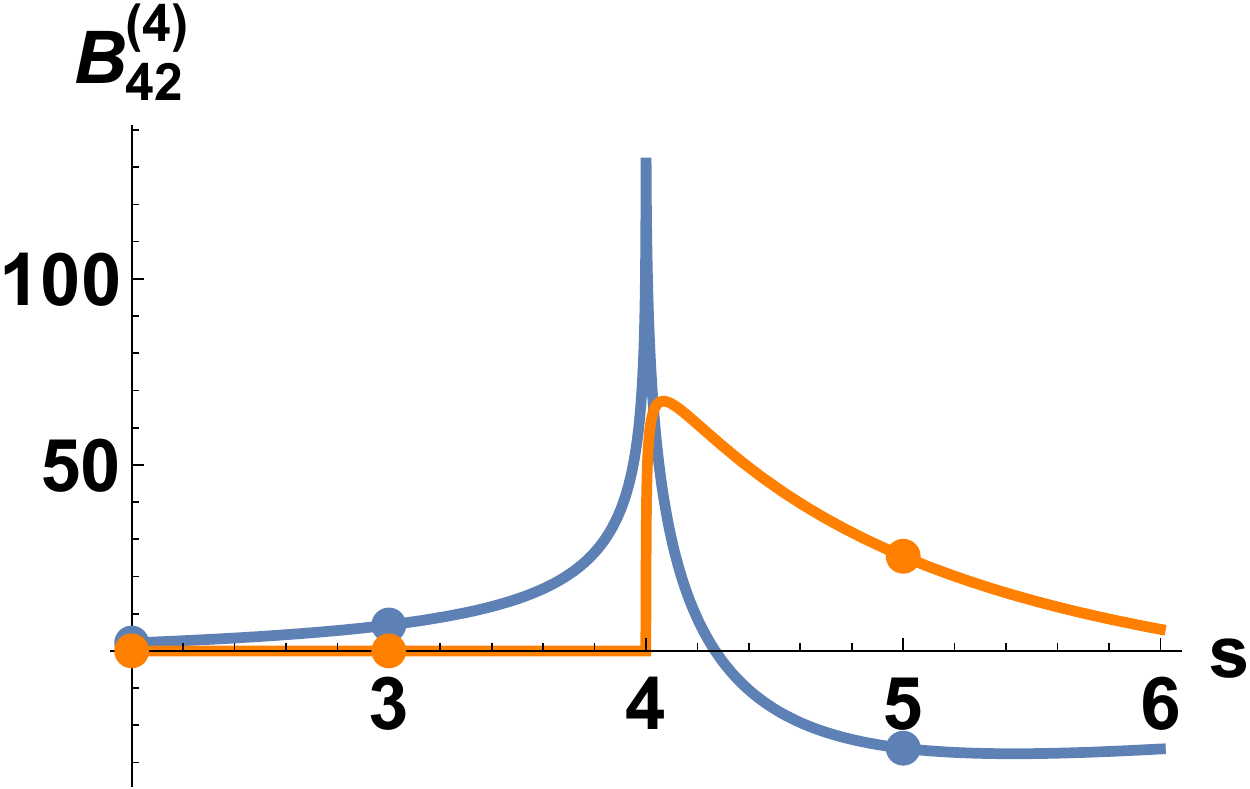} \\
\img{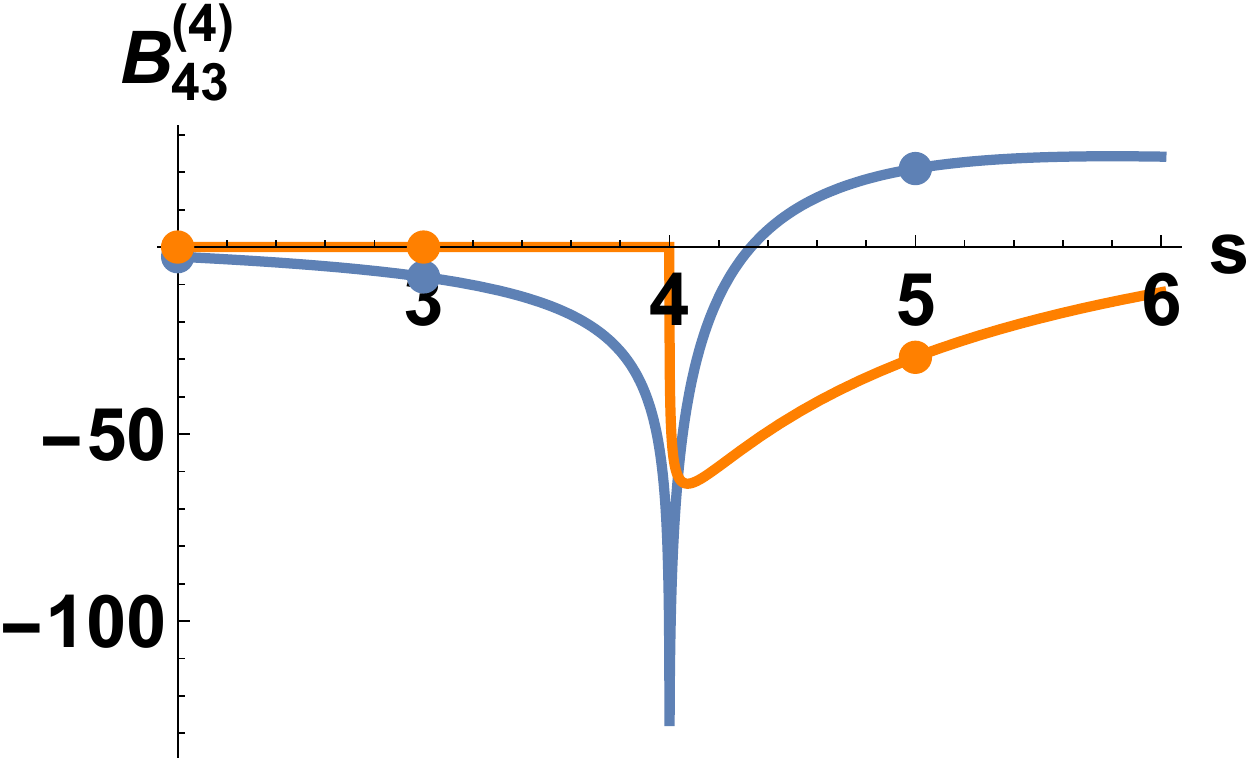} & \img{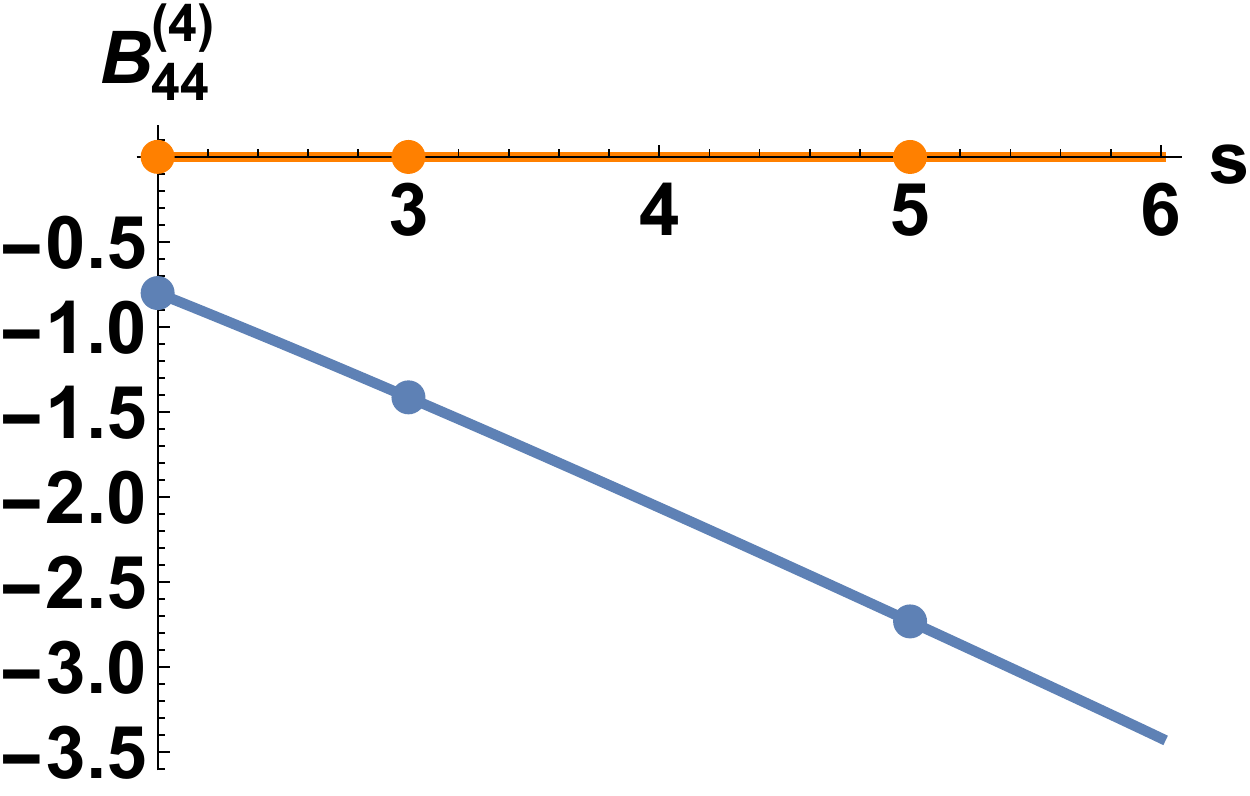} & \img{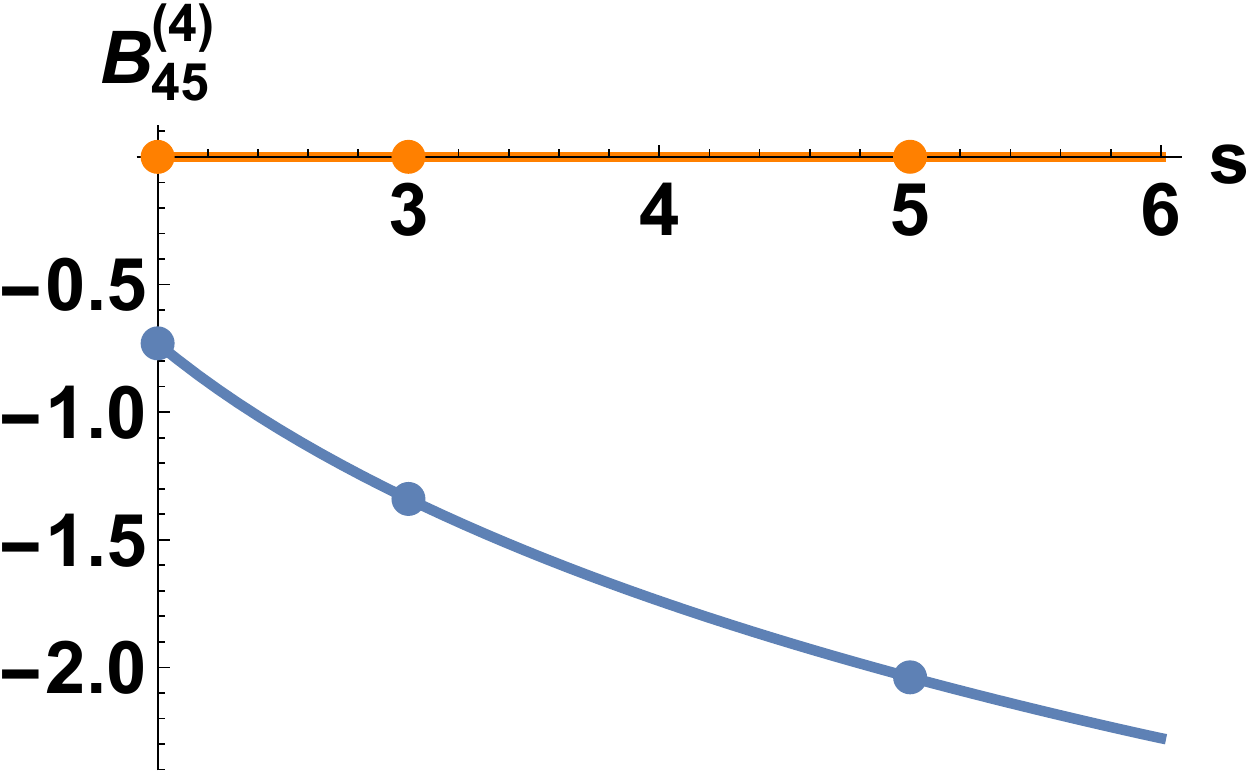} 
\end{tabular}
\end{table}

\begin{table}[t!]
	\begin{tabular}{ p{4.75cm}  p{4.75cm}  p{4.75cm} }
\img{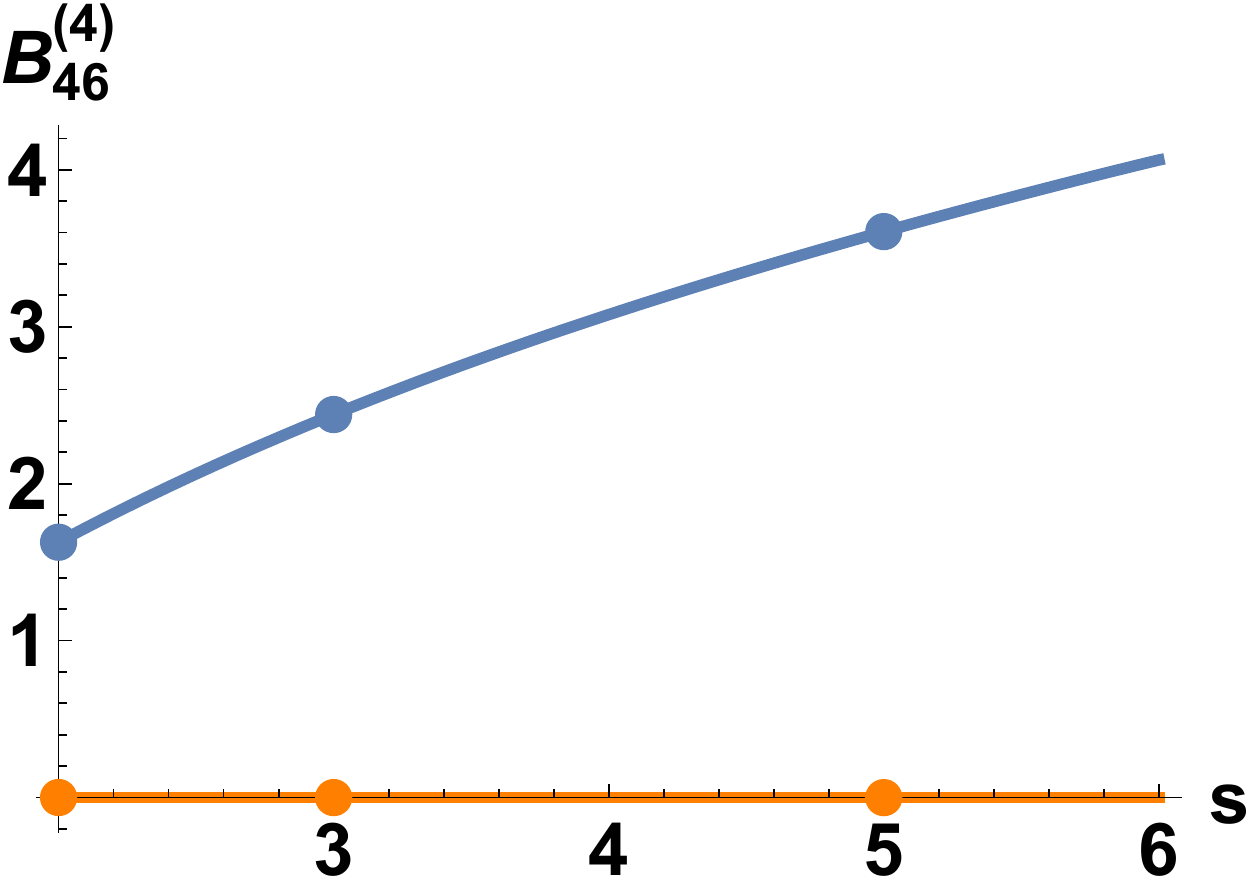} & \img{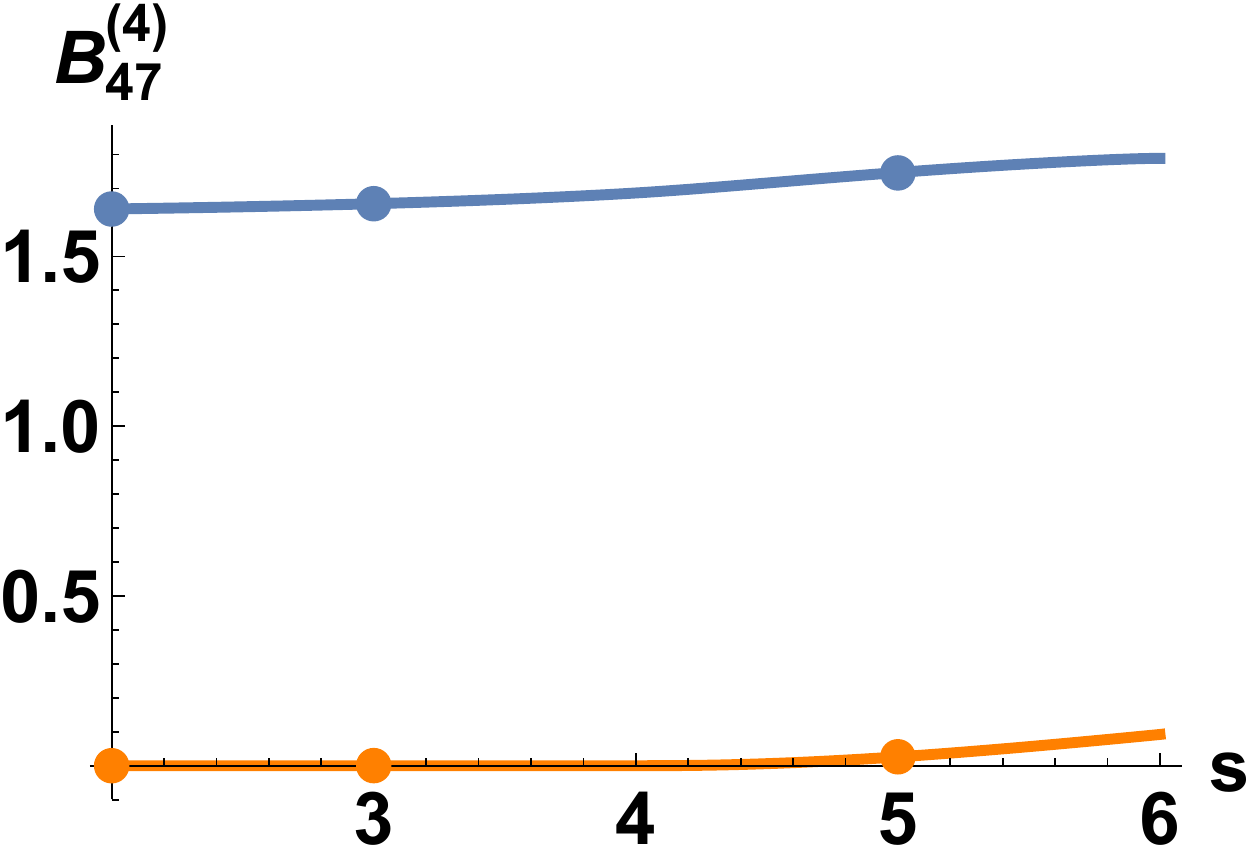} & \img{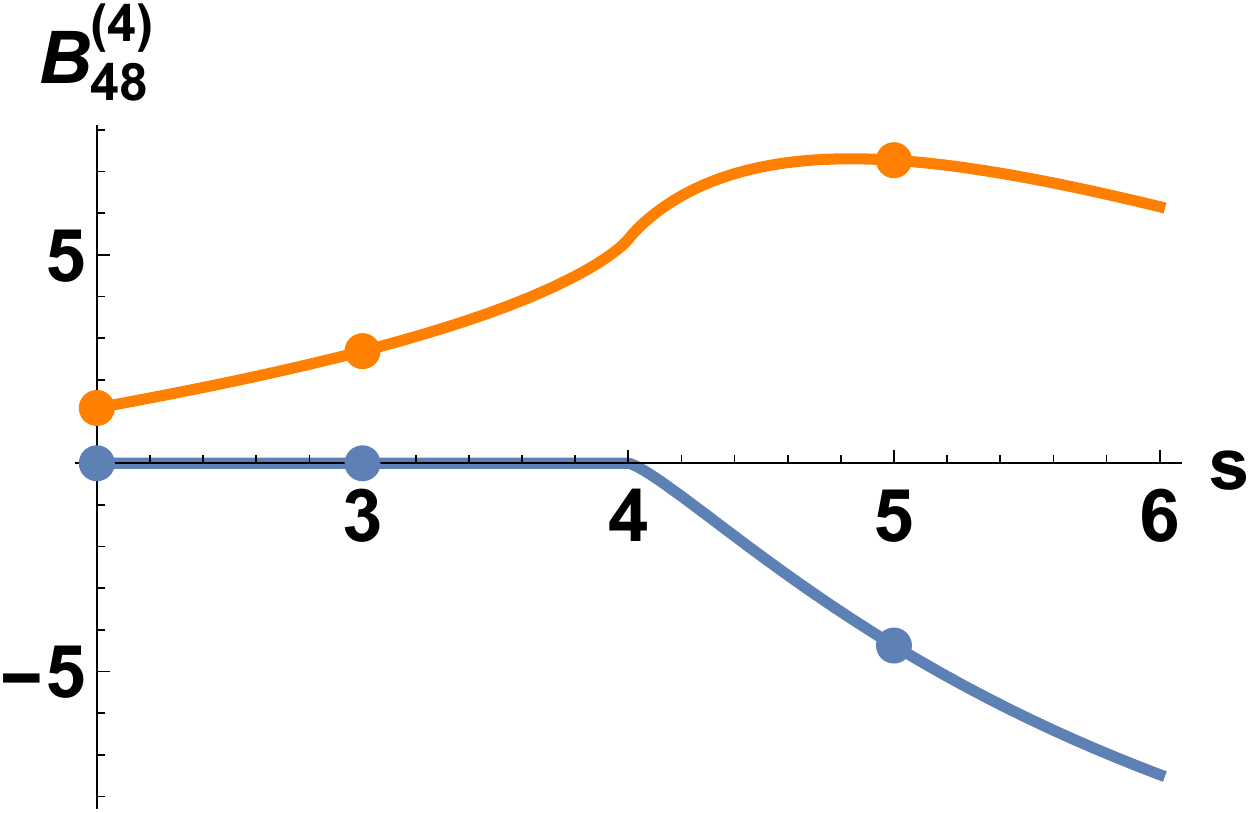} \\
\img{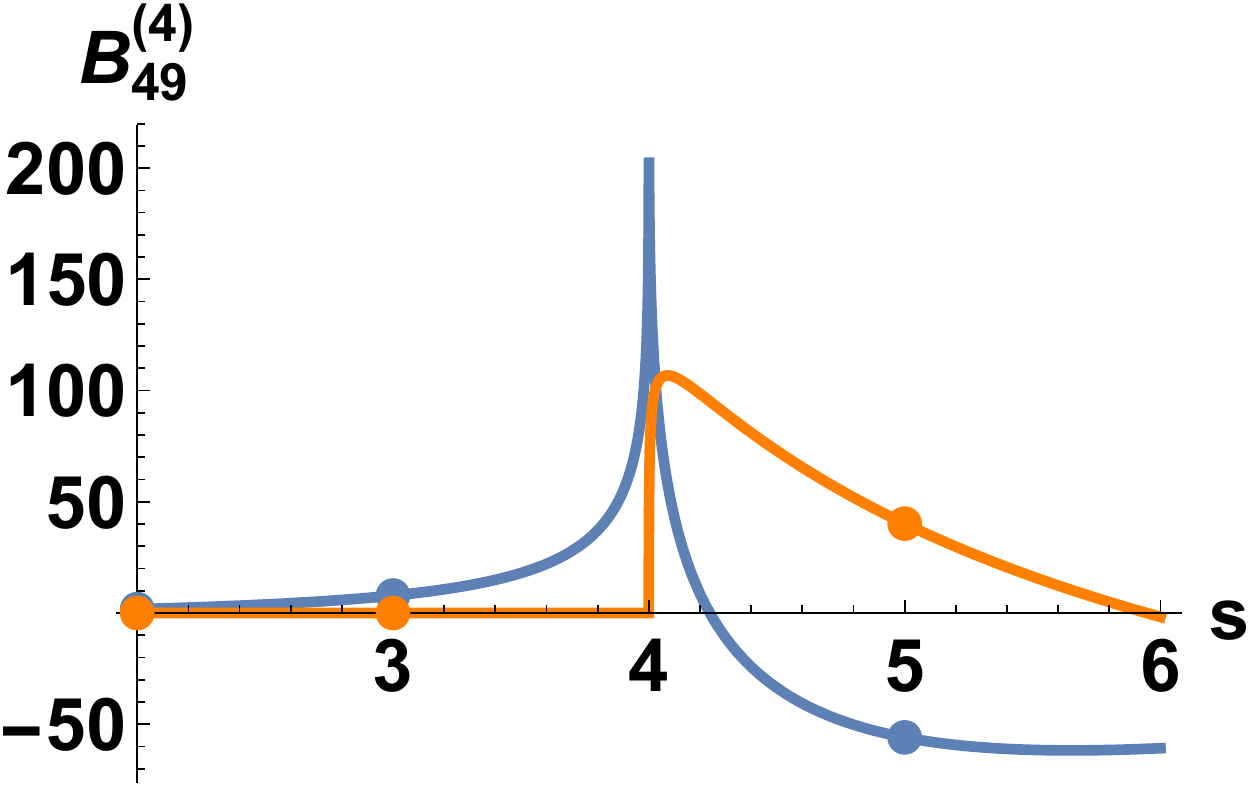} & \img{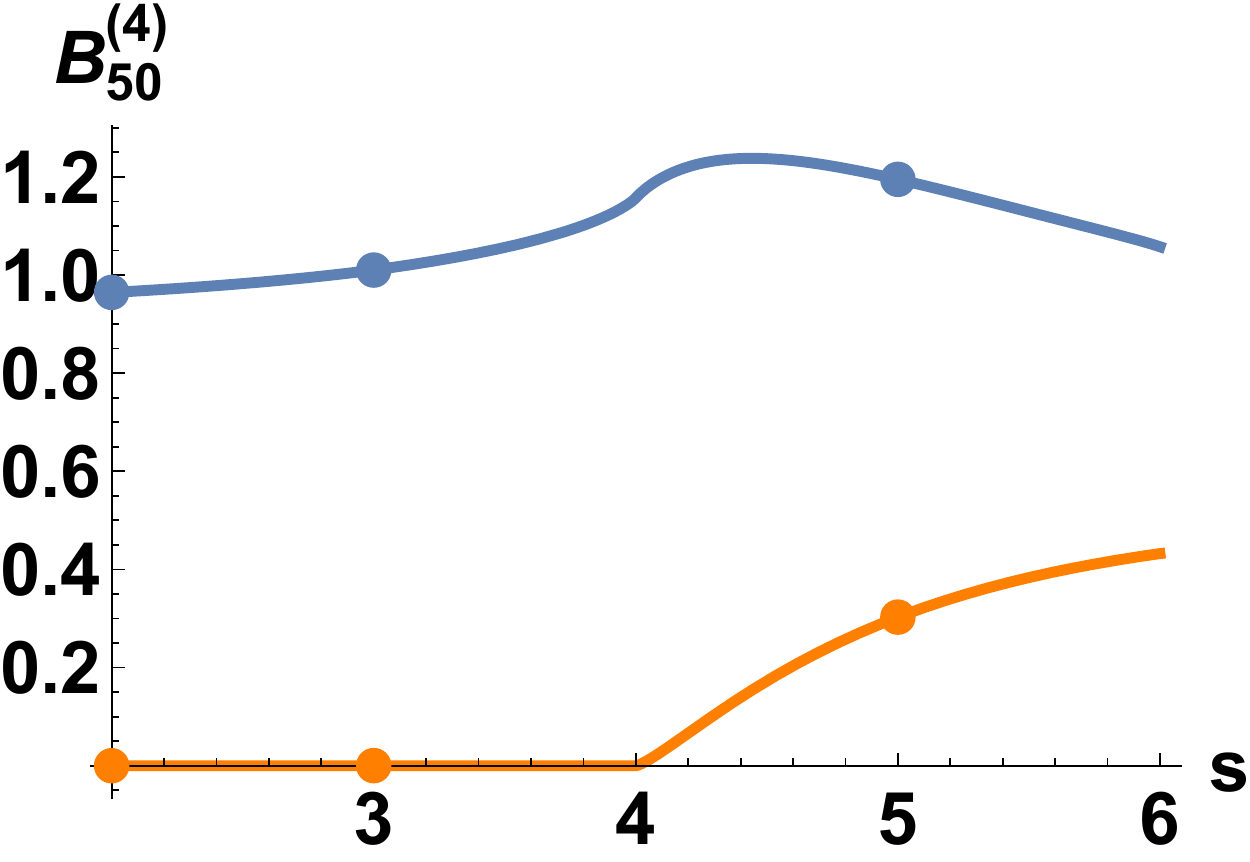} & \img{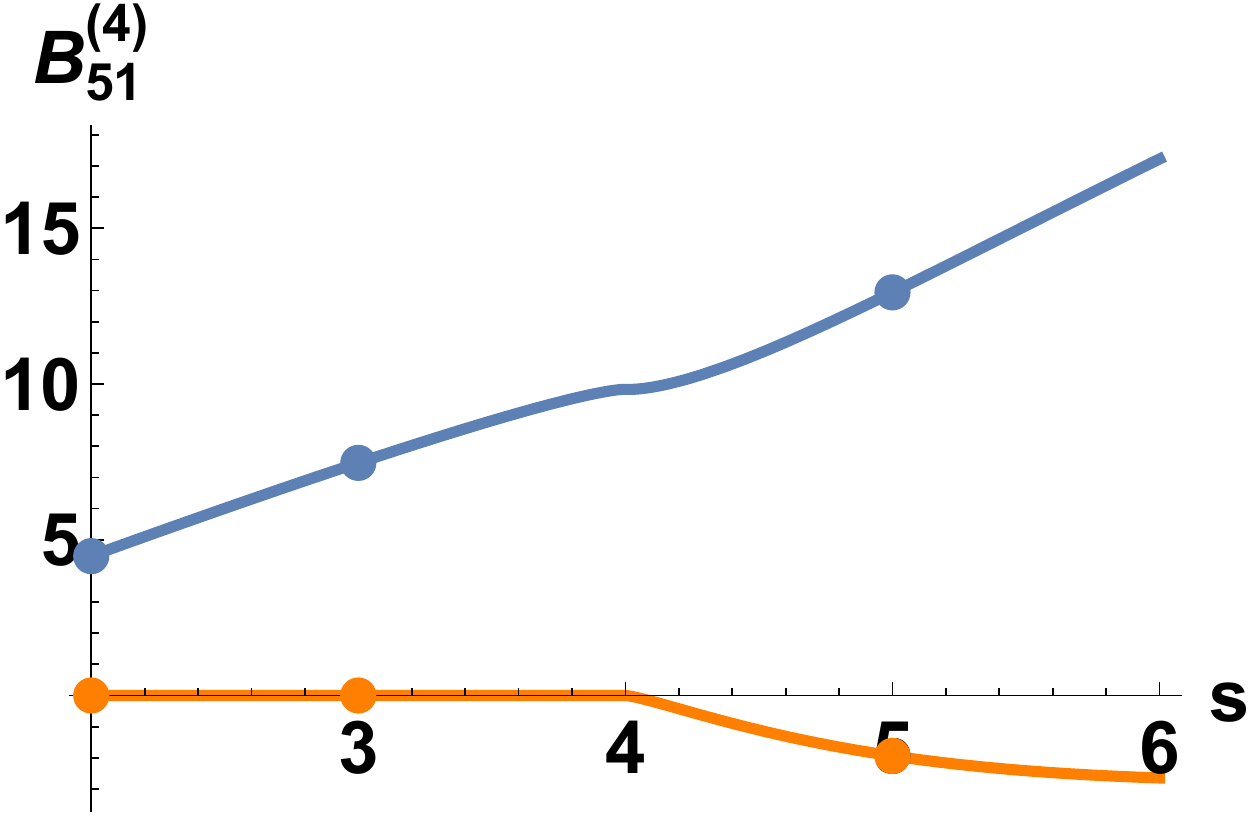} \\
\img{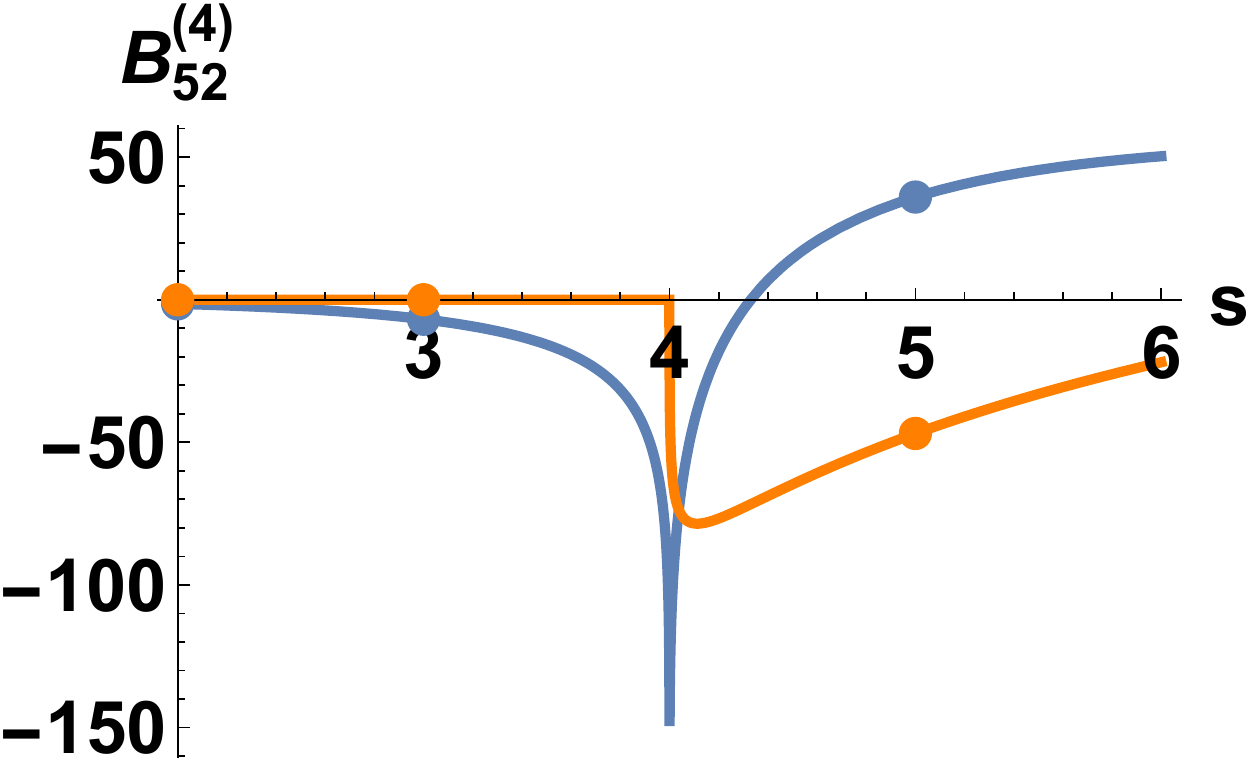} & \img{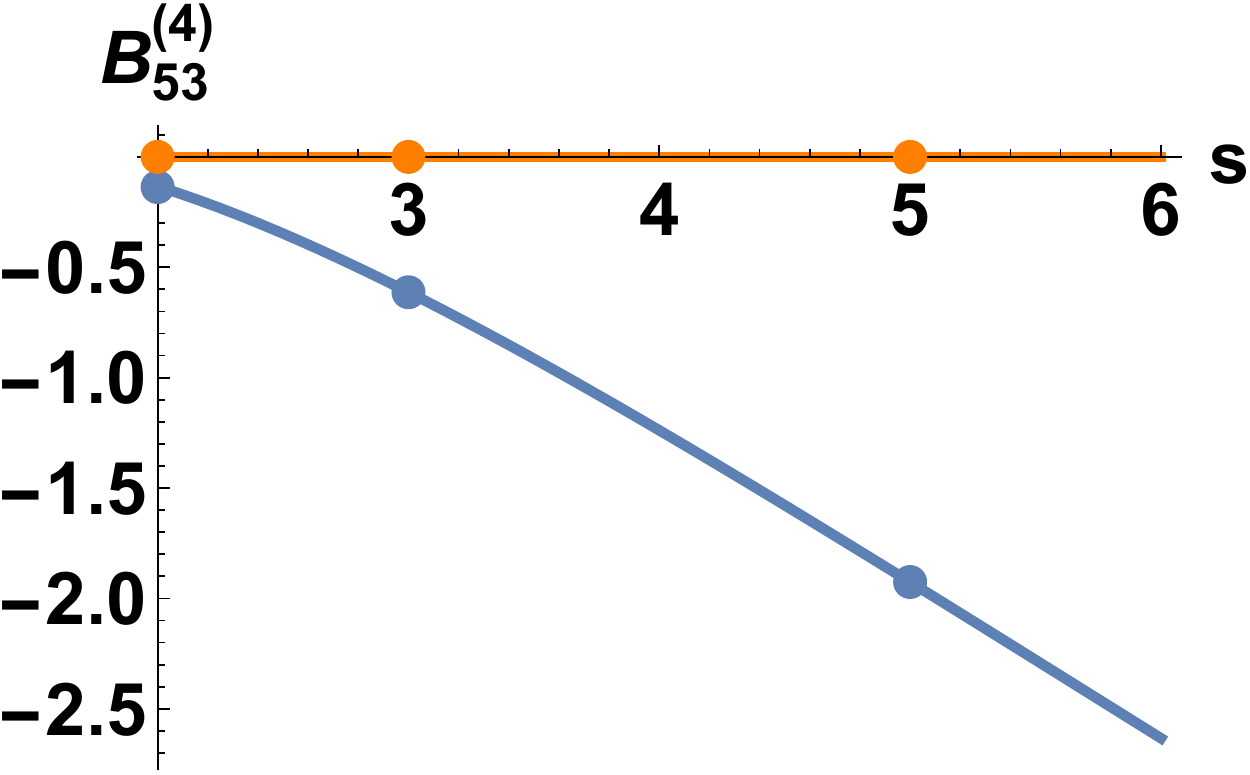} & \img{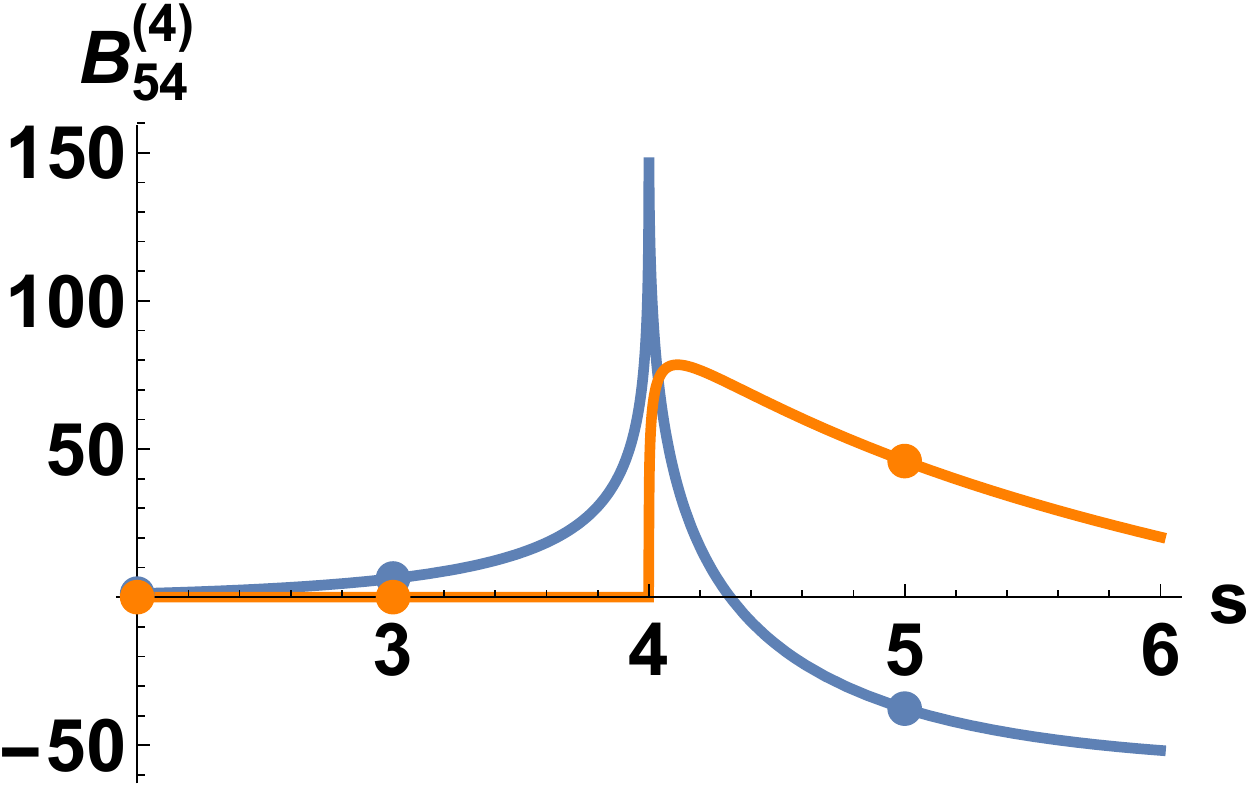} \\
\img{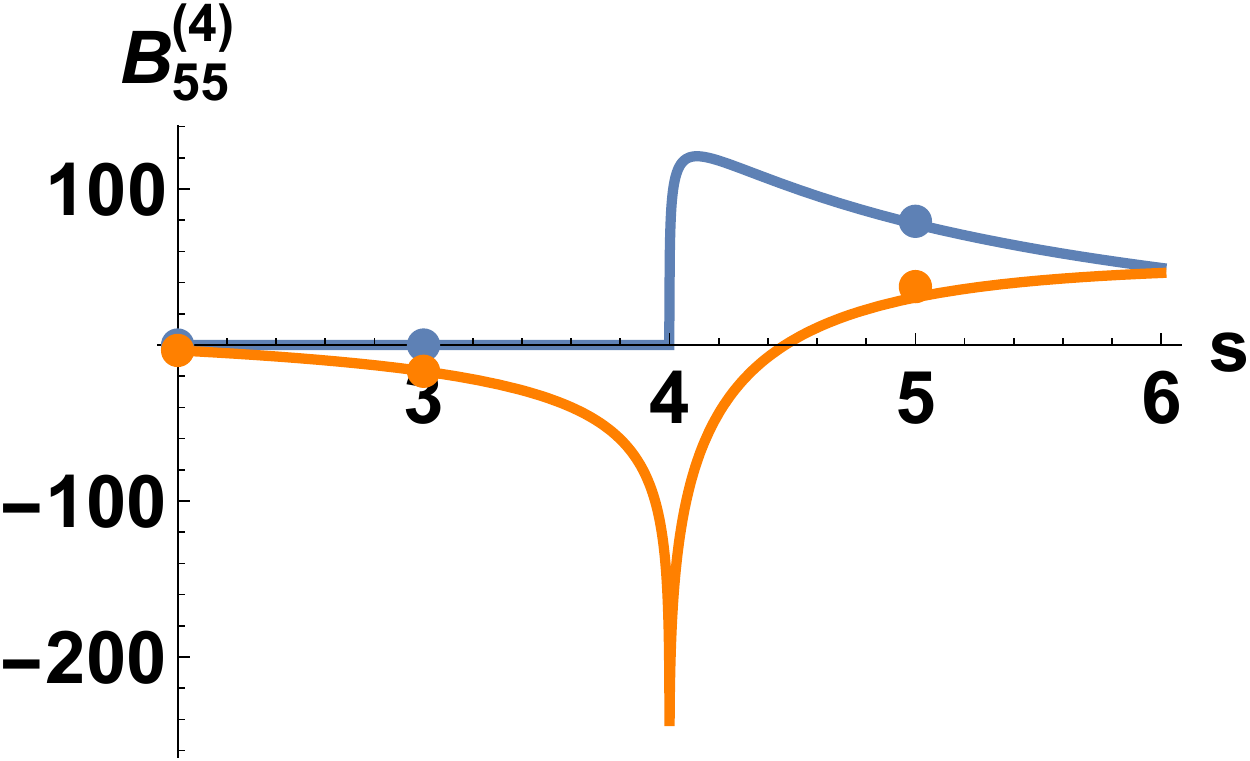} & \img{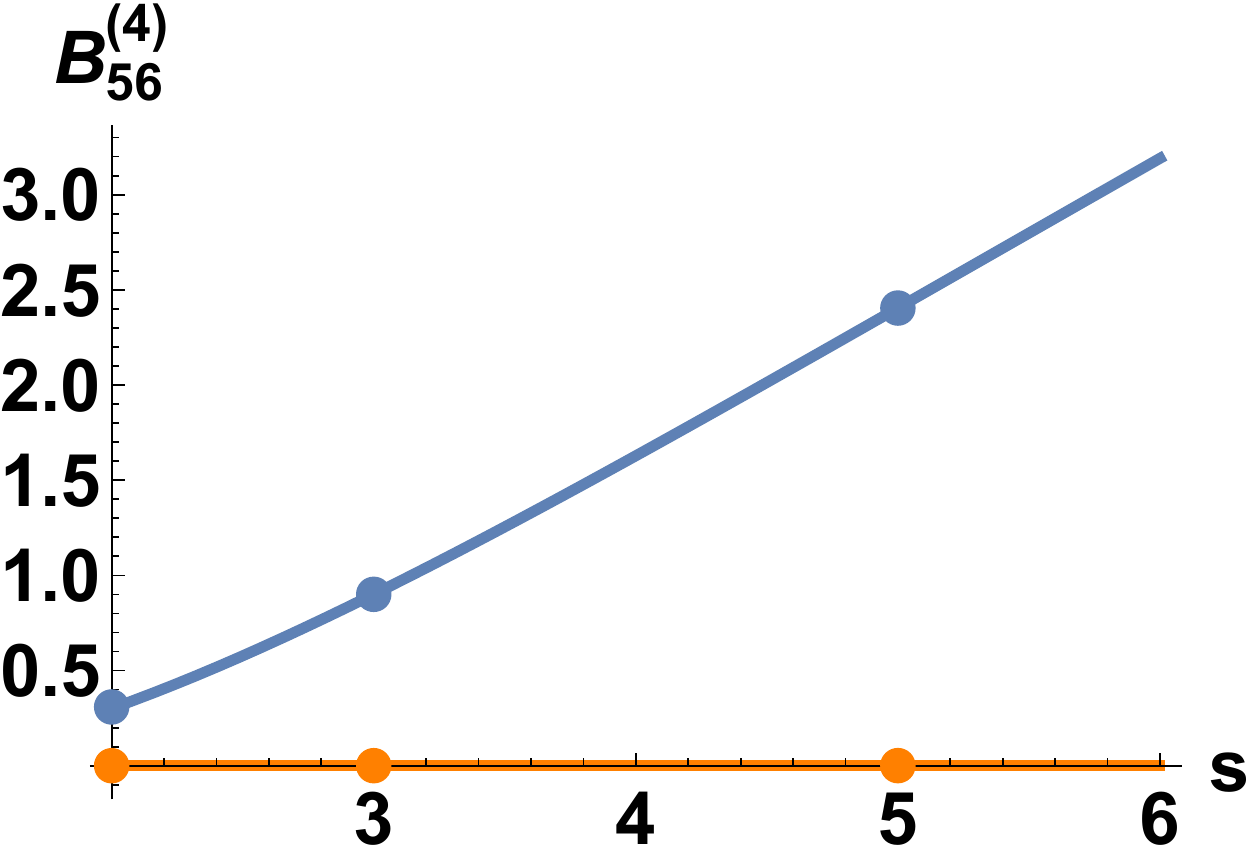} & \img{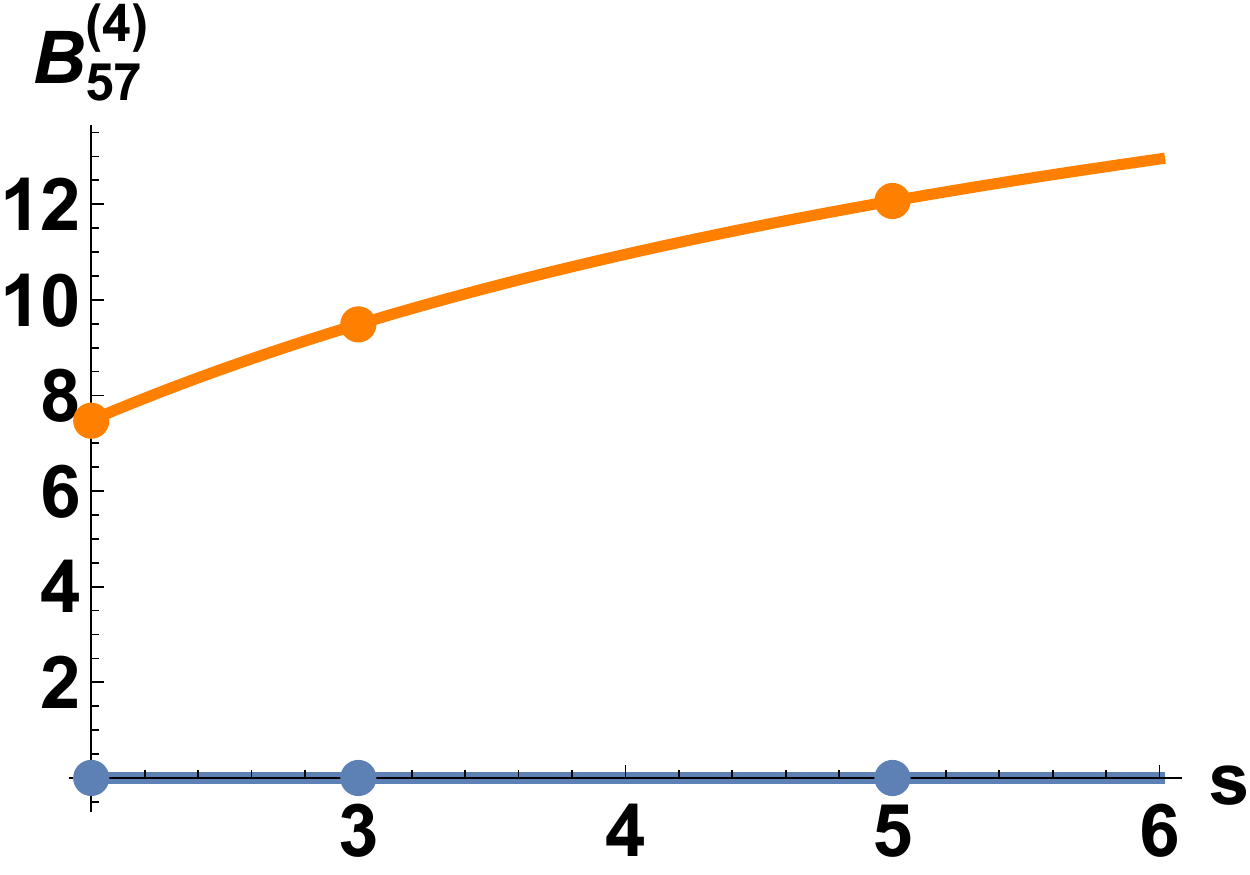} \\
\img{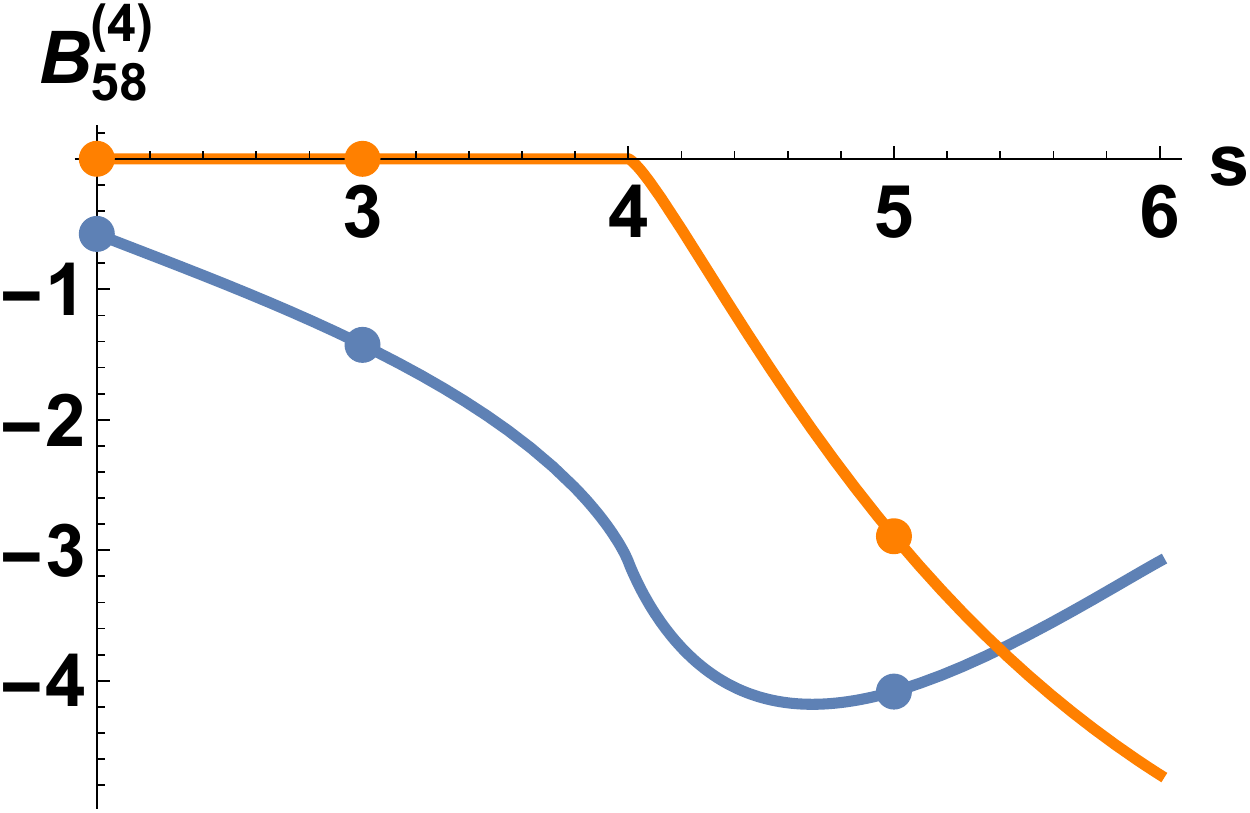} & \img{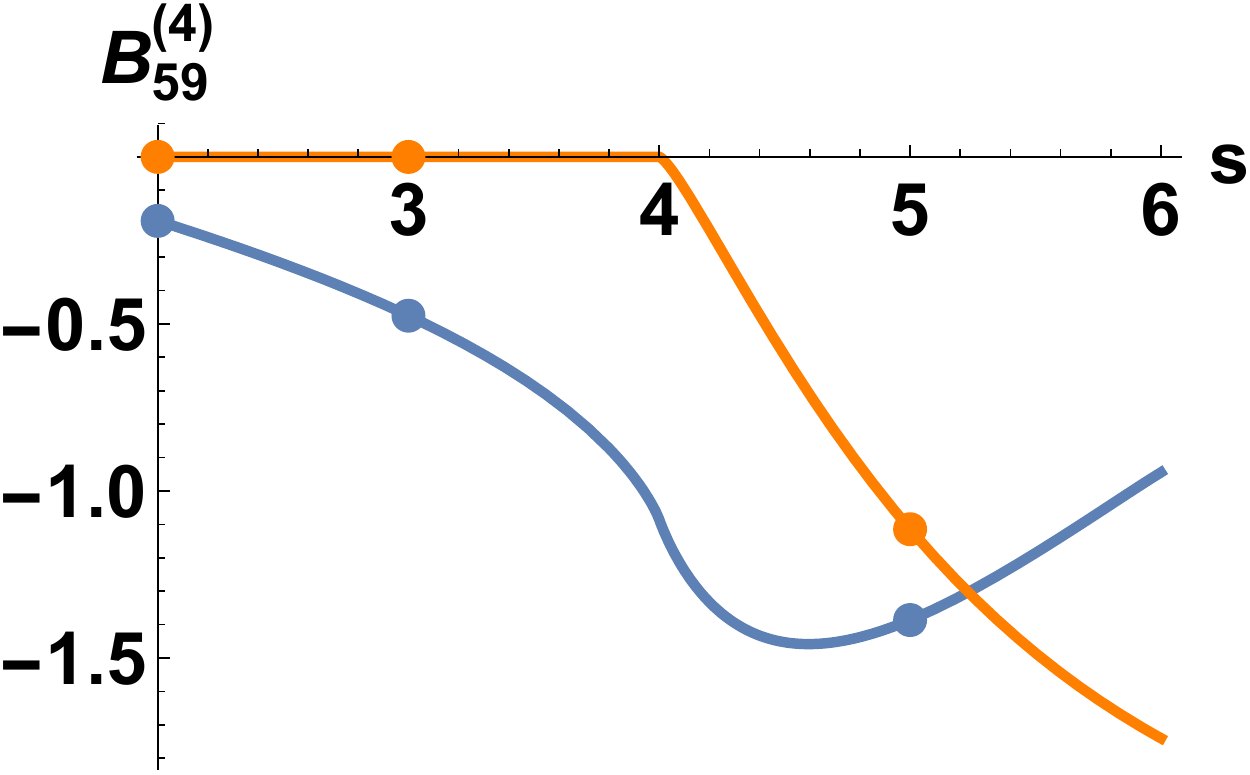} & \img{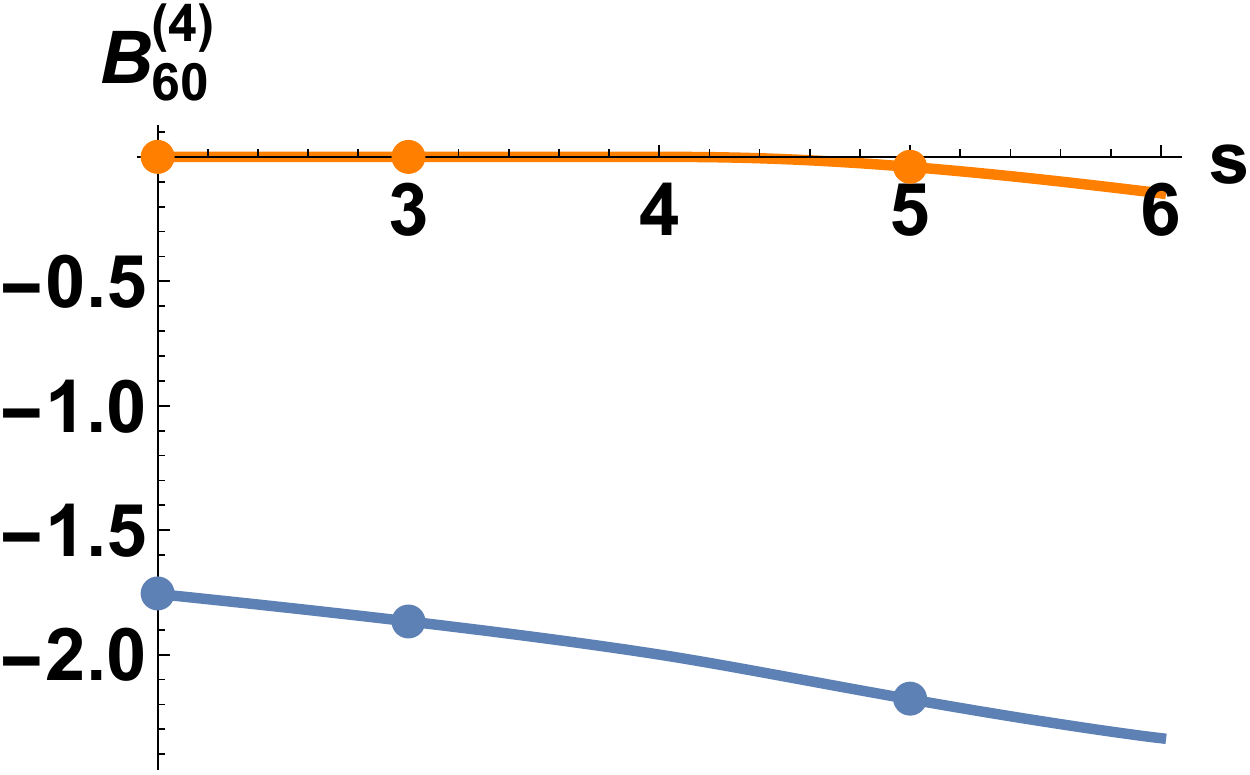} \\
\img{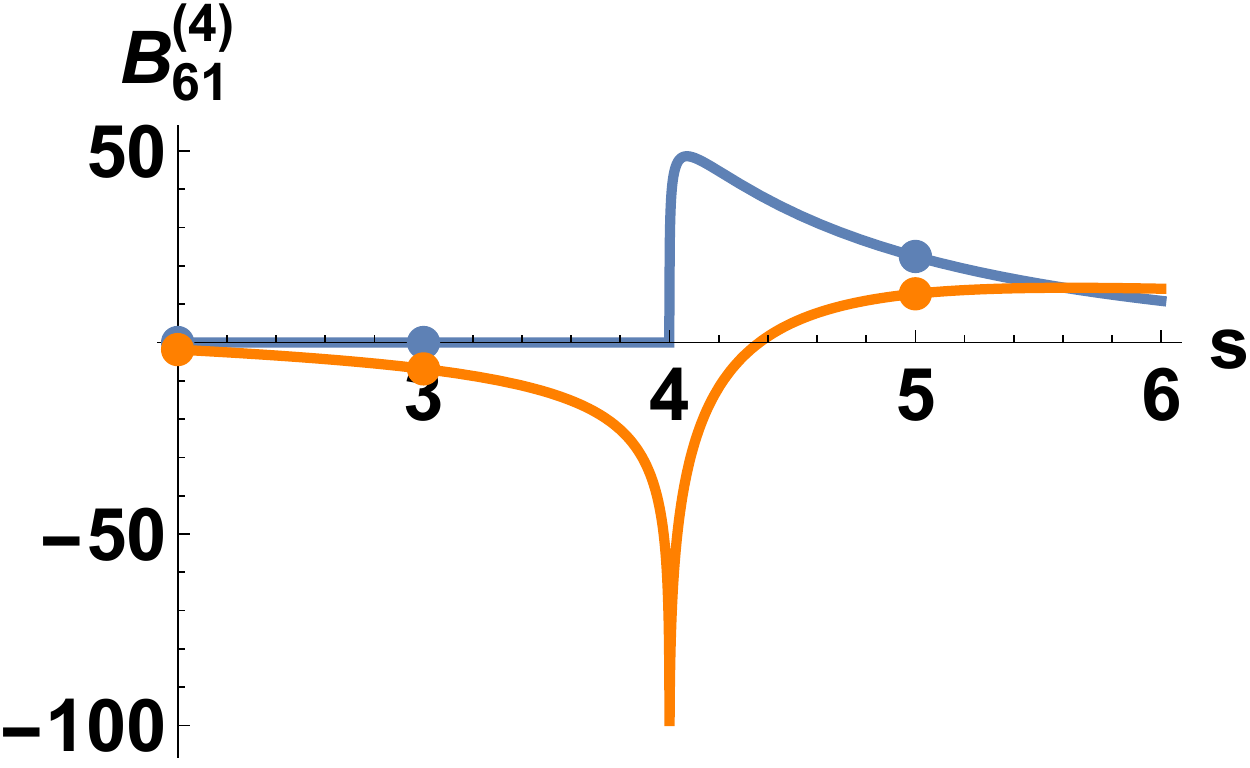} & \img{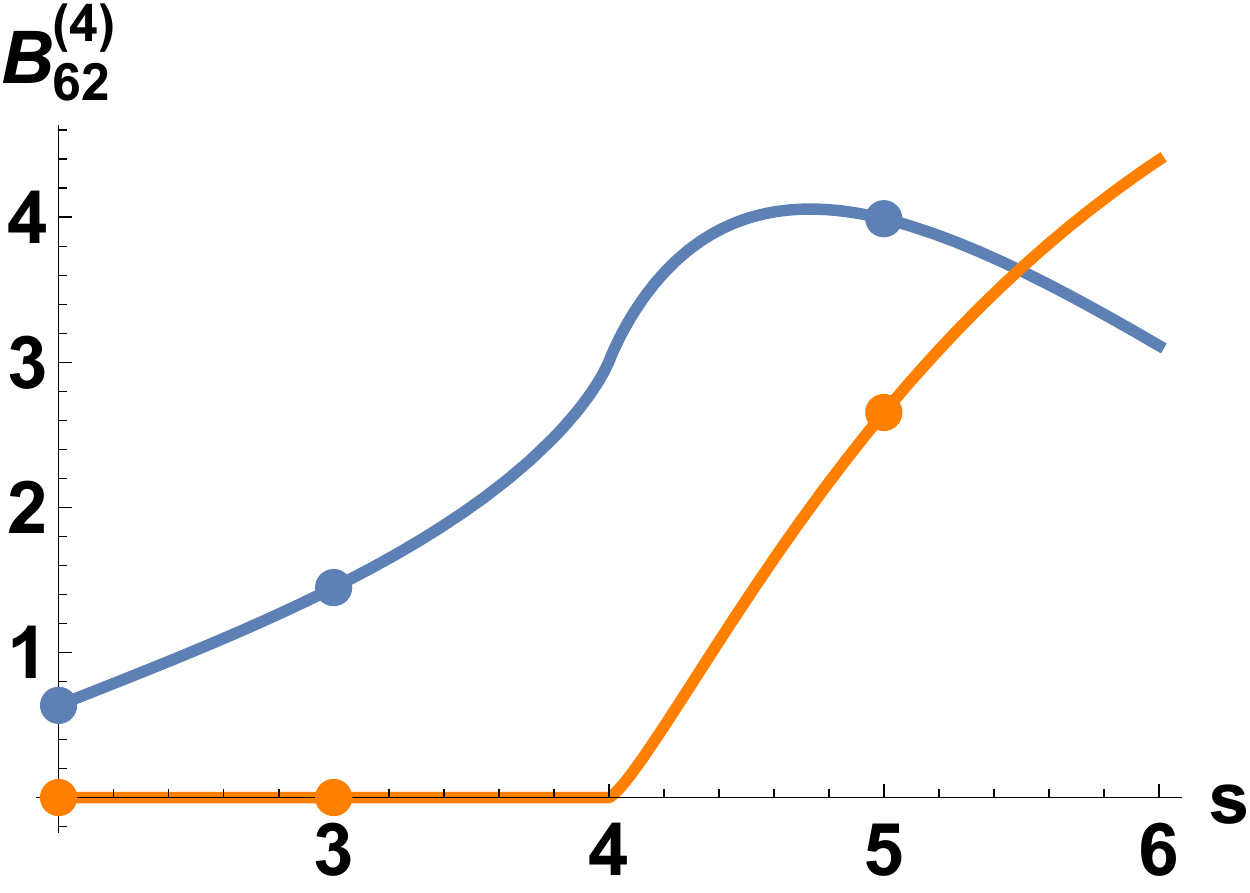} & \img{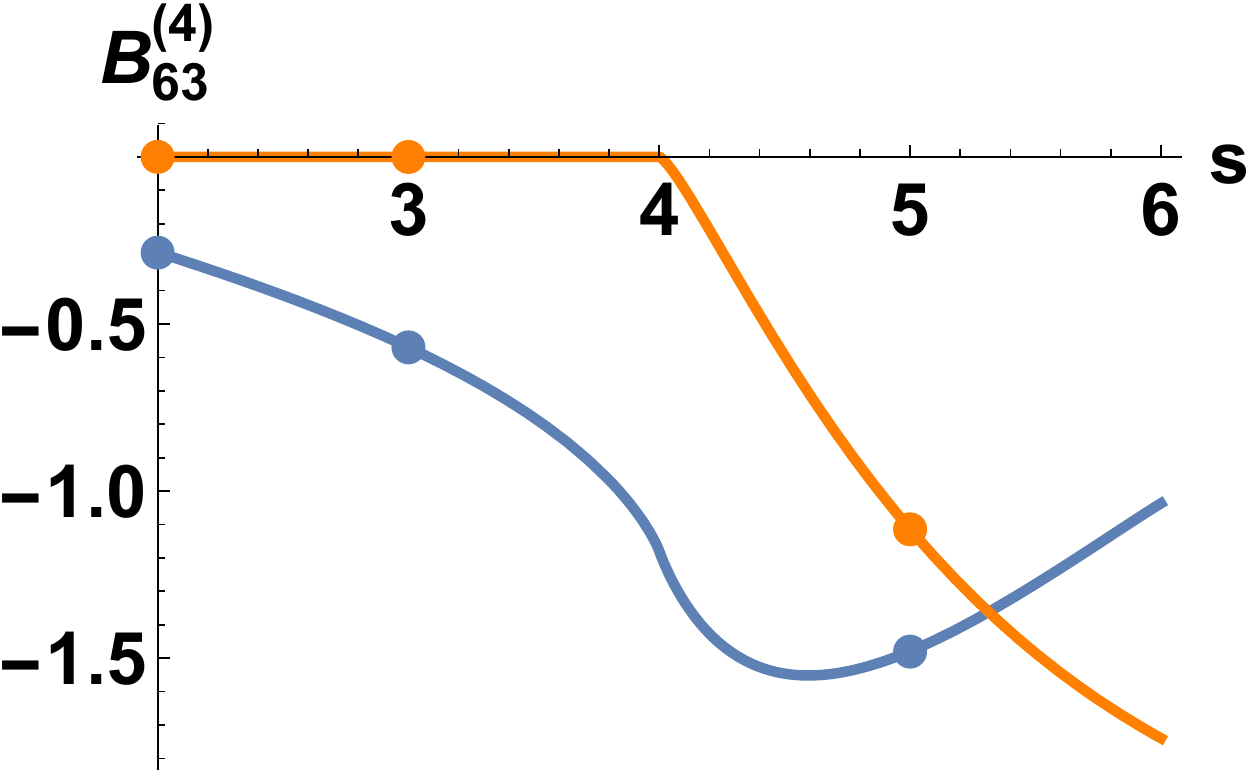} \\
\img{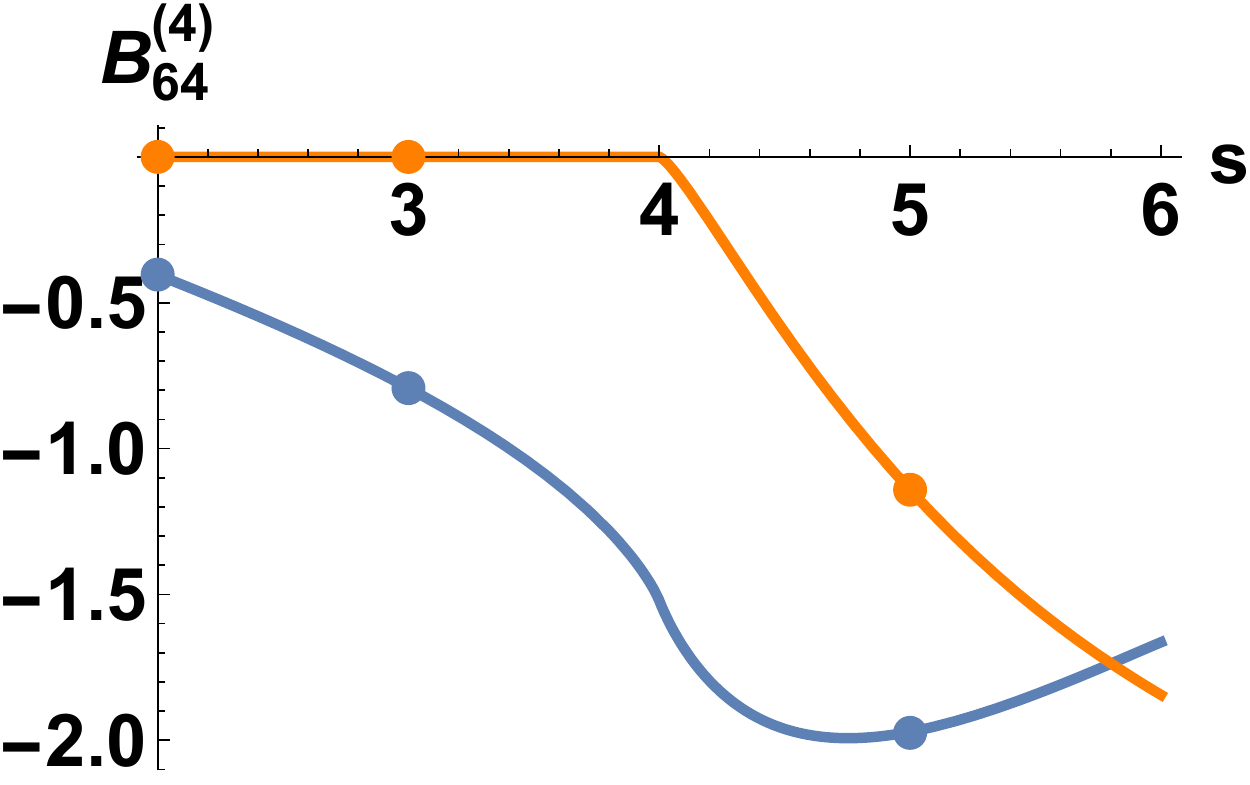} & \img{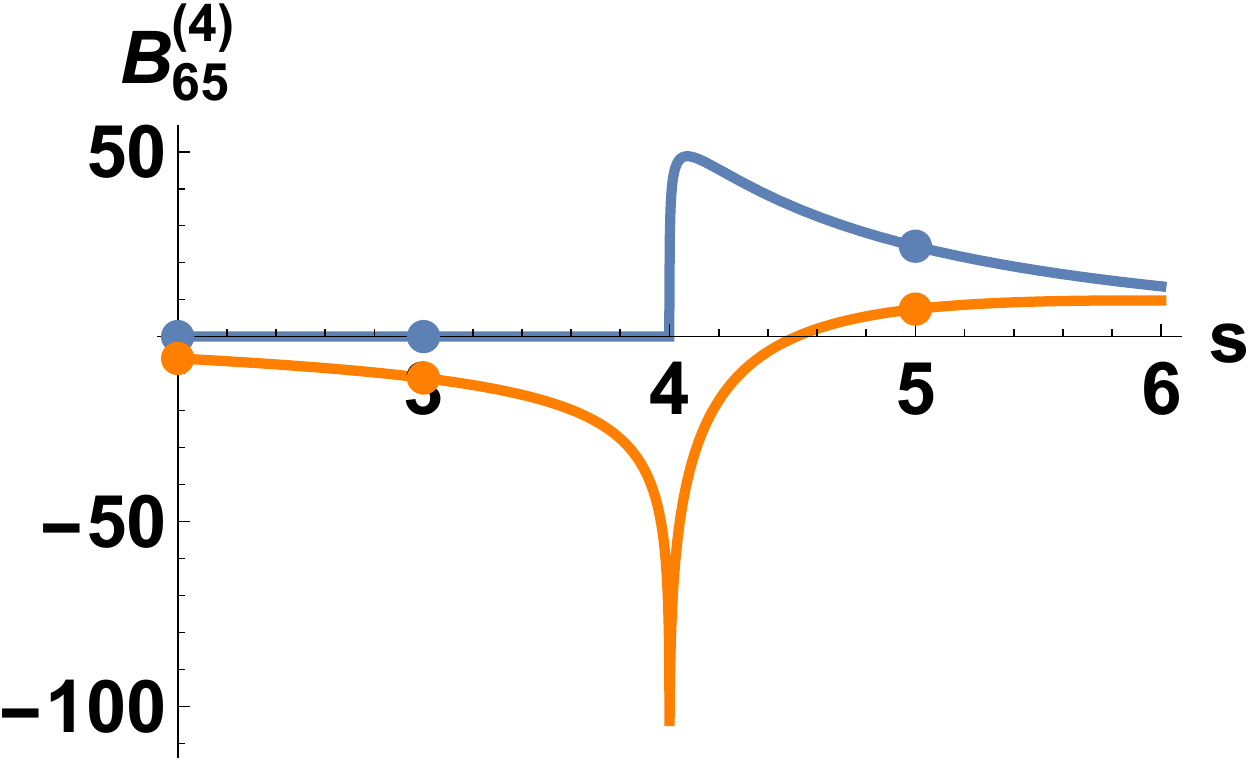} & \img{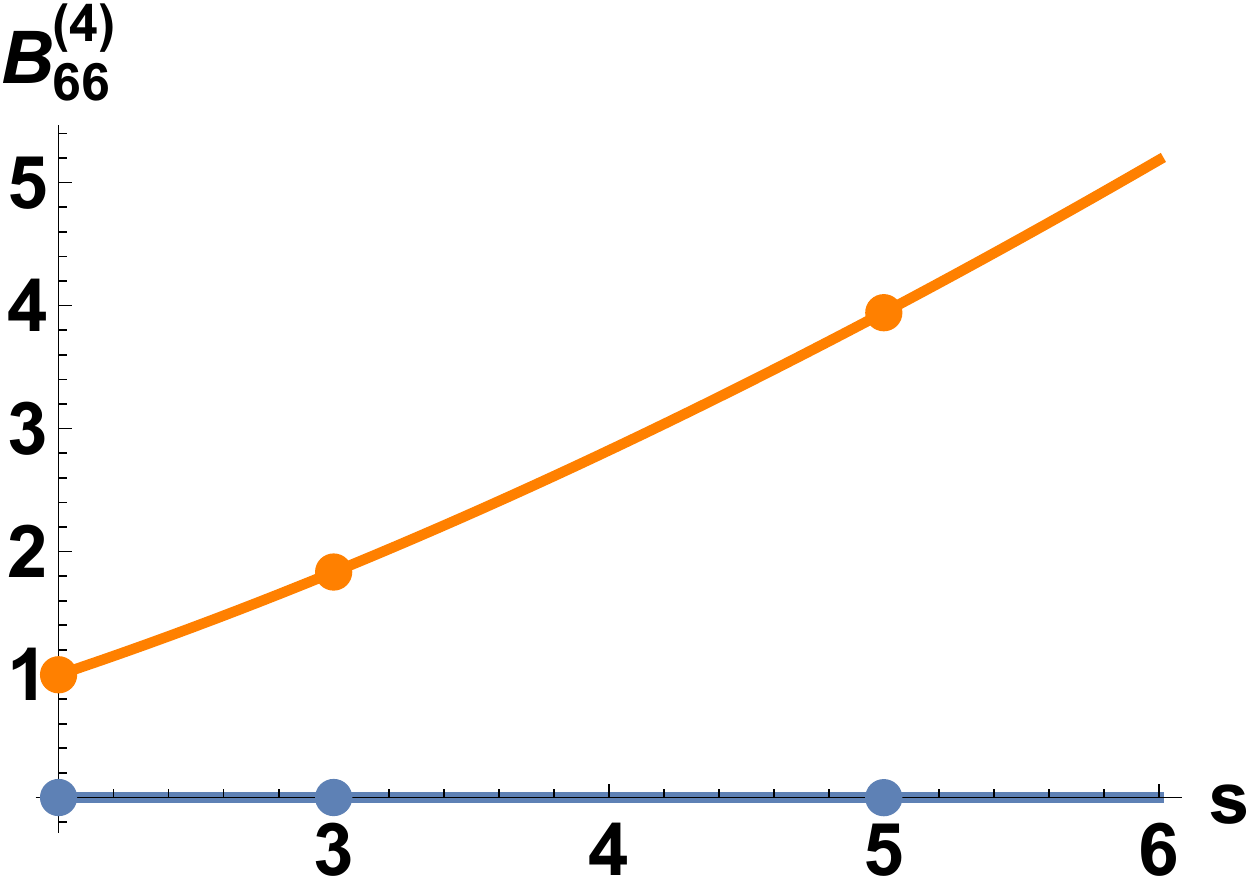} \\
\img{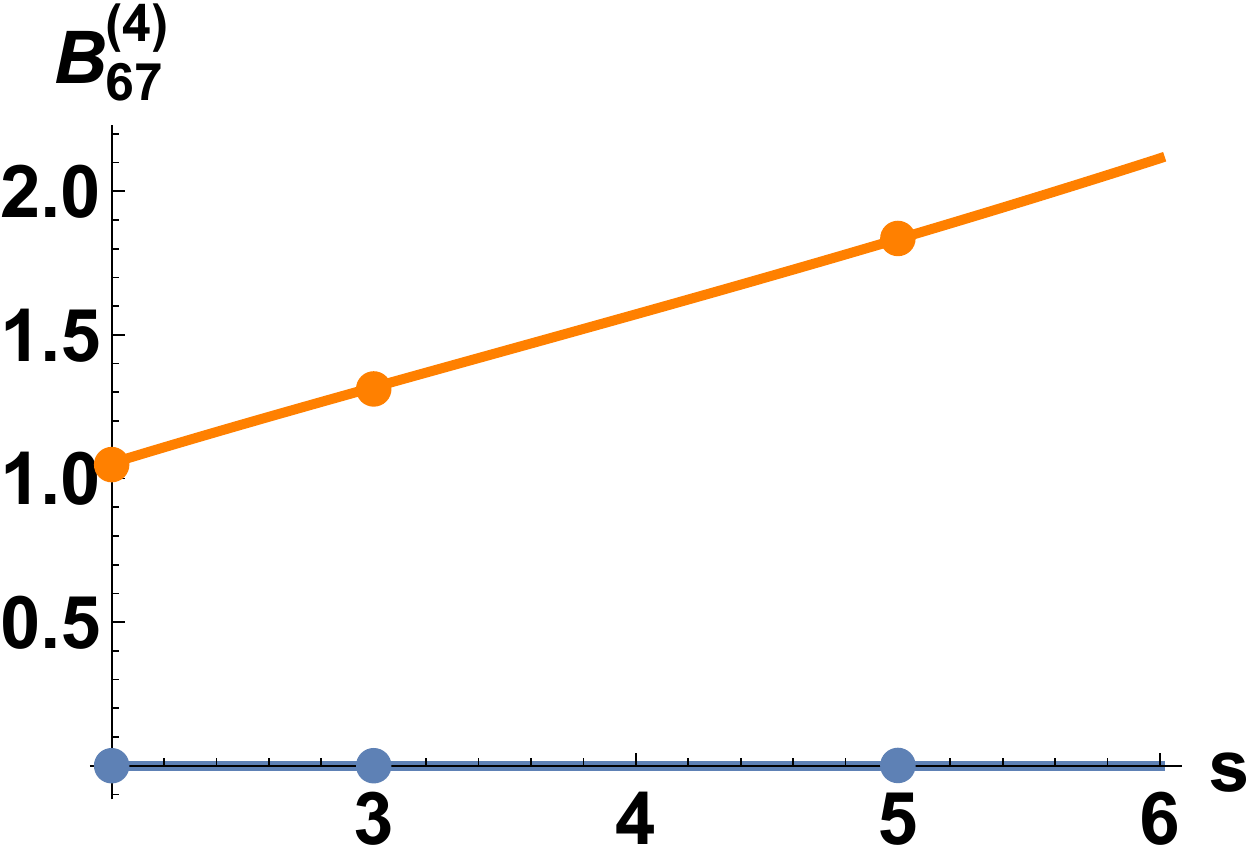} & \img{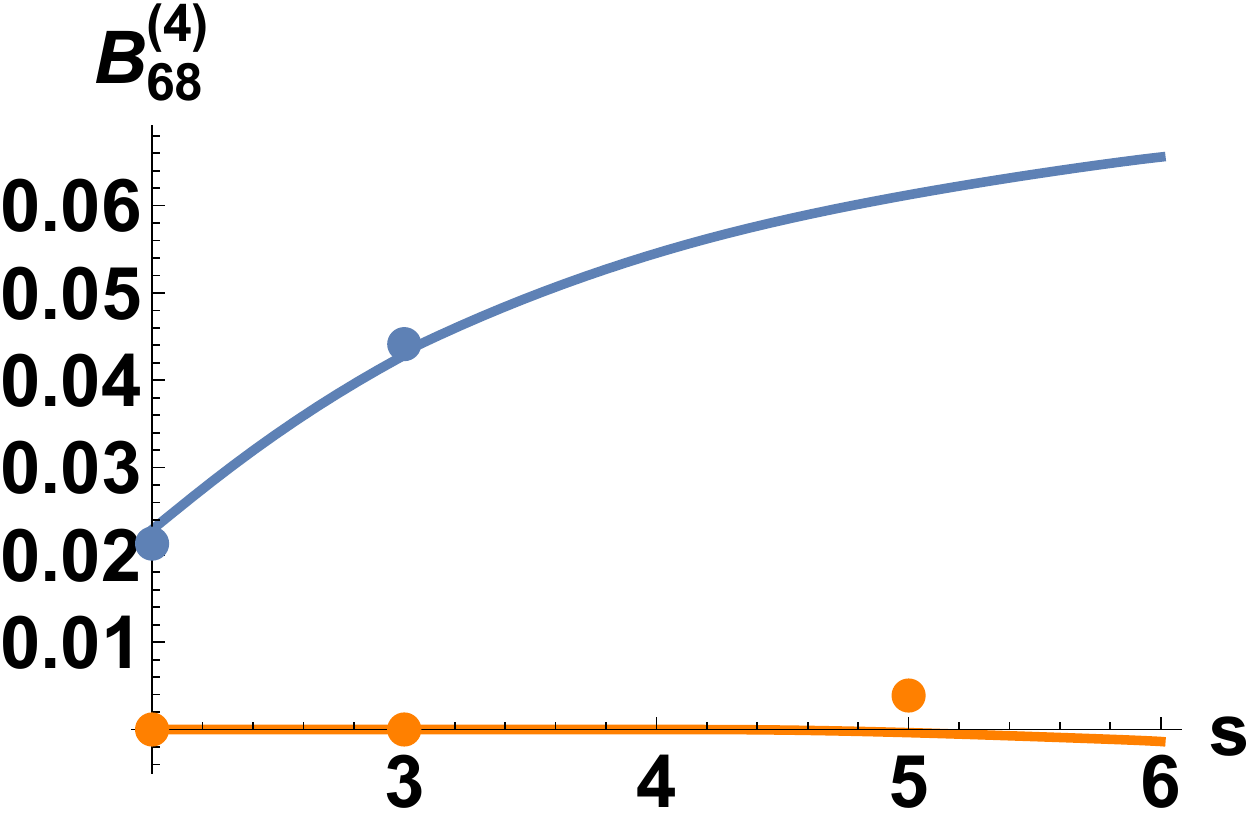} & \img{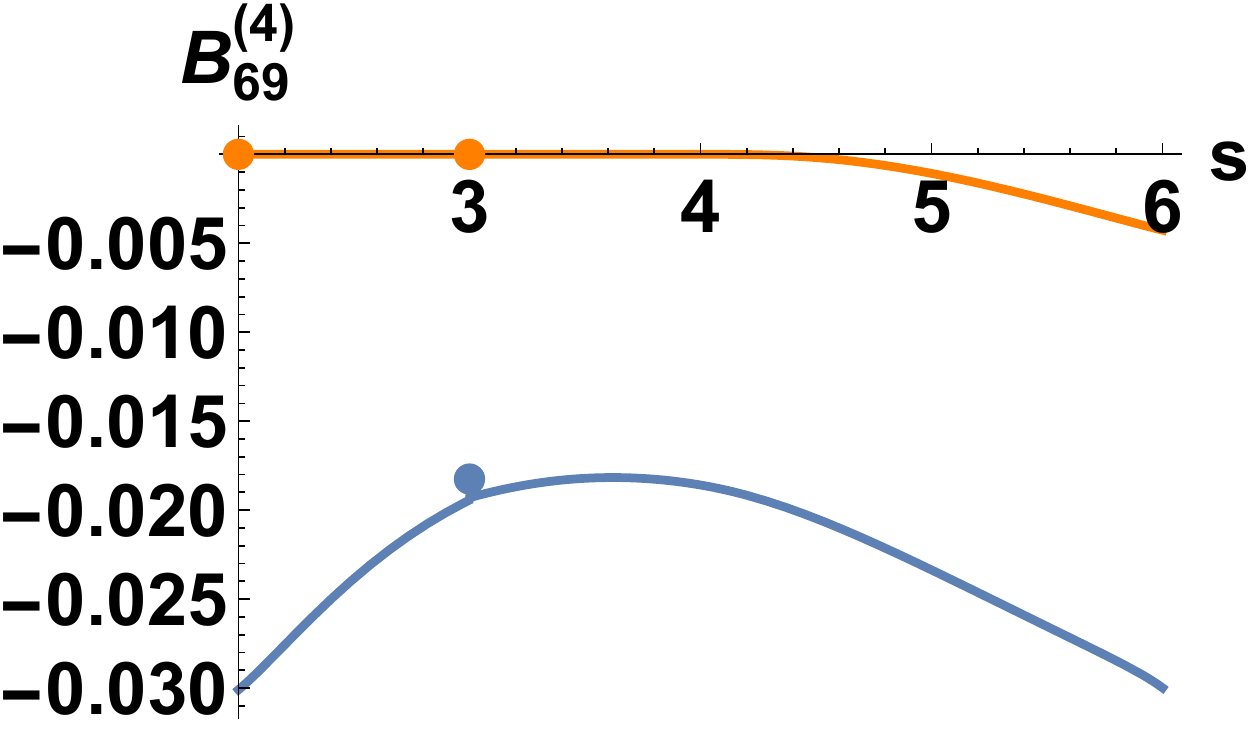} \\
\img{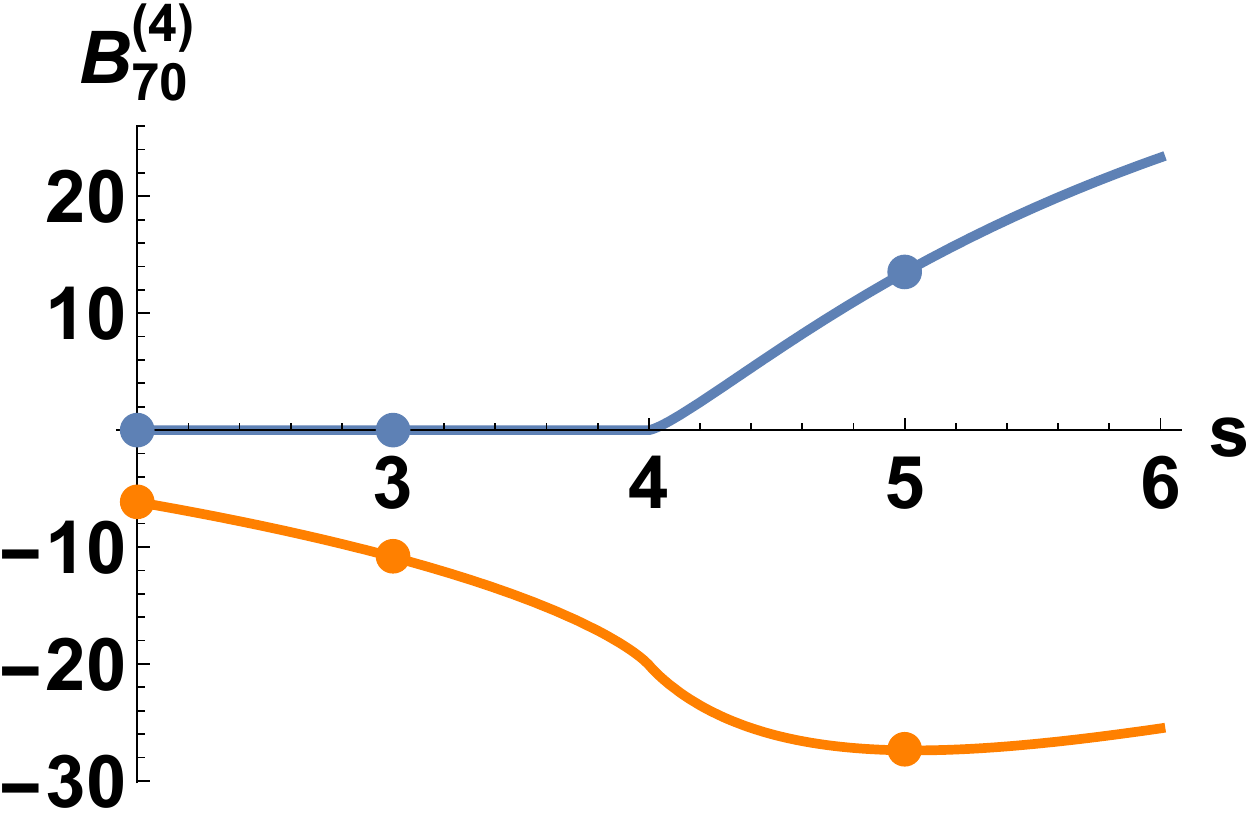} & \img{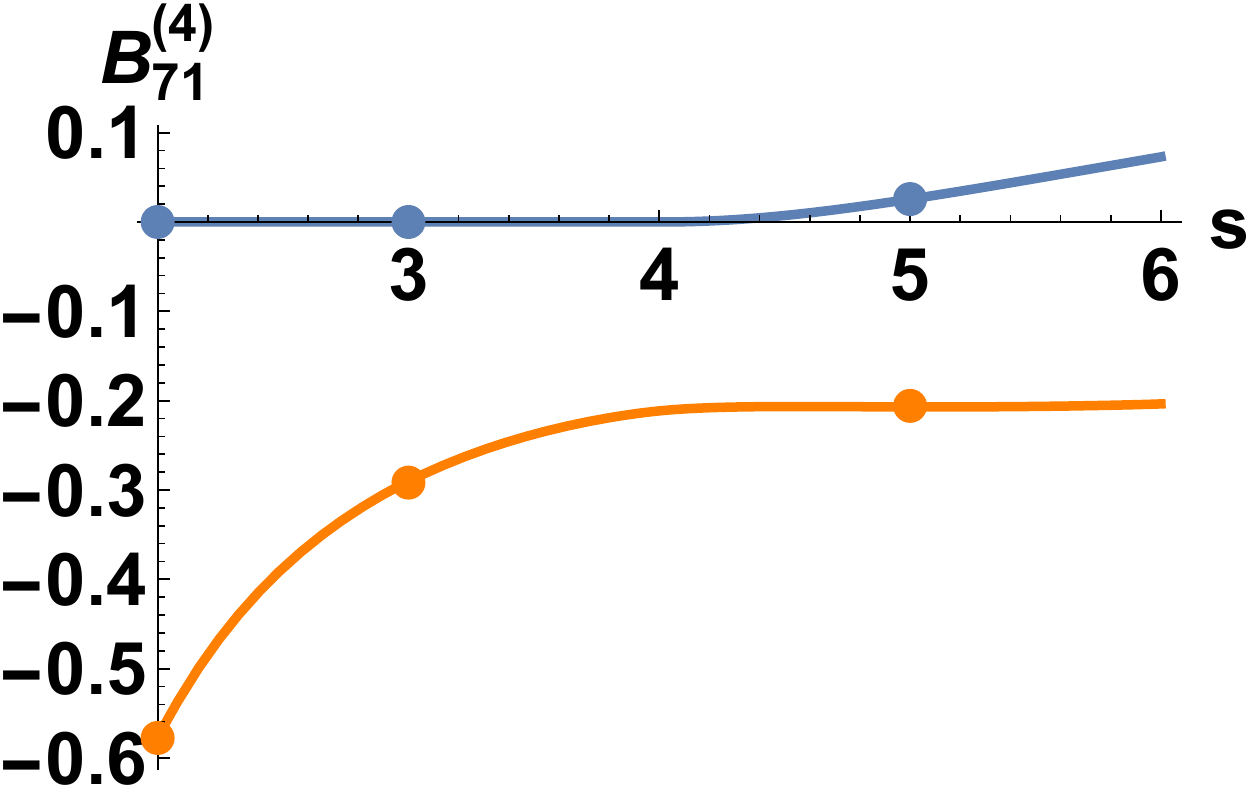} & \img{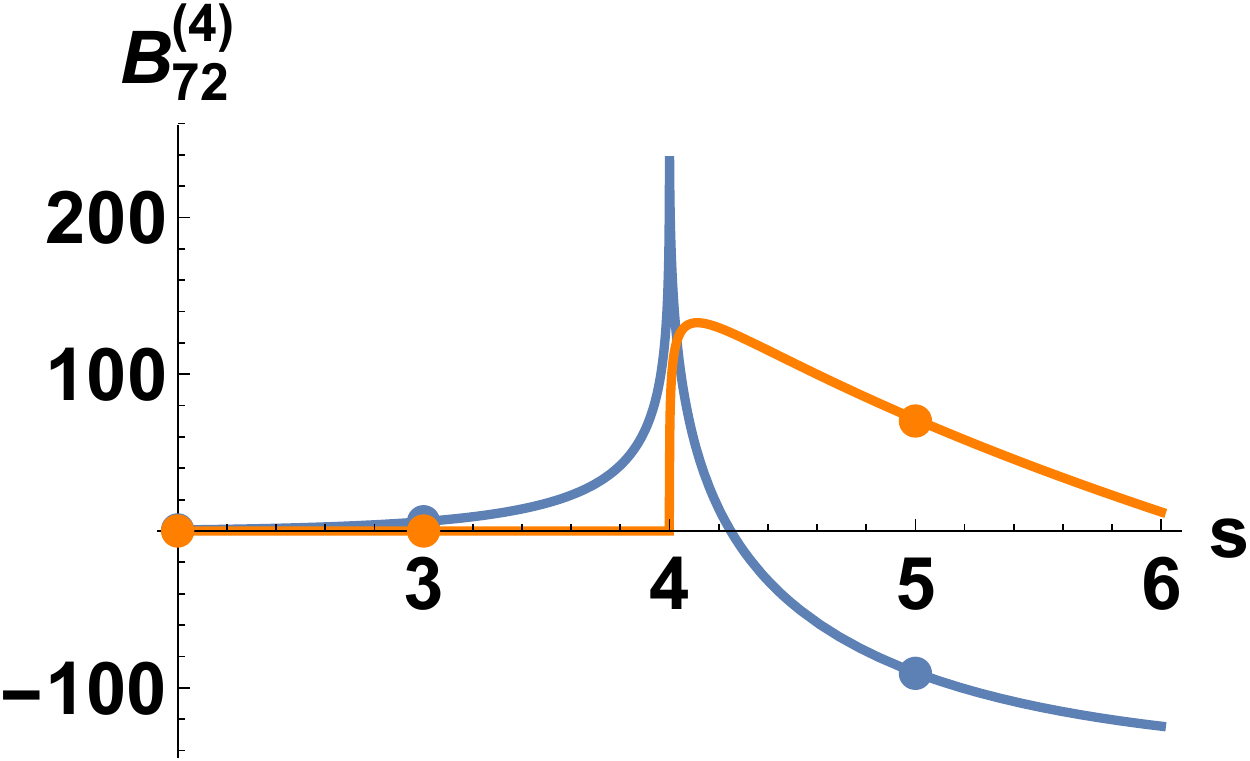} 
    
\end{tabular}
\end{table}
}
\FloatBarrier

\bibliographystyle{JHEP}
\bibliography{refs}

\end{document}